\documentclass[fleqn,usenatbib]{mnras}

\usepackage{newtxtext,newtxmath}

\usepackage[T1]{fontenc}

\DeclareRobustCommand{\VAN}[3]{#2}
\let\VANthebibliography\thebibliography
\def\thebibliography{\DeclareRobustCommand{\VAN}[3]{##3}\VANthebibliography}

\usepackage{graphicx}	
\usepackage{amsmath}	

\usepackage{amssymb}	
\usepackage{hyperref}
\usepackage{multirow}
\usepackage{rotating}
\usepackage[flushleft]{threeparttable}
\usepackage{subfloat}
\usepackage{xcolor}
\usepackage{longtable, makecell}

\usepackage{siunitx}
\usepackage{tablefootnote}
\usepackage{footnote}
\usepackage{float}
\usepackage{placeins}
\usepackage{afterpage}
\usepackage{perpage}
\usepackage{pdflscape}
\usepackage{threeparttablex}

\newcommand{\hii}{H~\textsc{ii}}
\newcommand{\msun}{$\rm M_\odot$}

\newcommand{\kms}{km~s$^{-1}$}

\newcommand{\hmole}{H$_2$}

\newcommand{\cm}{cm$^{-2}$}
\newcommand{\cmcube}{cm$^{-3}$}

\newcommand{\htcop}{H$^{13}$CO$^+$}
\newcommand{\htcopone}{H$^{13}$CO$^+$~$J=1-0$}

\newcommand{\hfourtyalpha}{H40$\alpha$}
\newcommand{\chtoh}{CH$_3$OH}

\newcommand{\cch}{CCH}
\newcommand{\cchline}{CCH~$N_{J,F}=1_{3/2,2}-0_{1/2,1}$}

\newcommand{\mclump}{$M_{\rm clump}$}

\newcommand{\RGC}{$R_{\rm GC}$}

\newcommand{\rvir}{$R_{\rm vir}$}

\newcommand{\mcore}{$M_{\rm core}$}
\newcommand{\sufdenscore}{$\Sigma_{\rm core}$}
\newcommand{\rcoreeff}{$r_{\rm core,eff}$}

\newcommand{\tkin}{$T_{\rm kin}$}
\newcommand{\tex}{$T_{\rm ex}$}

\newcommand{\tr}{$T_{\rm r}$}
\newcommand{\te}{$T_{\rm e}$}
\newcommand{\pressurei}{$P_{\rm i}$}

\newcommand{\RecRate}{$\dot{N}_{\rm rec}$}
\newcommand{\IonRate}{$\dot{N}_{\rm ion}$}
\newcommand{\nelectron}{$n_{\rm e}$}
\newcommand{\tdyn}{$t_{\rm dyn}$}
\newcommand{\ci}{$c_{\rm i}$}
\newcommand{\nini}{$n_{\rm ini}$}
\newcommand{\ahfourtya}{$A_{\rm H40\alpha}$}
\newcommand{\lhfourtya}{$L_{\rm H40\alpha}$}
\newcommand{\fhfourtya}{$\nu_{\rm H40\alpha}$}
\newcommand{\rhfourtya}{$r_{\rm H40\alpha,eff}$}

\newcommand{\htcopab}{$\chi_{\rm H^{13}CO^+}$}

\newcommand{\ceotwo}{C$^{18}$O $J=2-1$}

\newcommand{\dustt}{$T_{\rm dust}$}
\newcommand{\nhtnd}{$n_{\rm H_{2}}$}
\newcommand{\nhtcdpeak}{$N_{\rm H_{2}}^{\rm peak}$}

\title[dense fragments in the shell]{ATOMS: ALMA three-millimeter observations of massive star-forming regions -- XVIII. On the origin and evolution of dense gas fragments in molecular shells of compact \hii\ regions}

\author[Siju Zhang et al.]{Siju Zhang,$^{1,2}$\thanks{E-mail: sijuzhangastro@gmail.com (SZ)}
Tie Liu,$^{3,4}$
Ke Wang,$^{1}$\thanks{E-mail: kwang.astro@pku.edu.cn (KW)}
Annie Zavagno,$^{5,6}$
Guido Garay,$^{2,7}$
Hongli Liu,$^{8}$
Fengwei Xu,$^{1,9}$ 
\newauthor
Xunchuan Liu,$^{3}$
Patricio Sanhueza,$^{10,11}$
Archana Soam,$^{12}$
Jian-wen Zhou,$^{13}$
Shanghuo Li,$^{14}$
\newauthor
Paul F. Goldsmith,$^{15}$
Yong Zhang,$^{16}$
James O. Chibueze$^{17,18}$
Chang Won Lee,$^{19,20}$ 
Jihye Hwang,$^{19,20}$ 
\newauthor
Leonardo Bronfman,$^{2}$ 
and Lokesh K. Dewangan,$^{21}$
\\
Affiliations are listed at the end of the paper}

\date{Accepted 2024 October 18. Received 2024 October 18; in original form 2024 June 6}

\pubyear{2015}

\begin{document}
\label{firstpage}
\pagerange{\pageref{firstpage}--\pageref{lastpage}}
\maketitle

\begin{abstract}
Fragmentation and evolution for the molecular shells of the compact \hii\ regions are less explored compared to their evolved counterparts. We map nine compact \hii\ regions with a typical diameter of 0.4~pc that are surrounded by molecular shells traced by \cch. Several to a dozen dense gas fragments probed by \htcop\ are embedded in these molecular shells. These gas fragments, strongly affected by the \hii\ region, have a higher surface density, mass, and turbulence than those outside the shells but within the same pc-scale natal clump. These features suggest that the shells swept up by the early \hii\ regions can enhance the formation of massive dense structures that may host the birth of higher-mass stars.  We examine the formation of fragments and find that fragmentation of the swept-up shell is unlikely to occur in these early \hii\ regions, by comparing the expected time scale of shell fragmentation with the age of \hii\ region. We propose that the appearance of gas fragments in these shells is probably the result of sweeping up pre-existing fragments into the molecular shell that has not yet fragmented. Taken together, this work provides a basis for understanding the interplay of star-forming sites with an intricate environment containing ionization feedback such as those observed in starburst regions.
\end{abstract}

\defcitealias{ATOMSI}{ATOMS~I}
\defcitealias{ATOMSIII}{ATOMS~III}
\defcitealias{ATOMSIV}{ATOMS~IV}
\defcitealias{ATOMSVI}{ATOMS~VI}
\defcitealias{ATOMSXIII}{ATOMS~XIII}
\defcitealias{ATOMSXI}{ATOMS~XI}
\defcitealias{ATOMSVIII}{ATOMS~VIII}
\defcitealias{ATOMSXIV}{ATOMS~XIV}




\section{Introduction} \label{SECTION:INTRODUCTION}

Various energetic feedback processes including radiation, stellar winds, outflows, jets, and supernova explosions are associated with the process of high-mass star formation (HMSF). Among them, ionizing radiation and stellar winds powered by high-mass stars are thought to be the dominant contributors before being taken over by supernova events \citep{Walch2023}. These feedback processes remarkably modify the physical and chemical states of the natal molecular cloud, raising the question of whether and how future star formation in these feedback-driven clouds changes. To this end, the ALMA project ATOMS (ALMA three-millimeter observations of massive star-forming regions, project code: 2019.1.00685.S, \citealt[][hereafter \citetalias{ATOMSI}]{ATOMSI}) surveyed 146 massive and dense star-forming clumps \citep{Faundez2004} in ALMA 3~mm (Band 3) with the main array and 7~m array.  From the radio recombination line, \hfourtyalpha, of the survey, \citet[][hereafter \citetalias{ATOMSIII}]{ATOMSIII} demonstrated that the majority of the ATOMS clumps contain a series of early \hii\ regions from hyper compact (HC) to compact stages, showing the potential of the ATOMS sample to explore early ionization feedback.

The ATOMS project has revealed several significant results on how ionization feedback influences the surroundings. The first result is the predominance of ionization feedback in the overall state of the natal clumps. \citetalias{ATOMSIII} found that the luminosity-to-mass ratio (L/M) of the ATOMS clump, an empirical parameter of evolution \citep{Molinari2016}, has a noticeable increase when natal clumps host HC or ultra compact (UC) \hii\ regions. \citet[][hereafter \citetalias{ATOMSIV}]{ATOMSIV} further confirmed that ATOMS \hfourtyalpha\ emission samples well the upper part of the initial mass function and, therefore, underscores the connection between ionization and star formation of the natal clump. The second result is that the external ionization feedback influences not only the clump-scale conditions but also the core-scale conditions for HMSF. For example, \citet[][\citetalias{ATOMSVIII}]{ATOMSVIII} revealed that some candidate hot molecular cores identified using typical organic molecules are actually dense regions externally heated by \hii\ regions rather than real hot cores that are internally heated by embedded protostars. The third result is the active role of ionization feedback in compression and dispersal of dense gas of the natal clump. \citet[][\citetalias{ATOMSVI}]{ATOMSVI} suggested that large-scale compression flows driven by the expansion of nearby \hii\ regions probably formed the filaments embedded in the ATOMS clump G286.21$+$0.17. A recent analysis of \citet[][hereafter \citetalias{ATOMSXIII}]{ATOMSXIII} of the ATOMS clump IRAS 18290$-$0924 found that compression of the external \hii\ region is inducing HMSF in the clump. On the other hand, \citet[][\citetalias{ATOMSXI}]{ATOMSXI} revealed the destructive effect of expanding \hii\ regions on the hub-filament systems (HFSs), which can be the main HMSF sites. These authors found that under the effect of gas dispersal, the fraction of protoclusters containing HFSs decreases with the L/M ratio (evolutionary stage) of  natal clump.

How the transformation efficiency of gas-to-star and the birth speed of stars, known as the star formation efficiency and rate (SFE and SFR), are regulated by the ionization feedback is essential to understand HMSF and model the evolution of a galaxy \citep{Hopkins2014, Chevance2023,Zakardjian2023}.  Although there are numerous simulations targeting ionization feedback on SFR or SFE, it is still unclear whether, in general, ionization feedback plays a more positive role, which assists or accelerates the formation of dense gas structures by compression \citep{Bending2020, Wall2020,Ali2022, Hennebelle2022,Herrington2023, Suin2024}, or it plays a more negative role, which disperses the dense gas available for star formation \citep{Peters2017,Gonzalez2020,Rathjen2021, Dobbs2022, Fukushima2022, Verliat2022}. More and more studies show that a comprehensive picture of the feedback on star formation requires considering physical scales, evolutionary stages, and cloud structures of the observed or simulated star formation regions \citep{Beuther2022}.

The molecular shells at the edges of \hii\ regions, as specific dense cloud structures that have been profoundly shaped by ionization feedback, are ideal test beds for studying how feedback of high-mass stars influences the formation of next-generation stars. These shells could form from the gas collected during expansion of \hii\ regions, or from the pre-existing filamentary gas structures that contain and are being reshaped by expansion of \hii\ regions \citep{Elmegreen1998, Pineda2022}. Previous studies have used single-dish observations to investigate various properties of \textit{large-scale} molecular shells with a size of \textit{a few to a dozen parsecs}, such as their density distribution \citep{Tremblin2014}, turbulence \citep{Mazumdar2021}, magnetic field \citep{Tahani2023}, and star formation activities \citep{Sherman2012,Palmeirim2017}, confirmed that the dense gas of these large shells provides an initial condition for HMSF completely different from those in a feedback-free and quiescent environment \citep{PaperI}. The gas-collected shell may become dense enough to fragment and produce massive dense clumps that can host HMSF \citep{Elmegreen1977}. Alternatively, compression of ionized gas on clumps in pre-existing filamentary clouds could cause them to collapse and form new stars \citep{Bertoldi1989}. \citet{Palmeirim2017} detected an enhanced star formation in the vicinity of these \textit{large-scale shells}, suggesting that \hii\ regions may have a positive effect on star formation there. However, there is a lack of millimeter observations, especially for statistically meaningful surveys of an unbiased sample, which can resolve star formation down to the core scale ($<0.1$~pc) for \textit{smaller-scale} molecular shells produced by early compact \hii\ regions with a size of \textit{a few tenths of a parsec} and an age of \textit{a few tenths of a million years}. Investigating these ``young'' molecular shells is extremely important to better understand the evolution of feedback-driven structures and how ionization feedback shapes the subsequent formation of stars.  

The ATOMS project provides us with a unique opportunity to analyze these small-scale molecular shells systematically and statistically. In this paper, our objective is to study dense gas fragments in molecular shells driven by early compact \hii\ regions.  Our ALMA observations and \hii\ region sample are described in Section~\ref{SECTION:OBS-AND-SAMPLE}. The basic properties of the \hii\ regions are derived in Section~\ref{SECTION:HII-REGION}. Dense gas fragments are extracted and analyzed in Section~\ref{SECTION:DENSE-GAS}. The influence of \hii\ region on the properties of gas fragments is discussed in Section~\ref{SECTION:INFLUENCE}. The relationship between the molecular shell swept-up by the \hii\ region and the gas fragments is analyzed in Section~\ref{SECTION:FRAGMENTATION}. In Section~\ref{SECTION:SCENARIO}, we propose an evolutionary scenario to show the nature and origins of gas fragments in shells. A short summary is presented in Section~\ref{SECTIONS:CONCLUSIONS}.

\section{Observations and sample selections}\label{SECTION:OBS-AND-SAMPLE}
The ATOMS data reduction was performed using \texttt{Common Astronomy Software Applications 5.6} \citep[\texttt{CASA,}][]{Mcmullin2007}. The main array and 7~m array data were combined and then cleaned using the \texttt{CASA} task \texttt{TCLEAN} with natural weighting. The pixel size was set to 0.4\arcsec\ in the imaging process, which is about a fifth of the synthesized beam $\sim2\arcsec$. The imaged field was reduced to the 20\% power point of the primary beam response, equivalent to a circular region with a radius of 44\arcsec. Eight spectral windows (SPWs) were configured in the ATOMS ALMA observations, including six narrow SPWs and two wide SPWs. Table~\ref{TABLE:LINES} provides information on the SPWs that contain the transitions used in this work. The rms sensitivities of spectral observations are around 0.2~K for \cch\ and \htcop, and $<0.1$~K for \hfourtyalpha. Further details of the ATOMS data reduction can be found in the ATOMS series.

\begin{table}
\caption{\label{TABLE:LINES}Molecules and their transitions investigated in this work.} 
\begin{threeparttable}
\renewcommand{\arraystretch}{1.0}
\centering
\begin{tabular}{rrrrr}
\hline
\hline
transition                   &       $\nu$   &    channel width      &     rms\tnote{\textit{(a)}}   & SPWs\tnote{\textit{(b)}}      \\
                                       &       GHz      &     \kms     &       K    &           \\
\hline
H$^{13}$CO$^+$~$J=1-0$                 &       86.7543  &   0.21         &    0.15-0.23         &   SPW2      \\   
CCH~$N_{J,F}=1_{3/2,2}-0_{1/2,1}$      &       87.3169  &   0.21         &    0.15-0.24         &   SPW3       \\    
H40$\alpha$                            &       99.0230  &   1.48         &    0.05-0.10         &   SPW7       \\  
\hline
   \end{tabular}
      \begin{tablenotes}
      \item [\textit{(a)}] Cube without primary-beam (PB) correction is used in the rms estimate.
      \item [\textit{(b)}] More detailed information on SPWs is provided in \citetalias{ATOMSI}.
      \end{tablenotes}
      \end{threeparttable}
\end{table}

For the purpose of studying molecular shells and their dense gas surrounding compact \hii\ regions, a thorough examination on the spatial distributions of \cch\ and \hfourtyalpha\ emission was performed toward the ALMA images of all 146 ATOMS sources. ATOMS observations are capable of resolving distant \hii\ regions at the compact stage ($\gtrsim0.1$~pc) up to around 10~kpc with a wide field of view (FoV).  The selection of molecular shells in this article is based on two criteria: 1) A prominent \cch\ shell with a high signal-to-noise ratio (SNR) is at the edges of \hfourtyalpha\ emission of \hii\ region. \cch\ is thought to be a sensitive tracer for photodissociation region \citep[PDR,][]{Cuadrado2015, Kirsanova2021} and therefore well represents the borders where UV photons actively interact with the gas of molecular shells; 2) \hfourtyalpha\ emission is well resolved in the FoV of ATOMS data.  Too extended or too small \hii\ regions such as HC\hii\ regions presented in \citetalias{ATOMSIII}, are not taken into account in this work as their molecular shell structures are too noisy to analyze or too small to resolve. A total of nine compact \hii\ regions are selected and information on their natal clumps is presented in Table~\ref{TABLE:SAMPLE}.

\begin{table*}
\caption{\label{TABLE:SAMPLE}Natal clump information of selected sources.} 
\begin{threeparttable}
\setlength{\tabcolsep}{4.5pt}
\renewcommand{\arraystretch}{1.0}
\centering
\begin{tabular}{cccccccccccc}
\hline
\hline
name\tnote{\textit{(a)}} &   \mclump    &    \dustt & \nhtcdpeak  &  $r$\tnote{\textit{(b)}}  & GCSC FWHM \tnote{\textit{(c)}} & $D$ &  $c_{\rm s}$\tnote{\textit{(e)}}  &  $\sigma_{\rm H^{13}CO^+}$\tnote{\textit{(f)}}  &  $\sigma_{\rm C^{18}O}$\tnote{\textit{(f)}}            &     \multicolumn{2}{c}{$n_{\rm ini}$\tnote{\textit{(g)}}}   \\
                         & log \msun  &      K    & log $\rm cm^{-2}$ &    pc                              & pc                & kpc &   \kms   &  \kms  & \kms   &     \multicolumn{2}{c}{$10^4~\rm cm^{-3}$}  \\
\hline
                &           &           &             &          &                   &         &   & &     &  \textit{i} & \textit{ii} \\ 
\hline
I13080$-$6229   &   3.2    &       34.1 &   22.81     &      1.03&     0.63          & 3.80  & 0.34(0.02) & 1.36(0.16)    &  2.62(0.09)   & 1.00(0.36)  & 3.32(0.74)    \\   
I15411$-$5352   &   2.7    &       30.5 &   23.08     &      0.57&     0.24          & 1.82  & 0.33(0.02) & 1.30(0.14)    &   -   & 1.85(0.67)  & 16.3(3.7)   \\  
I15570$-$5227   &   3.4    &       28.7 &   22.48     &      1.66&     0.99          & 5.99  & 0.32(0.02) &  -   &  1.07(0.04)    & 0.35(0.13)  & 0.99(0.22)   \\   
I15584$-$5247   &   3.1    &       23.9 &   22.74     &      0.86&     0.75          & 4.41  & 0.29(0.01) & 0.98(0.09)   &  1.29(0.04)    & 1.48(0.53)  & 2.36(0.53)  \\   
I16362$-$4639   &   2.5    &       24.0 &   22.38     &      0.54&     0.58          & 3.01  & 0.29(0.01) & 1.18(0.51)    &  1.73(0.09)    & 1.26(0.46)  & 1.34(0.30)  \\   
I17006$-$4215   &   2.8    &       27.7 &   22.99     &      0.50&     0.28          & 2.21  & 0.31(0.02) &  1.25(0.13)   &  -    & 3.33(1.20)  & 11.5(2.6) \\   
I17160$-$3707   &   3.6    &       28.5 &   23.03     &      1.00&     0.87          & 6.20\tnote{\textit{(d)}} & 0.32(0.02) &  1.60(0.12)   &  2.17(0.06)    & 3.14(1.13)  & 3.99(0.89)    \\   
I18116$-$1646   &   3.1    &       33.8 &   22.83     &      0.99&     0.57          & 3.94  & 0.34(0.02) & 1.19(0.26)    &  1.22(0.04)    & 0.88(0.32)  & 3.85(0.86) \\   
I18317$-$0757   &   3.4    &       30.4 &   22.60     &      1.93&     0.88          & 4.79  & 0.33(0.02) &  -    &   -    &  0.25(0.09) & 1.45(0.32) \\   
\hline
   \end{tabular}
      \begin{tablenotes}
      \item [\textit{(a)}] Clump mass \mclump, dust temperature \dustt, peak column density \nhtcdpeak, and distance $D$ are given by \citet{Urquhart2018} based on ATLASGAL source catalog, with typical uncertainties of 20~\%, 20~\%, 10~\% and 0.3~kpc, respectively, according to the estimate of \citet{Urquhart2018}.
      \item [\textit{(b)}] \texttt{SExtractor} radius $r$ of the ATLASGAL clump measured by \citet{Urquhart2014} with the algorithm \texttt{SExtractor}, equal to $2.4\times$ deconvolved standard deviation size given by \texttt{SExtractor}.
      \item [\textit{(c)}] GCSC (\texttt{GaussClump} Source Catalogue) FWHM is the ATLASGAL clump FWHM measured independently with algorithm \texttt{GaussClump} by \citet{Csengeri2014}.
      \item [\textit{(d)}] Clump properties of I17160$-$3707 are corrected by a more reliable distance from the detailed case studies of \citet{Nandakumar2016}.
      \item [\textit{(e)}] Sound speed derived using \dustt.
      \item [\textit{(f)}] Clump-scale velocity dispersion measured from Millimetre Astronomy Legacy Team 90~GHz survey \citep[MALT90, ][]{Jackson2013} \htcopone\ and Structure, Excitation, and Dynamics of the Inner Galactic InterStellar Medium survey \citep[SEDIGISM, ][]{Schuller2017} \ceotwo.  
     \item [\textit{(g)}] Clump initial undisturbed number densities of hydrogen nuclei in situations \textit{i} and \textit{ii} are estimated based on two ATLASGAL source extraction techniques described in Section~\ref{SECTION:HII-REGION}.
      \end{tablenotes}
      \end{threeparttable}
\end{table*}

\section{properties of compact \hii\ regions} \label{SECTION:HII-REGION}
\begin{figure*}
\centering
\includegraphics[width=0.99\textwidth]{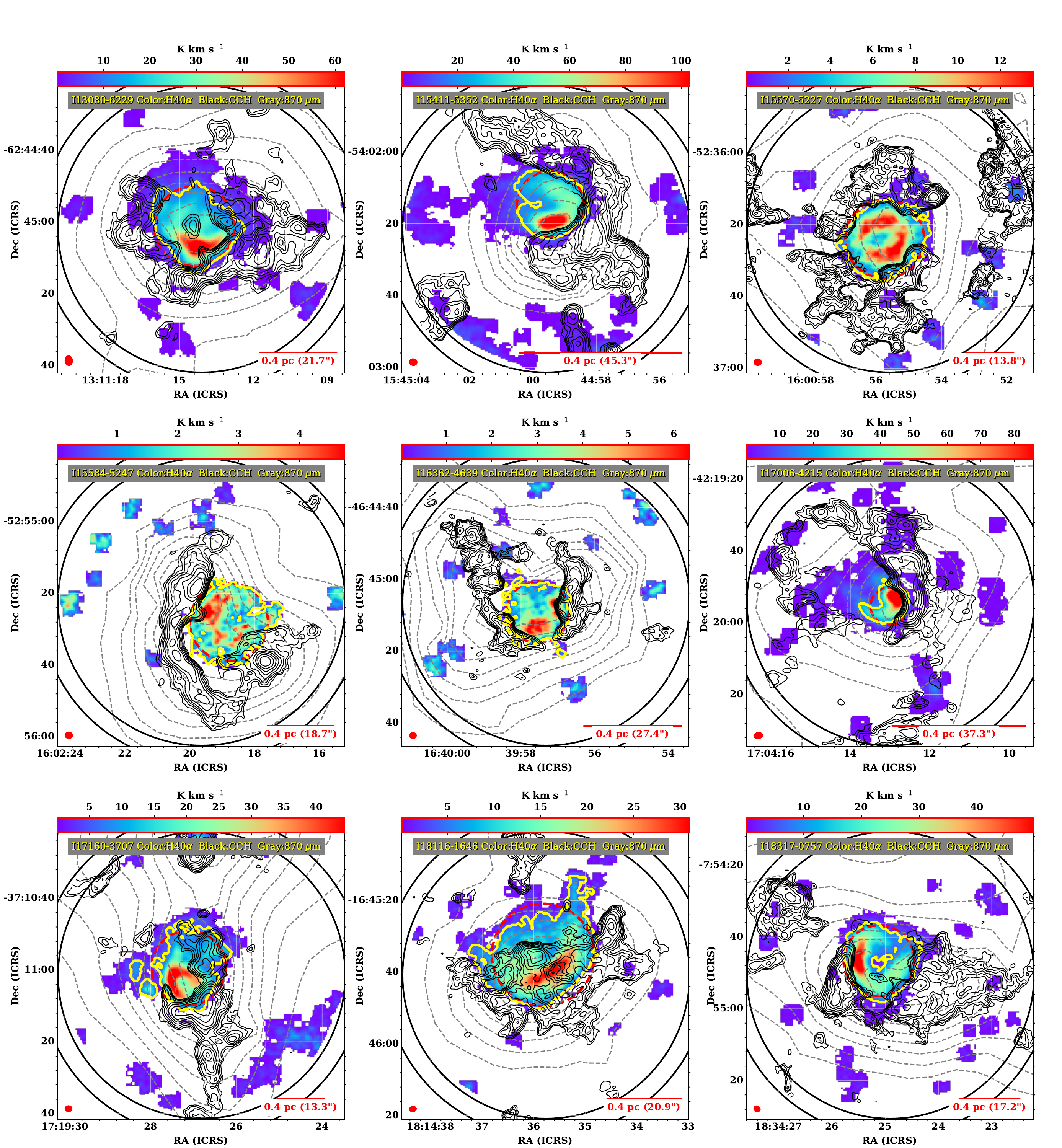}
       \caption{Compact \hii\ regions in this work. The color images display the \hfourtyalpha\ moment 0 maps created using the technique described in \citet{Dame2011}, with a channel clipping level of $4\times\rm rms$. The black solid contours and the gray dashed contours represent the \cch\ emission and the ATLASGAL 870~\micron\ emission \citep{Schuller2009}, respectively. The contour levels start from 20\% of the maximum to the maximum with eight evenly spaced steps on a linear scale and on a logarithmic scale, respectively. The yellow contours and the red circles indicate the edges of \hii\ regions and their corresponding circular sizes $r_{\rm H40\alpha,eff}$, as described in Section~\ref{SECTION:HII-REGION}. All moment 0 maps shown here have been corrected by primary beam (PB). Two black circles indicate the 20\% and 30\% power points of PB, respectively. Note that the \cchline\ line of I15570$-$5227 and I17160$-$3707 is located on the border of the spectral window and therefore only a part of the \cchline\ emission channels are actually integrated.}
        \label{FIGURE:NORMAL1}
\end{figure*}

Moment 0 maps of \cchline\ and \hfourtyalpha\ for nine selected regions are shown in Fig.~\ref{FIGURE:NORMAL1}. All regions, apart from I18116$-$1646, have evident molecular shells. The well-defined cometary shape of \hfourtyalpha\ emission in I18116$-$1646 strongly suggests the existence of a molecular shell, but this shell is not visible probably due to projection effect. To obtain a preliminary impression of our selected regions, we first measure basic properties of ionized gas using \hfourtyalpha\ emission.

\textit{1)} Projected size of the \hii\ region (\ahfourtya). A simple Gaussian fit is not appropriate to measure the size of our \hii\ regions in the \hfourtyalpha\ moment 0 maps due to the intricate morphology of \hfourtyalpha\ emission, which is classified as cometary or shell-like according to the definitions of \citet{Wood1989}. Furthermore, some regions show UC \hfourtyalpha\ cores with an extended and diffuse background, such as I15411$-$5352 \citep{delaFuente2020}, indicating a highly inhomogeneous distribution of ionized gas within the \hii\ region. The \cch\ molecular shell that is partially containing and interacting with \hii\ region provides important constraints on the size of \hii\ region. In the \hfourtyalpha\ moment 0 maps produced from the data cube with a clipped threshold of ${4\rm rms}$, we further mask the pixels with a value below 10\% of the image peak. The resulting edges of the \hfourtyalpha\ emission regions are shown as yellow contours in Fig.~\ref{FIGURE:NORMAL1}. They are well correlated with the edges of the interacting molecular shells. Therefore, we use the area corresponding to these yellow contours as the projected sizes of the \hii\ region \ahfourtya. Small sub \hii\ regions near the central one are not considered, such as the two tiny \hii\ regions located at the east of the central \hii\ region of I17160$-$3707.

\textit{2)} Ionization rate (\IonRate). With the hypothesis that the equilibrium between ionization and recombination has been reached and helium ionization is ignored, the recombination rate \RecRate\ is equal to the ionization rate \IonRate\ and then \IonRate\ can be estimated using the methods in \citetalias{ATOMSIV} and \citet[][\citetalias{ATOMSXIV} hereafter]{ATOMSXIV}
\begin{equation} \label{EQU:RECOMBINATION}
    \dot{N}_{\rm ion} = \dot{N}_{\rm rec} = L_{\rm H40\alpha} \frac{\nu_{\rm H40\alpha}}{c} \frac{\alpha_{\rm B}}{\epsilon},
\end{equation}
where \lhfourtya, \fhfourtya, $\alpha_{\rm B}$ and $\epsilon$ are the luminosity and frequency of \hfourtyalpha, the Case B total recombination coefficient, and the efficiency factor, respectively. Case B means that the ionizing photons emitted during recombination are immediately reabsorbed and then produce another ion and free electron through photoionization \citep{Barker1938}.The efficiency factor $\epsilon$ is calculated with
\begin{equation} \label{EQU:EPSILON}
    \epsilon = b_{ul} f_{ul}\left({\rm LTE}\right) A_{ul} h \nu_{\rm H40\alpha},
\end{equation}
where $b_{ul}$ is the departure coefficient, $A_{ul}$ is the spontaneous emission coefficient, $f_{ul}\left({\rm LTE}\right)$ is a factor related with the fractional population of upper level and it can be determined from Saha-Boltzmann ionization equation \citep{Gordon2002}. For an assumed electron temperature $T_{\rm e} = 10^4~{\rm K}$ and an electron number density $n_{\rm e} = 10^4~{\rm cm^{-3}}$, the corresponding $\epsilon = 1.99\times10^{-32}~{\rm cm^{3}}~{\rm erg}~{\rm s}^{-1}$ when taking $b_{ul} = 0.788$, $f_{ul}\left({\rm LTE}\right) = 7.03\times10^{-19}~{\rm cm^3}$, and $A_{ul} = 54.7~{\rm s^{-1}}$. At the same time, the corresponding $\alpha_{\rm B}= 2.54\times10^{-13}~{\rm cm}^{3}~{\rm s}^{-1}$ \citepalias{ATOMSIV}.  \citetalias{ATOMSIV} found that $\alpha_{\rm B}$ and $\epsilon$ are unlikely to significantly alter resultant \RecRate\ because there is only a variation of $\simeq10\%$ when \nelectron\ and \te\ are as low as 100~\cmcube\ and 5000~K, respectively. The spectral type of ionizing stars is then determined from \IonRate\ using the relations in \citet{Sternberg2003}. It should be noted that severe dust absorption of ionizing photons in early \hii\ regions may lead to an underestimation of spectral type of ionizing stars.

\textit{3)} Averaged number density of electron (\nelectron) and thermal pressure of ionized gas (\pressurei). Although the distribution of \nelectron\ within our \hii\ regions is highly inhomogeneous, we attempt to estimate an averaged \nelectron\ using 
\begin{equation} \label{EQU:NE}
    n_{\rm e} = \left(\frac{\dot{N}_{\rm ion}}{\alpha_{\rm B} V_{\rm H40\alpha}}\right)^{1/2},
\end{equation}
 where effective volume $V_{\rm H40\alpha} = 4 \pi r_{\rm H40\alpha,eff}^3/3$ (details of equation~\ref{EQU:NE} are in \citetalias{ATOMSIV}). The effective radius of \hii\ region \rhfourtya\ is derived from \ahfourtya\ by $r_{\rm H40\alpha,eff} = \sqrt{A_{\rm H40\alpha}/\pi}$. Then the ionized gas pressure \pressurei\ ignoring He ionization is derived using
 \begin{equation}
    P_{\rm i} =  n_{\rm e} m_{\rm H} c_{\rm i}^2 = n_{\rm e} m_{\rm H} \left(\sqrt{\frac{2 k_{\rm B} T_{\rm e}}{ m_{\rm H}} } \right)^2,
\end{equation}
where \ci\ is the sound speed of ionized gas \citep{Tremblin2014b,Krumholz2017}. \te\ is taken from the measure of \citetalias{ATOMSXIV} in general. We used the Galactocentric distance \RGC\ - \te\ relation derived by \citetalias{ATOMSXIV} when \te\ is not measured:
\begin{equation}
      T_{\rm e} = \left( 6271\pm481\right) + \left(135\pm83\right) \left(\frac{R_{\rm GC}}{{\rm kpc}}\right)~{\rm K}.
\end{equation}

 \textit{4)} Dynamical age of \hii\ regions (\tdyn). The \tdyn\ is estimated using \citep{Dyson1997}
\begin{equation} \label{EQU:TDYN}
    t_{\rm dyn} \approx {\rm 0.05587}~\left(\frac{r_{\rm S}}{\rm pc}\right) \left[\left(\frac{r_{\rm H40\alpha,eff}}{r_{\rm S}}\right)^{7/4} -1\right]~{\rm Myr},
\end{equation}
 here $r_{\rm S}$ is Str{\"o}mgren radius where an ionization-recombination equilibrium reaches at Str{\"o}mgren time $t_{\rm S}$, given as
 \begin{equation} \label{EQU:STROMGREN}
 \begin{aligned}
  r_{\rm S} &\approx {\rm 0.74}  \left(\frac{\dot{N}_{\rm ion}}{\rm 10^{49}~s^{-1}}\right)^{1/3} \left(\frac{n_{\rm ini}}{\rm 10^3~cm^{-3}}\right)^{-2/3} {\rm pc} \\
  t_{\rm S} &\approx 0.0002 \left(\frac{n_{\rm ini}}{\rm 10^3~cm^{-3}}\right)^{-1}~{\rm Myr}.
  \end{aligned}
 \end{equation}

Calculations of $r_{\rm S}$, $t_{\rm S}$, and $t_{\rm dyn}$ require taking into account the initial undisturbed number density of hydrogen nuclei \nini\ \citep{Whitworth1994}. The density of clump at which high-mass protostars just started to ionize their natal clump is difficult to determine from our present-day observations because mass accretion and dispersal could exist at the same time during the early evolution of clump. \citet{Urquhart2019, Urquhart2022} studied few thousand massive clumps from the ATLASGAL project \citep[APEX Telescope Large Area Survey of the Galaxy,][]{Schuller2009} according to their evolutionary stages (quiescent, protostellar, YSO, \hii\ region). The difference in the typical density of the clumps (column density and volume density) between different evolutionary stages is negligible ($\sim 20~\%$) compared to the density dispersion of the clumps at a certain evolutionary stage in their large sample (see Table 2 of \citealt{Urquhart2019} and Table 7 of \citealt{Urquhart2022}), indicating that the average density of a massive clump may not change much during evolution from the quiescent stage to the \hii\ region stage. We therefore estimate \nini\ using the present-day properties of natal clumps measured by ATLASGAL observations. ATLASGAL extracted and cataloged clumps using two independent algorithms that are sensitive to structures on different angular scales: \textit{(i)} \texttt{SExtractor} \citep{Urquhart2014} and \textit{(ii)} \texttt{Gaussclump} \citep{Csengeri2014}. The first algorithm is better at determining global properties of a clump, whereas the second algorithm is more sensitive to embedded compact sources of a clump. The difference between these two algorithms is verified by the different clump sizes in Table~\ref{TABLE:SAMPLE}. We estimate the \hmole\ number density \nhtnd\ in two cases \textit{(i)} the total clump mass and \texttt{SExtractor} clump radius (${n_{\rm H_{2}} = 3 M_{\rm clump}}/4\pi r^3 m_{\rm H} \mu_{\rm H_{2}}$, $\mu_{\rm H_{2}} = 2.8$), and \textit{(ii)} the peak column density and \texttt{GaussClump} FWHM ($n_{\rm H_{2}}$ = $N_{\rm H_{2}}^{\rm peak}/2\rm FWHM$). \\

\begin{table*}
\caption{\label{TABLE:HII-PROPERTIES}\hii\ region properties.} 
\begin{threeparttable}
\renewcommand{\arraystretch}{1.0}
\centering
\begin{tabular}{rrrrrrrrrrrr}
\hline
\hline
name        &   RA     &    DEC  &$r_{\rm H40\alpha,eff}$& int. \hfourtyalpha\ \tnote{\textit{(a)}} & \RecRate\ \& \IonRate\   & sp. type   & \nelectron\        & \ci  &  \pressurei/$k_{\rm B}$   & \multicolumn{2}{c}{\tdyn\tnote{\textit{(b)}}}    \\
            &    deg   &    deg  &     pc                &   Jy~\kms\            & 10$^{49}$ s$^{-1}$   &        &  10$^{3}$~cm$^{-3}$  &  \kms & 10$^7$ K cm$^{-3}$  & \multicolumn{2}{c}{0.01~Myr} \\ 
\hline
            &        &       &                      &               &   &        &    &  & & \textit{i} & \textit{ii} \\ 
\hline
I13080$-$6229 & 197.810 & -62.750 & 0.21                  & 66.5                  & 0.49(0.11)             & O8.0   & 4.2(0.8)      &   11.1(0.6)     &  6.3(1.4)     & 1.0(0.6) &  2.7(0.7)      \\
I15411$-$5352 & 236.248 & -54.037 & 0.08                  & 81.0                  & 0.14(0.03)              & B0.0   & 8.9(1.7)       &  10.5(0.6)     &  11.8(2.6)     & 0.3(0.2) &  1.8(0.4)    \\
I15570$-$5227 & 240.232 & -52.607 & 0.32                  & 18.0                  & 0.33(0.07)              & O9.0   & 1.8(0.3)       &   10.3(0.4)     &  2.3(0.5)     & 1.1(0.9) &  3.3(0.9)    \\
I15584$-$5247 & 240.578 & -52.924 & 0.23                  & 6.1                   & 0.06(0.01)              & B0.5   & 1.2(0.2)       &   10.7(1.1)     &  1.7(0.5)     & 3.9(1.1) &  5.2(1.2)   \\
I16362$-$4639 & 249.990 & -46.753 & 0.13                  & 4.6                   & 0.02(0.01)              & B0.5   & 1.9(0.3)       &   10.8(1.1)     &  2.6(0.7)     & 1.5(0.5) &  1.6(0.4)    \\
I17006$-$4215 & 256.054 & -42.332 & 0.06                  & 31.4                  & 0.08(0.02)              & B0.0   & 12.1(2.3)      &   10.1(0.5)    & 14.8(3.2)     & 0.3(0.2) &  0.8(0.2)     \\
I17160$-$3707 & 259.863 & -37.183 & 0.31                  & 39.7                  & 0.77(0.17)              & O7.5   & 2.9(0.5)       &   10.1(0.5)     &  3.5(0.7)     & 5.1(1.5) &  5.8(1.4)   \\
I18116$-$1646 & 273.649 & -16.760 & 0.27                  & 56.1                  & 0.44(0.10)              & O8.5   & 2.6(0.5)       &   10.2(0.4)     &  3.4(0.7)     & 1.9(0.8) &  5.2(1.2)   \\
I18317$-$0757 & 278.605 &  -7.913 & 0.23                  & 42.2                  & 0.49(0.11)              & O8.0   & 3.6(0.7)       &   9.6(0.4)     &  4.0(0.8)     & -\tnote{\textit{(c)}}& 1.9(0.6)           \\
\hline
   \end{tabular}
      \begin{tablenotes}
      \item [\textit{(a)}] Integrated intensity of \hfourtyalpha\ emission. Its typical uncertainty is less than 10\%.
      \item [\textit{(b)}] \textit{i} and \textit{ii} are dynamical ages estimated using the initial density \nini\ described in Section~\ref{SECTION:HII-REGION}.
      \item [\textit{(c)}] Dynamic age of this region cannot be reliably estimated.
      \end{tablenotes}
      \end{threeparttable}
\end{table*}

The estimated properties of \hii\ regions are listed in Table~\ref{TABLE:HII-PROPERTIES}. The typical uncertainties of \RecRate\ or \IonRate\ are 20\% when dust absorption is ignored. The typical uncertainties of \nelectron, \ci, and \pressurei\ are at level of 20\%, 5\% and 20\%, respectively, if $V_{\rm H40\alpha}$ well describes the real volume of the \hii\ region. The \rhfourtya\ and \nelectron\ range from 0.1 to 0.3~pc and $1.2\times10^{3}$ to $1.2\times10^{4}$~\cmcube, respectively. Using physical size and \nelectron\ as the compactness probes of \hii\ regions, \citet{Kurtz2005} proposed an evolutionary sequence for \hii\ regions: (1) Hyper compact stage ($\lesssim0.03$~pc, $\gtrsim10^6~\rm cm^{-3}$), which represents the earliest time of \hii\ regions  \citep{Sewilo2011,Yang2019,Yang2021, Patel2024}, (2) Ultra compact stage \citep[${\lesssim0.1~\rm pc}$, ${\gtrsim10^4~\rm cm^{-3}}$,][]{Churchwell2002,Hoare2007}. Some of the well-studied samples are presented in \citet{Wood1989,Kalcheva2018}. (3) Compact stage ($\lesssim0.5$~pc, $\gtrsim5\times10^3~\rm cm^{-3}$) such as the regions explored by \citet{Garay1993}. (4) Classical / giant stages ($\gtrsim10$~pc, $\lesssim100~\rm cm^{-3}$).  Our work represents a unique sample of \hii\ regions that are at the early time of the third stage (``compact'' stage) in the sequence of \citet{Kurtz2005} because our sample is slightly more evolved than the typical UC\hii\ regions that are thought to be smaller and denser. Therefore, in the following discussion the evolutionary stages of our sample are referred to as ``early compact''.

All regions should already be in equilibrium of ionization and recombination due to the very short time to reach $r_{\rm S}$ ($t_{\rm S}$ $\sim$ several to a few tens of years) according to equation~\ref{EQU:STROMGREN}. The dynamic age is estimated to be around $10^4$ years, but should be considered as a lower limit for the lifetime of these \hii\ regions. The actual lifetime could be an order of magnitude longer than the estimated dynamical age for early \hii\ region, which is already shown in some attempts of estimating UC\hii\ region lifetime by massive star counting \citep{Wood1989b, Mottram2011}. The typical lifetime of UC\hii\ region is $\sim0.3$~Myr \citep{Churchwell2002} and therefore, it is reasonable to propose that the typical age of our early compact \hii\ regions is likely $\gtrsim0.3$~Myr. An independent test using simulations of expansion of the \hii\ region in a turbulent molecular cloud carried out by \citet{Tremblin2014b} suggests a typical age around 0.2~Myr with large error for our \hii\ regions, roughly consistent with the estimated lifetime.

The pressure of ionized gas in our sample, \pressurei/$k_{\rm B}\sim 10^7$ to $10^8~\rm K~cm^{-3}$,  is one to three orders of magnitude higher than that of more evolved classical \hii\ regions measured by a single-dish telescope \citep{Mookerjea2022, Pandey2022}. This high pressure is expected to profoundly change the physical state, particularly the density structure, of the surrounding medium. In the following sections, we extract and analyze dense molecular gas structures to understand how interactions work between the \hii\ region and the dense gas that forms stars.

\defcitealias{ATOMSII}{ATOMS~II}

\section{dense gas fragments traced by \htcop\ emission} \label{SECTION:DENSE-GAS}
\begin{figure*}
\centering
\includegraphics[width=0.99\textwidth]{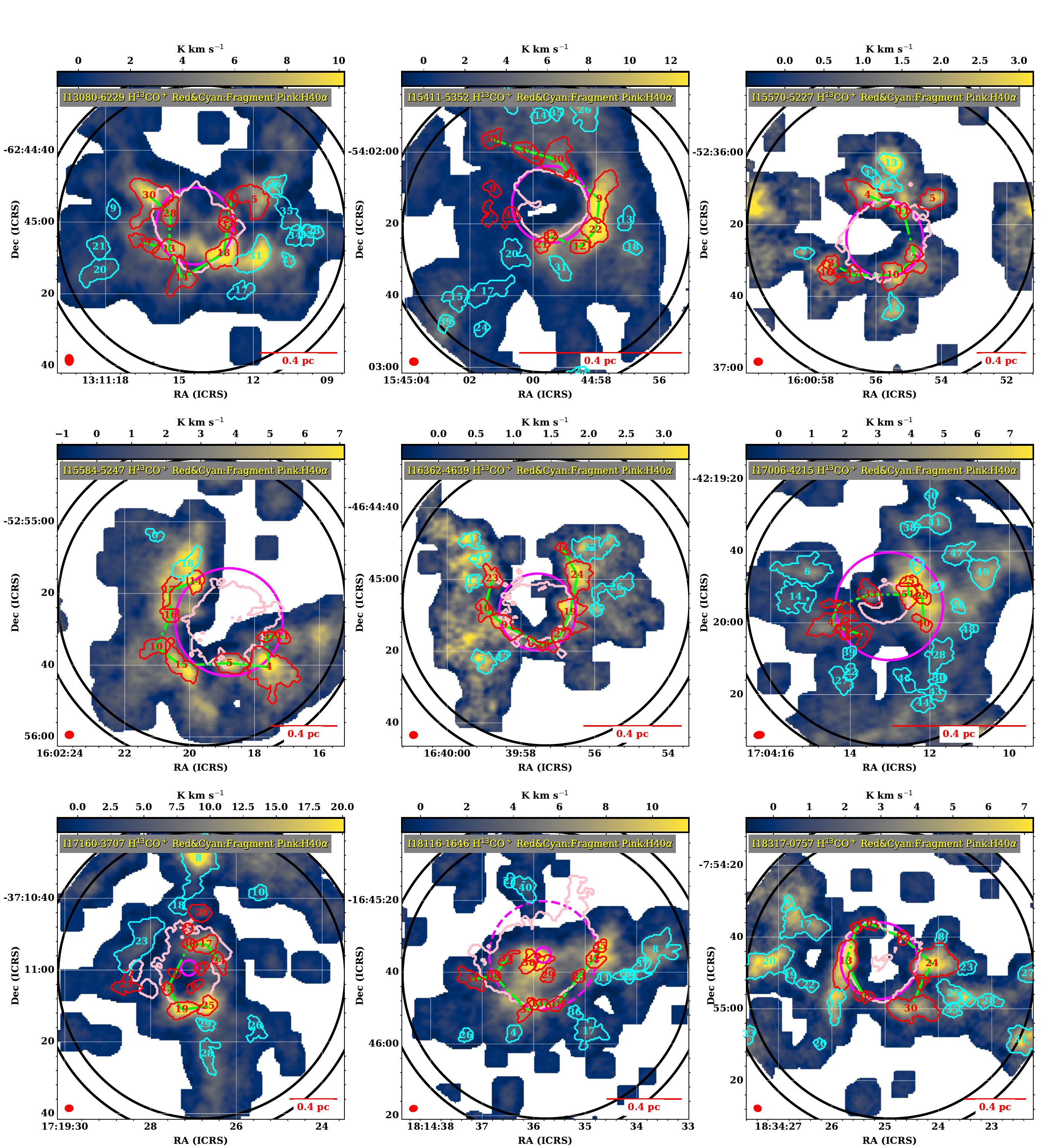}
       \caption{\htcopone\ emission and extracted fragments. The color images show the \htcopone\ moment 0 maps (PB corrected) produced by the technique described in \citet{Dame2011} with a channel clipping level of $4\times\rm rms$. The red and cyan contours outline the extracted fragments close to and far from the central \hii\ regions (N and F fragments; see Section~\ref{SECTION:INFLUENCE}), respectively. Pink contours indicate the edges of \hii\ regions. Magenta circles indicate the peaks of angular two-point correlation function (solid: first peak, dashed: second peak) discussed in Section~\ref{SUBSECTION:FRAGMENTATION-DISTRIBUTION}. The lime lines are the \texttt{minimum spanning tree (MST)} connections for the N fragments probably located in shell. The two black circles indicate the 20\% and 30\% power points of PB.}
        \label{FIGURE:NORMAL2}
\end{figure*}

We use \htcopone\ transition to detect and analyze dense molecular gas \citep[][\citetalias{ATOMSII}]{ATOMSII} because \htcop\ is an effective tracer of high-density molecular gas, especially for the dense gas involved in star formation \citep{Gao2004}. Figure~\ref{FIGURE:NORMAL2} displays \htcopone\ moment 0 maps for the nine \hii\ regions, and it reveals numerous bright fragments at the edges and off the edges of \hii\ regions. These fragments are representative of relatively dense structures within the regions and, therefore, closely correlate with star formation activities. The continuum emission at 3~mm for our ATOMS target regions is dominated by free-free emission, and thus it is very difficult to analyze these fragments with dust emission at 3~mm (see details in \citetalias{ATOMSIV}).

\subsection{Extraction of dense gas fragments} \label{SUBSECTION:CORE-EXTRACTION}
Dense gas fragments were extracted from \htcopone\ cubes by employing \texttt{astrodendro} that is capable of recognizing intricate and hierarchical structures of molecular clouds \citep{Rosolowsky2008}.  The 3D dendrogram calculations were performed on cubes without primary beam (PB) correction to ensure a uniform noise field during structure extraction \citep{Redaelli2021, Redaelli2022}. After a deliberate test, the parameters in the calculation of the dendrogram are established as follows: the minimum intensity for a voxel to be considered $\texttt{min{\_}value} = 3.5~\rm rms$, the minimum step to differentiate an independent structure $\texttt{min{\_}delta} = 2~\rm rms$, the minimum number of voxels for a structure to be included $\texttt{min{\_}npix}$ equivalent to three channels $\times$ half of the beam size. We pruned the \texttt{astrodendro} calculation results by additionally requiring that the minimum peak for an indecomposable structure $\texttt{min{\_}peak} = 4.5~\rm rms$, the minimum channel number equal to three, and the minimum spatial size for a structure is half of the beam. 

We take the ``leaves'' structures in the dendrogram results as dense gas fragments, since they represent the smallest structures without substructures. The reliability of extraction is reflected in Fig.~\ref{FIGURE:NORMAL2} where the extracted fragments agree well with the bright emissions in the \htcopone\ moment 0 maps. After removing fragments with low SNR and complicated \htcopone\ spectral profiles (see Section~\ref{SUBSECTION:ESTIMATION-BIASES}), a total of 164 dense fragments were extracted and considered in the following analysis.

\subsection{Fragment properties} \label{SUBSECTION:CORE-PROPERTIES-CALCULATION}
The velocity dispersion $\sigma$, mass \mcore, density, and virial parameters are essential to determine the physical state of a fragment. We estimate these properties using \htcopone\ cubes that are corrected by the primary beam:

\textit{1)} The value of $\sigma$ given by \texttt{astrodendro} is usually a lower estimate of the actual $\sigma$ associated with fragments due to channel loss in the pruning process, particularly for weak structures whose line wing channels are ignored. To better reconstruct $\sigma$ of the fragment, we extracted in practice the average spectra from the dendrogram spatial area of fragments and then fitted them with Gaussian within the dendrogram velocity range of fragments. The recalculated $\sigma$ is around 0.2 to 1.1~\kms, which is larger than the dendrogram-produced one by 0.1 to 0.3~\kms.

\textit{2)} With the assumption that \htcopone\ is optically thin, the associated \htcop\ column density $N_{\rm H^{13}CO^+}$ is calculated using
\begin{equation}   \label{EQUATION:THINCOLUMNDENSITY}
\frac{\Delta N_{\rm H^{13}CO^+}}{\Delta{\rm v}}|_{\rm thin} \simeq \left(\frac{8\pi k_{\rm B}\nu_{ul}^2}{h c^3 A_{ul} g_u}\right)
 Q_{\rm rot}(T_{\rm ex}) \exp\left(\frac{E_u}{k_{\rm B}T_{\rm ex}}\right)\frac{T_{\rm r}({\rm v})}{f_{\rm beam}},
\end{equation}
where $\nu_{ul}$, $f_{\rm beam}$, \tr, \tex, $E_u$, $g_u$, and $Q_{\rm rot}$ are the transition frequency, beam filling factor, brightness temperature, excitation temperature, energy and degeneracy of the upper level, and partition function, respectively. Taking $A_{ul} = 10^{-4.41416}~\rm s^{-1}$, $f_{\rm beam} = 1$, $E_u = 4.16~\rm K$, and $g_u = 3$, the fragment mass can be estimated with equation~\ref{EQUATION:THINCOLUMNDENSITY} by integrating the spectra and the area of fragment when \tex\ and \htcop\ abundance \htcopab\ are known. It is assumed that \tex\ is equal to the dust temperature \dustt\ on the clump scale measured by \citet{Urquhart2018}. The compact \hii\ regions in this work span a range of the Galactocentric distance from 2.4 to 6.9~kpc and therefore \htcopab\ can differ for each region due to the Galactocentric gradient of the isotope \citep{Milam2005}. \citet{Kim2020} explored \htcopone\ emission towards $\simeq400$ massive clumps with IRAM 30~m telescope, and by taking advantage of their result we find that in their sample the relationship between the clump-scale \htcopab\ and the Galactocentric distance is quite flat within the Galactocentric distance range of our sources. Therefore, a uniform $\chi_{\rm H^{13}CO^+} = 3.6\times10^{-11}$ is taken for our sample, which is the average for UC\hii\ regions in the observation of \citet{Kim2020}. The resultant fragment mass \mcore\ ranges from 1 to 200~\msun\ with a typical uncertainty of 20\%.

\textit{3)} Mass surface density \sufdenscore\ and \hmole\ number density \nhtnd. \sufdenscore\ is key to characterizing massive star formation \citep{Krumholz2008,Kauffmann2010} and we estimate it using 
\begin{equation}
    \Sigma_{\rm core} = M_{\rm core}/A_{\rm core} =  M_{\rm core}/\left(\pi r^2_{\rm core,eff}\right),
\end{equation}
where the effective area $A_{\rm core}$ and its corresponding effective radius \rcoreeff\ are derived from the 2D spatial extent of the dendrogram resulted leaves \citep{Takekoshi2019,Redaelli2021,Takemura2023}. The typical radius ranges from 0.02 to 0.1~pc, corresponding to the core-scale structures. The surface density of the dense gas fragments ranges from 0.2 to 7~g~cm$^{-2}$, with a typical uncertainty of 30\%. The fragment \hmole\ number density \nhtnd\ is estimated using
\begin{equation}
 n_{\rm H_2} = \frac{3 M_{\rm core}}{4 \pi r^3_{\rm core,eff} \mu_{\rm H_2} m_{\rm H}},
\end{equation}
and its values range from $2\times10^5$ to $5\times10^6$~cm$^{-3}$ with uncertainties around 40\%.

\textit{4)} Virial ratio characterizes whether the gas fragments are bound by gravity. Regardless of magnetic fields, rotation, and external pressure and assuming a constant density profile of fragments, the virial ratio of fragments is estimated using \citep{Bertoldi1992,Redaelli2021,Wong2022}
\begin{equation}
    R_{\rm vir} = \frac{M_{\rm vir}}{M_{\rm core}} = \frac{5\sigma^2 r_{\rm core,eff}}{M_{\rm core} G},
\end{equation}
and the $R_{\rm vir}$ range from 0.3 to 5, with typical uncertainties of about 30\%.

The estimated properties of each fragment are listed in the supplementary online materials. In Section~\ref{SUBSECTION:CORE-DIFFERENCE} we discuss these properties in more detail to study how ionization feedback regulates fragments and associated star formation.

\subsection{Biases in fragment extraction and properties estimation.}
\label{SUBSECTION:ESTIMATION-BIASES}

One of the biases that influence the 3D extraction of fragments is the self-absorption of \htcopone. It can create fake independent velocity components separated by the absorption dip in an extremely dense region, leading to misclassification or distortion of the extracted fragments. The fact that only a few fragments in the nine regions have notable spatial overlaps supports that self-absorption is not severe in the regions studied here, because more extracted fragments would overlap each other if the fake velocity components created by the self-absorption dip took over. Assuming the following properties for a typical fragment: kinetic temperature $T_{\rm kin} = 30~\rm K$, background temperature $T_{\rm bg} = 2.73~\rm K$, $n_{\rm H_{2}} = 10^6~\rm cm^{-3}$, and \htcop\ column density $N_{\rm H^{13}CO^+} = 3.6\times10^{13}~\rm cm^{-2}$ (corresponding to a \hmole\ column density of $\sim 10^{24}~\rm cm^{-2}$), the optical depth $\tau$ given by \texttt{RADEX}\footnote{\url{http://var.sron.nl/radex/radex.php}} online tool is $\simeq0.46$ \citep{vanderTak2007}. It shows that \htcopone\ emission in most fragments should have a $\tau<1$ and self-absorption is unlikely to have a major impact on 3D extraction. To ensure a reliable estimate of fragment properties, we thoroughly checked the spectral profile of \htcopone\ and its Gaussian fit for each fragment. A total of 16 massive and dense fragments are neglected in property estimation due to their complicated spectral profiles. 

Variation of \tex\ and \htcopab\ within the clump-scale environment can be another important uncertainty in property calculations.  Different strengths of heating and UV radiation for fragments at different locations can cause different \htcopab\ for each fragment, although \citet{Kim2020} found no significant differences in the clump \htcopab\ between those that host \hii\ regions and those without \hii\ regions. This is also supported by observations of \citet{Sanhueza2012} that explore more than 100 clumps and confirm that the clump-scale \htcopab\ does not change much over different evolutionary stages. The dissociative recombination with electrons, which is the main destruction mechanism of \htcop, probably strengthens when fragments are close to \hii\ region, while the warm environment causes the mantles of icy grains to evaporate and then more CO and H$_2$O are produced to join the formation reactions of \htcop\ \citep{Miettinen2014, Kim2020}. Therefore, the general effect of \hii\ region on \htcopab\ is difficult to determine without careful chemical modeling \citep{Stephan2018}, which is beyond the scope of this article.

\section{Influence of \hii\ region on fragments}
\label{SECTION:INFLUENCE}
How external evolved \hii\ regions alter the initial conditions of HMSF on the clump scale has been extensively studied \citep{Schneider2020} such as compression on clump density structure and magnetic field \citep{Eswaraiah2020, PaperI}, injecting turbulence \citep{Mazumdar2021,Shen2024}, radiation heating \citep{PaperI}, changing clump fragmentation mode \citep{Liu2017,Rebolledo2020}. To further investigate the ionization feedback of the compact \hii\ regions on the core-scale conditions of star formation, we classified the \htcop\ fragments into two categories based on their relationship with the central \hii\ region: \textit{1) fragments at the edges or near the edges of \hii\ region (N fragments hereafter) and 2) fragments relatively far away from \hii\ region (F fragments hereafter)}. Basically, we extend the spatial extents of the fragments with a beam width and then count the spatial overlaps between the extended fragment and the \hii\ region. The extending spatial extent procedure is to reduce the bias of losing spatial overlaps caused by PDR \citep{Goicoechea2016} that exists between the ionized region traced by \hfourtyalpha\ and the dense molecular region traced by \htcop\ emission. A fragment is classified as N type when the overlap is greater than 50\% size of the extended fragment or when the mean \hfourtyalpha\ intensity in the overlap is greater than 10\% of the \hfourtyalpha\ emission peak of the entire \hii\ region. The purpose of setting the \hfourtyalpha\ intensity threshold is to include the situation in which a N fragment has a small overlap with \hii\ region but a significant \hfourtyalpha\ emission indicating ionization interaction is detected in this small overlap. These thresholds are carefully tested. The identified number of N fragments decreases with the overlap threshold, and the trend of decrease becomes flatter when the overlap threshold reaches 50\%, which means that the classification results become relatively stable when the overlap threshold reaches 50\%.

These two types of fragments are shown in red and cyan contours in Fig.~\ref{FIGURE:NORMAL2}. The chosen threshold leads to that the classified fragments are in good agreement with the eye check, supporting the appropriateness of the classification. There are 80 N and 84 F fragments with estimates of physical properties. The N fragments are generally closer to the \hii\ region and most of them are embedded in the \cch\ shell partly enclosing the \hii\ region. The influence of ionization feedback on fragments can be revealed directly by comparing the properties of these two types of fragments, because fragments closer to the \hii\ region are expected to be more affected. 

\subsection{Nature of fragments under impacts of \hii\ region} \label{SUBSECTION:CORE-DIFFERENCE}

\begin{figure*}
\centering
\includegraphics[width=0.99\textwidth]{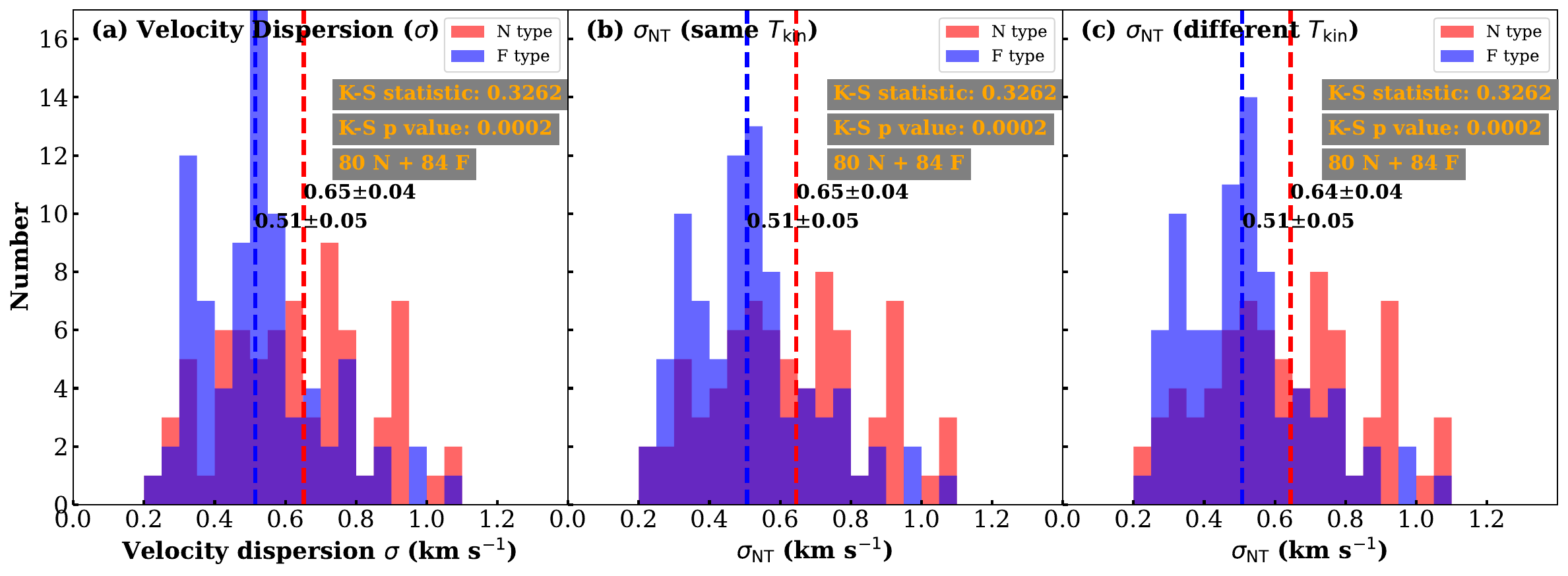}
   \caption{Number distribution of fragment \htcopone\ velocity dispersion $\sigma$ and its non-thermal component $\sigma_{\rm NT}$. Panels (b) and (c) show the $\sigma_{\rm NT}$ estimated with the same \tkin\ and different \tkin\ between N and F fragments, respectively. Orange texts show the statistic and $p$ value of the K-S test. The red and blue dashed lines marked with the nearby numbers show the median values for N and F fragments, respectively. }
        \label{FIGURE:COMPARISON_VSIGMA}
\end{figure*}

We first discuss fragment velocity dispersion $\sigma$ because its measurement is more reliable compared to other properties derived from \htcop\ column density. The panel (a) of Fig.~\ref{FIGURE:COMPARISON_VSIGMA} shows that the $\sigma$ distributions are different between the N and F fragments and the difference is confirmed by a Kolmogorov–Smirnov (K-S) test with a \textit{p}-value threshold of 0.05. The K-S test rejects the null hypothesis that the two types of fragments have the same distributions. The N fragments have a larger $\sigma$ than the F fragments, considering the small error in the $\sigma$ measurement.  To study the strength of turbulence, the non-thermal component $\sigma_{\rm NT}$ is tentatively extracted by using 
\begin{equation}
\label{EQUATION:SIGMA_NT}
\sigma_{\rm NT} = \left(\sigma^2 - k_{\rm B} T_{\rm kin}/m_{\rm H^{13}CO^+}\right)^{1/2},
\end{equation}
where \tkin\ is the gas kinetic temperature. The heating effect of \hii\ region is expected to create a different \tkin\ between the N and F fragments. To partially mitigate this bias, we estimate the fragment $\sigma_{\rm NT}$ in situations where \tkin\ of the N and F fragments are equal to clump-scale $T_{\rm dust} + {\rm 5~K}$ and clump-scale $T_{\rm dust} - \rm {5~K}$, respectively. The assumed \tkin\ difference is reasonable, considering that the observations of \citet{PaperI} reveal that the temperature of massive infrared dark clumps around the evolved \hii\ region is around 6 to 10~K higher than the ones far away from \hii\ region. The stronger turbulence for N fragments is well demonstrated in panels (b) and (c) of Fig.~\ref{FIGURE:COMPARISON_VSIGMA} which include the possible difference in \tkin\ between N and F fragments.

\defcitealias{ATOMSIX}{ATOMS~IX}

\begin{figure}
\centering
\includegraphics[width=0.47\textwidth]{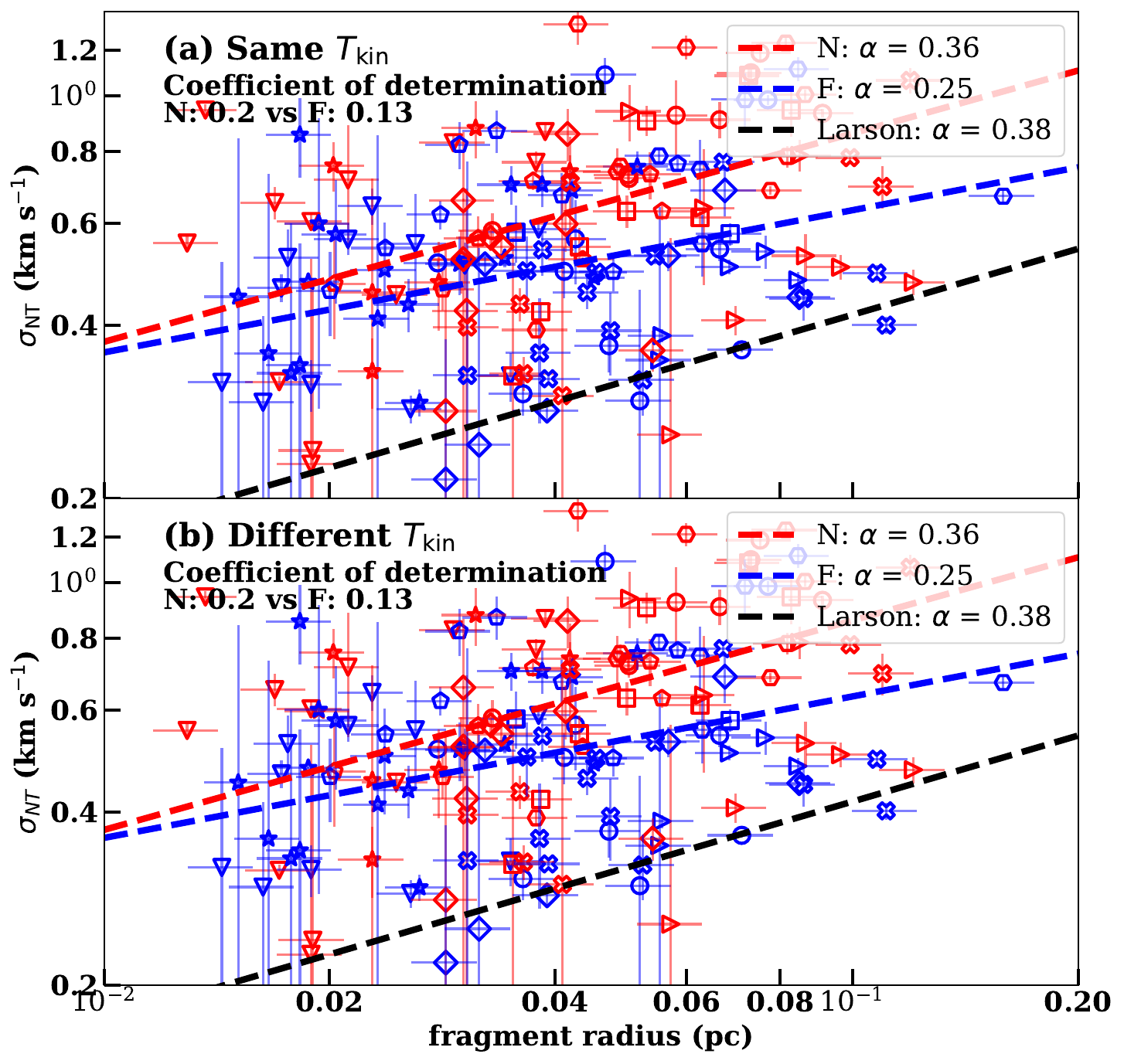}
   \caption{Relation between fragment \htcopone\ $\sigma_{\rm NT}$ and effective radius \rcoreeff. Panels (a) and (b) show the $\sigma_{\rm NT}$ estimated with the same \tkin\ and different \tkin\ between N and F fragments, respectively. Red and blue markers with different shapes represent N and F fragments of different regions, respectively. Red and blue lines show the regressions for N and F fragments, respectively. Black line shows the extension of Larson power-law relation to the physical scales of this work. Note that Larson law is valid only within scales of 0.1 to 100~pc.}
        \label{FIGURE:LARSON}
\end{figure}

The nature of turbulence can be partially reflected by the $\sigma_{\rm NT}$ - size relation of molecular gas. \citet{Larson1981} proposed a power-law relation with a power index of 0.38 for the $\sigma_{\rm NT}$-size relation of molecular clouds with a size of $\sim0.1$ to $\sim100$~pc, but the following studies show that this $\sigma_{\rm NT}$-size relation becomes more complicated when moving to smaller scales. \citet[][hererafter \citetalias{ATOMSIX}]{ATOMSIX} found that the $\sigma_{\rm NT}$ - size relation breaks down into two parts with different power law indices on a transition scale of $\sim0.1$~pc in infrared dark cloud (IRDC) G34. \citetalias{ATOMSIX} suggests that gravity-driven chaotic collapse may be a major factor in driving turbulence and the reason for the steeper power law with greater scatter seen on the scale of $<0.1$~pc. Figure~\ref{FIGURE:LARSON} shows a positive correlation between $\sigma_{\rm NT}$ and the radius of our fragments, but with a very large scatter. Regression analysis gives a power law index of 0.36 and 0.25 with a determination coefficient of 0.2 and 0.13 for N and F fragments, respectively. The large scatter hinders us from any further detailed analysis, but sheds light on the point that ionization feedback can be one of the dominant mechanisms that drives turbulence in fragment \citep{Nakano2017,Saldano2019}. The highly inhomogeneous injection of turbulence that stems from the inhomogeneous ionization feedback makes the $\sigma_{\rm NT}$-size relation deviate from the power law and become more irregular.

\begin{figure*}
\centering
\includegraphics[width=0.99\textwidth]{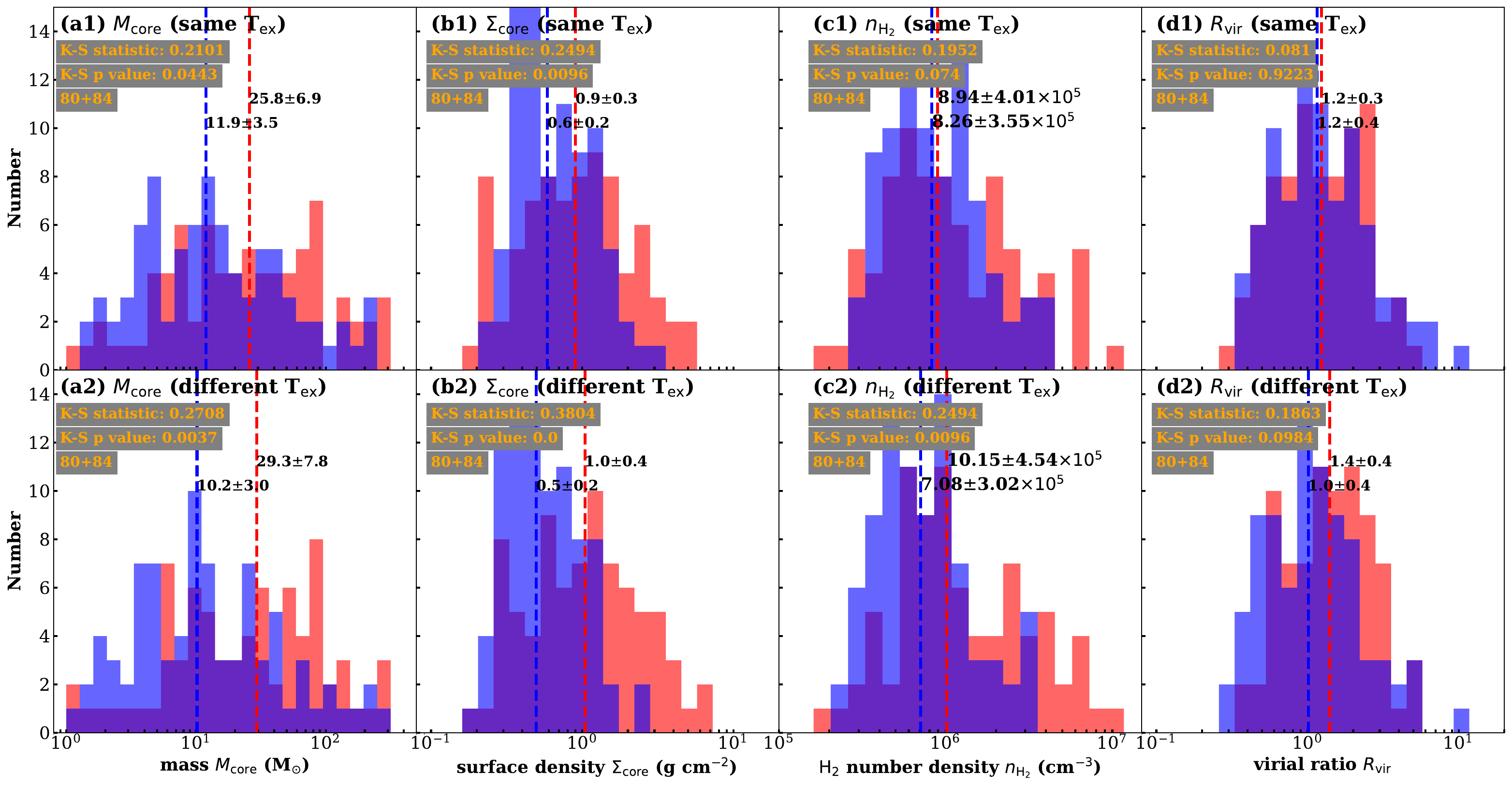}
       \caption{Number distributions of fragment mass \mcore, surface density \sufdenscore, \hmole\ number density \nhtnd, and virial ratio $R_{\rm vir}$ (red one for N fragments and blue one for F fragments). The top and bottom panels show the corresponding properties derived with the assumption that \tex\ is the same for N and F fragments and that \tex\ is different between N and F fragments, respectively. The red and blue dashed lines marked with nearby numbers show the median values for N and F fragments, respectively.}
        \label{FIGURE:COMPARISON}
\end{figure*}

Then we compare \mcore, \sufdenscore, \nhtnd, and \rvir\ between the N and F fragments. In the primary estimate of \mcore\ and its related parameters, \tex\ of \htcop\ is set to be equal to clump-scale \dustt\ for both the N and F fragments in a \hii\ region. Here we additionally estimate \mcore\ and its related properties with the assumption that the N and F fragments have a \tex\ equal to the clump-scale $T_{\rm dust} + 5$~K and the clump-scale $T_{\rm dust} - 5$~K, respectively, similar to the calculations of $\sigma_{\rm NT}$. The extracted fragments span a wide range of \mcore, \sufdenscore, \nhtnd, and \rvir. The number distributions plus results of K-S test in Fig.~\ref{FIGURE:COMPARISON} show that the N fragments have higher \mcore\ and \sufdenscore\ than F fragments when considering uncertainties. With the assumption of a different \tex, the differences in \mcore\ and \sufdenscore\ are even more remarkable. It is worth to note that \sufdenscore\ of the N fragment has a median value of $\sim1~{\rm g~cm^{-2}}$ that meets the minimum \sufdenscore\ requirement of HMSF proposed by \citet{Krumholz2008}, indicating that the N fragment may be the main locations for HMSF compared to the F fragment. The difference in \nhtnd\ is not clear, although the K-S test suggests a different distribution when \tex\ is different. The \htcop\ emission is less optically thin for the N fragment due to its higher \sufdenscore, indicating that the \nhtnd\ difference may appear if the optical depth is considered in the calculation.

\begin{figure*}
\centering
\includegraphics[width=0.99\textwidth]{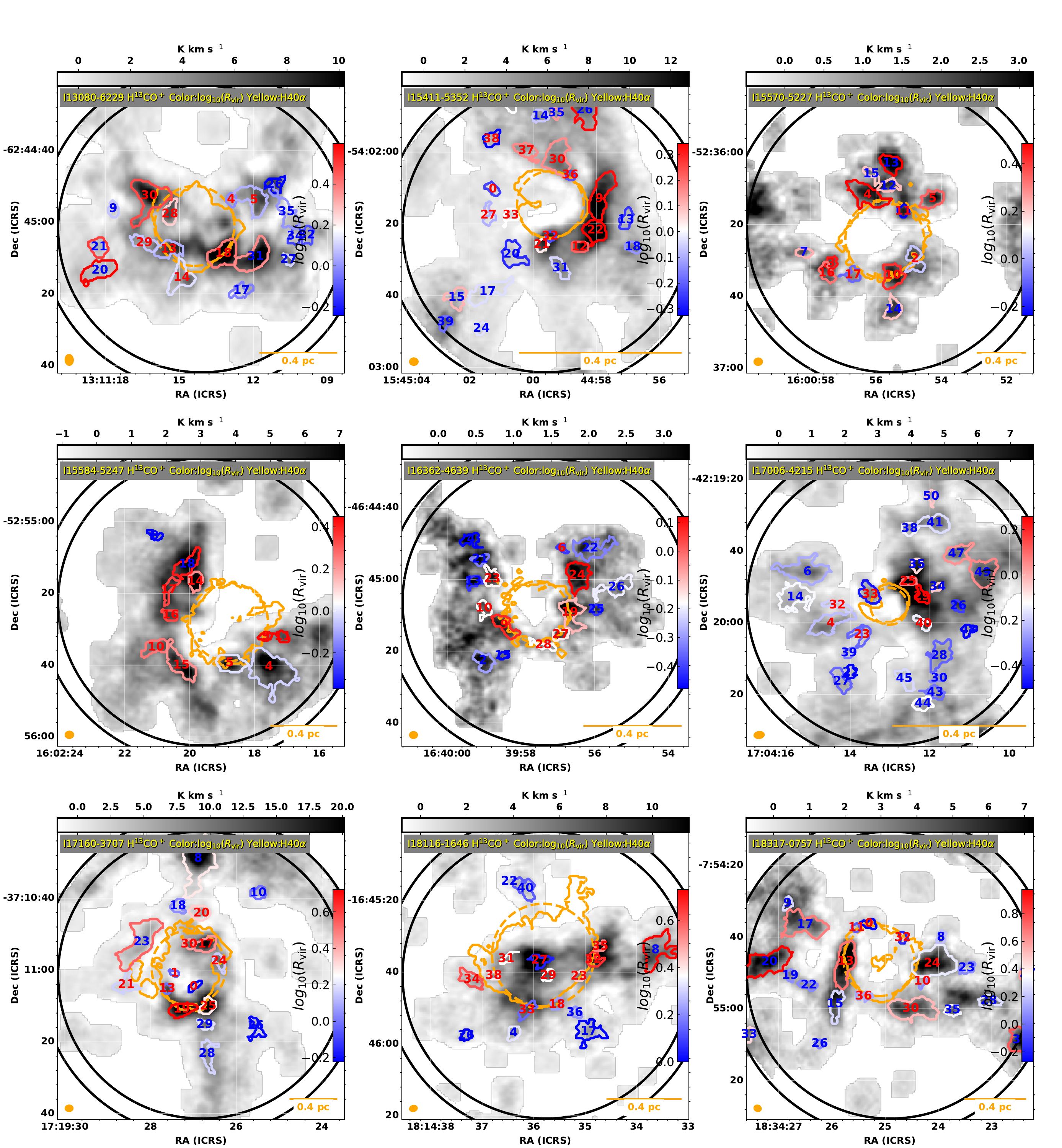}
       \caption{Virial ratio $R_{\rm vir}$ of fragments. Gray images show the \htcopone\ moment 0 maps the same as Fig.~\ref{FIGURE:NORMAL2}. The contours outline the extracted fragments with a color-coding which represents the virial ratio $R_{\rm vir}$ in logarithmic scale. The N and F fragments are marked with red and blue fragment ID labels, respectively. Orange contours and circles indicate the edges and the effective sizes of \hii\ regions, respectively.}
        \label{FIGURE:NORMALVIRIAL}
\end{figure*}

We do not find a difference in $R_{\rm vir}$ between the N and F fragments no matter under assumptions of the same or different \tex, especially considering the uncertainty of 30\% in the virial estimate. The majority of the fragments are not far from the marginally bound status ($R_{\rm vir}\lesssim2$) as shown in Fig.~\ref{FIGURE:COMPARISON}. We used the value of $R_{\rm vir}$ as a color-coding to explore whether there are any relations between the specific fragment locations and $R_{\rm vir}$ in Fig.~\ref{FIGURE:NORMALVIRIAL}. It shows that the most massive N fragments have a greater $R_{\rm vir}$ but this does not mean that these N fragments are definitely less bound than the F fragments. The external pressure exerted by the ionized gas \pressurei\ is ignored in our calculation of $R_{\rm vir}$ due to the high inhomogeneity of \pressurei. The pressure of ionized gas may help the massive N fragment to maintain a bound status \citepalias{ATOMSXIII}.\\

The more massive and turbulent nature, plus a higher surface density of fragments affected by \hii\ regions, suggests that the core-scale conditions for star formation differ from those in a quiescent environment. The global trend of difference revealed here does not rely on detailed definitions of fragment size such as the ones given by \texttt{astrodendro} \citep{Rosolowsky2008} and 2D Gaussian fit \citepalias{ATOMSIII}, which are sensitive to calculations of density and virial ratio. The distinct characteristics of the N fragments should associate with the particular environment of the molecular shell. In fact, most of the N fragments are located in shells traced by \cch\ and we highlight them by connecting them with lime lines in Fig.~\ref{FIGURE:NORMAL2} using the technique of \texttt{minimum spanning tree} \citep[\texttt{MST}, ][]{Naidoo2019}.  The molecular shells of compact \hii\ regions, as a direct consequence of gas collection and compression, probably create an environment conducive to the formation or survival of dense and massive fragments. In the next section, we demonstrate that the \hii\ regions have a major influence on the spatial distributions and kinematics of fragments and shells in the regions.

\subsection{Regulated spatial distribution of fragments} \label{SUBSECTION:FRAGMENTATION-DISTRIBUTION}
Spatial distributions of the fragments around \hii\ regions in the FoV could be regulated by expansion of overpressured ionized gas. To study the spatial connection between fragments and \hii\ regions, the angular two-point correlation function (ATPCF) between the positions of \hii\ regions and fragments is calculated \citep{Landy1993}. The resulting ATPCF is a function of the separation between the central \hii\ region and the surrounding fragments, and a higher value in ATPCF indicates a higher probability of finding fragments at this separation. This approach has been applied in the analysis of the relationships between \textit{evolved} bubble \hii\ regions and their surrounding massive star-forming clumps, and revealed an overdense activity of star formation at the edges of \hii\ region \citep{Thompson2012,Kendrew2016}.

\begin{figure*}
\centering
\includegraphics[width=0.95\textwidth]{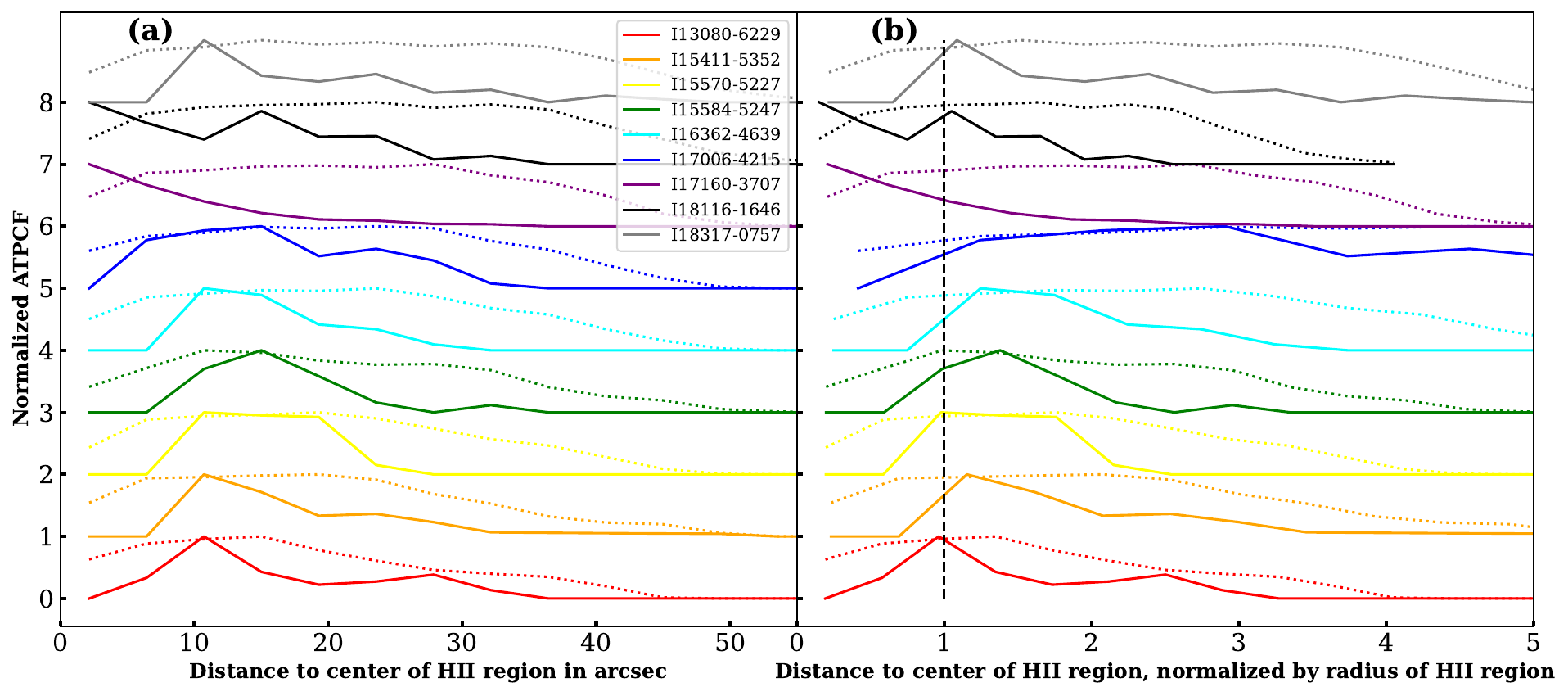}
       \caption{Angular two-point correlation function (ATPCF) between dense fragments and \hii\ regions, normalized by ATPCF peak. The distances of fragments to the center of \hii\ region are shown with the unit of arcsec in panel (a) and with the unit of effective radius of \hii\ region $r_{\rm H40\alpha,eff}$ in panel (b). Solid and dashed lines show the ATPCFs produced from the observed positions of fragments and produced from an average of 500 random spatial distributions of fragments with the same number as the observed ones, respectively.}
        \label{FIGURE:ATPCF}
\end{figure*}

Figure~\ref{FIGURE:ATPCF} presents the ATPCF for our \hii\ regions and fragments, calculated with the ``Landy-Szalay'' estimator. ATPCF is normalized by its peak and radius of \hii\ region \rhfourtya\ to unify the analysis. Most of the regions in our study have an ATPCF that peaks around or slightly farther than \rhfourtya, except for I17160$-$3707 and 18116$-$1646. The secondary bump of the ATPCF for 18116$-$1646 is well correlated with $r_{\rm H40\alpha,eff}$, as shown in the magenta dashed circle in Fig.~\ref{FIGURE:NORMAL2}. Although its \hfourtyalpha\ emission shows a nice cometary morphology, \cch\ and \htcop\ emission is not only in the directions of the edges of \hii\ region but also in the central direction of \hii\ region, as shown in Figs.~\ref{FIGURE:NORMAL1} and \ref{FIGURE:NORMAL2}. This observed distribution of molecular gas is probably a consequence of the projection for a 3D shell structure. We further test the reliability of the presented ATPCF peak by calculating the average ATPCF of 500 2D random distributions that are generated from the same number of fragments. The ATPCF of the random distribution is much flatter and its peak matches the radius of the \hii\ region worse, again demonstrating that the spatial distribution of fragments is not random in most regions.

There are two potential explanations for the agreement between the ATPCF peak and the edges of \hii\ region. An explanation is that expansion of \hii\ region rearranges the spatial distribution of pre-existing fragments at the same time some of the pre-existing fragments protruding from the shell are photoevaporated. Another explanation is that fragments at the edges of the \hii\ region were formed by fragmentation of the molecular shell collected during expansion of the \hii\ region.

In the first case, if dense fragments were formed by initial spontaneous (rather than \hii\ region-induced) fragmentation of their natal clump prior to the development of the \hii\ region and the initial fragments were somehow ``randomly'' distributed in space, in later evolution these fragments would be displaced by expansion of the \hii\ region or photoevaporated by UV radiation, making their distribution peak at the edges of \hii\ region. The shells are relatively smooth and protruding structures are not common, as shown in Figs.~\ref{FIGURE:NORMAL1} and \ref{FIGURE:NORMAL2}, suggesting that photoevaporation is probably efficient. Some of the irradiated low-mass fragments may be photoevaporated during expansion from the UC stage to the compact stage. For an irradiated fragment, its mass loss rate due to photoevaporation can be estimated using the equation in \citet{Haworth2012},
    \begin{equation}
    \dot{M}_{\rm eva} = 4.4\times10^{-3}\left(\frac{\Phi}{\rm cm^{-2}~s^{-1}}\right)^{1/2}\left( \frac{r_{\rm core,eff}}{\rm pc} \right)^{3/2}~\rm M_{\odot}~Myr^{-1},
    \end{equation}
where $\Phi$ is the flux of ionizing photon. Our fragments have a typical radius of 0.04~pc, the resultant $\dot{M}_{\rm eva}$ are $\sim26$ and $\sim60~\rm M_{\odot}~Myr^{-1}$ when they are 0.1 to 0.2~pc (typical radius of our \hii\ regions is 0.2 pc) away from the ionizing stars that release ionizing photons at a rate of $10^{48}~\rm s^{-1}$. It illustrates that fragments with several solar masses will be photoevaporated within 0.1~Myr if there is no efficient shelter between ionizing stars and fragments.

In the second case, excess fragments produced in shell fragmentation lead to an ATPCF peaking at the edges of \hii\ region. However, in Section~\ref{SECTION:FRAGMENTATION} we show that this possibility is rather low. Most of the fragments in the shell are very likely to be pre-existing fragments produced by clump fragmentation before development of the \hii\ region, and thus the observed spatial distribution mainly comes from photoevaporation and regulation effect of the \hii\ region.

\subsection{Influence on fragment kinematics}
\label{SUBSECTION:EXPANSION}
\begin{figure*}
\centering
\includegraphics[width=0.95\textwidth]{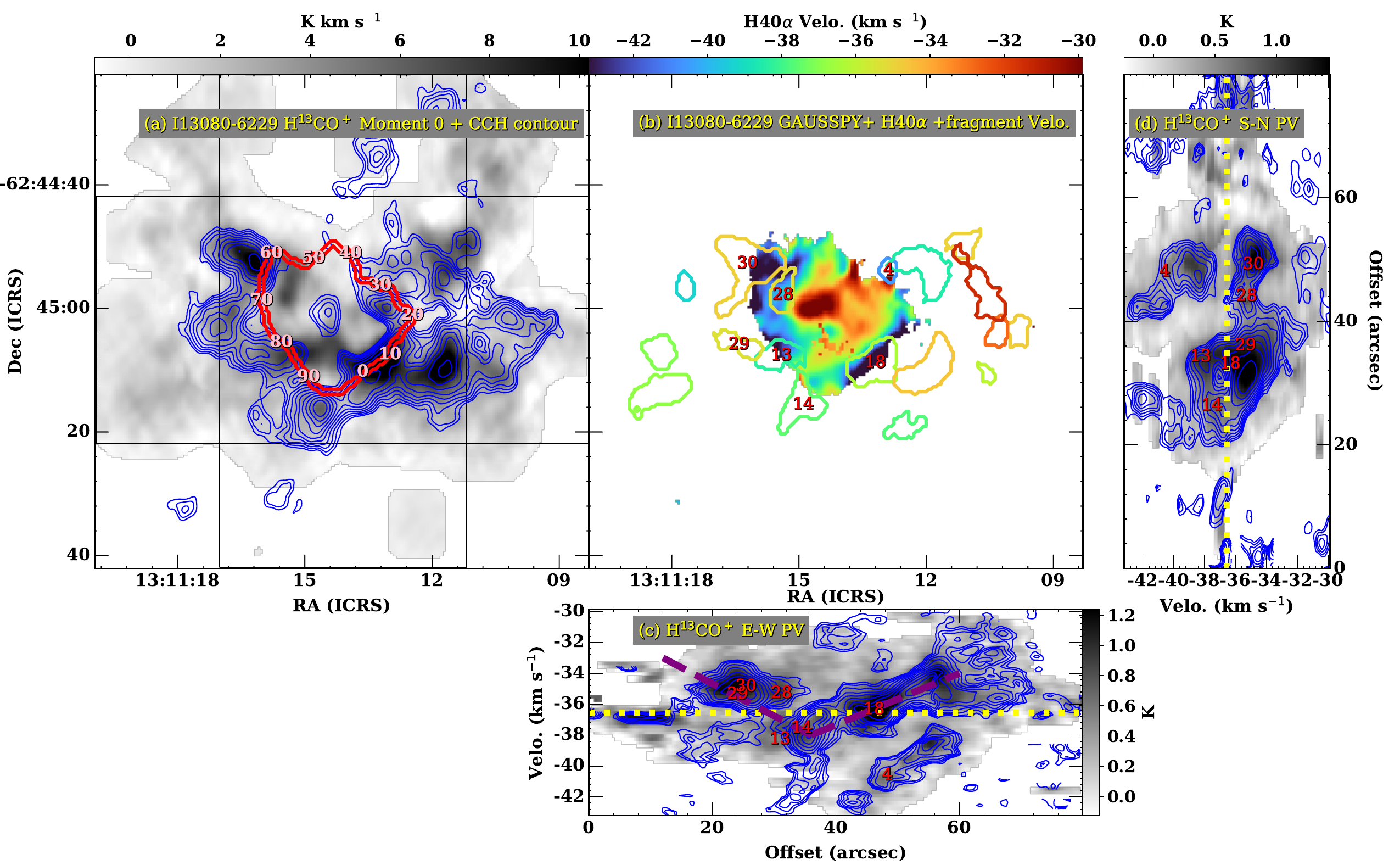}
       \caption{Velocity structure of I13080$-$6229. Panel (a) \htcopone\ moment 0 map overlaid with the contours (blue) of \cch\ emission. The contour levels are the same as in Fig.~\ref{FIGURE:NORMAL1}. The red contour outlines the position-velocity (PV) cut path of \htcopone, \cchline, and \hfourtyalpha\ cubes shown in Fig.~\ref{FIGURE:I13080-6229_EDGE_PV}. The numbers in pink indicate the offsets in Fig.~\ref{FIGURE:I13080-6229_EDGE_PV}. The two vertical and horizontal lines indicate the width and path of PV cuts along the south-north (S-N) and east-west (E-W) directions, respectively. Panel (b) Velocity field of \hfourtyalpha\ (background) and \htcop\ velocity of dense fragments (contour). \hfourtyalpha\ velocity field is generated using \texttt{GAUSSPY$+$} \citep{Riener2019}. Panels (c) and (d) show the PV cuts of \htcop\ cube along the E-W and S-N directions, respectively. The yellow dotted lines mark the system velocity of \hfourtyalpha\ emission. The V-shape velocity structure of the shell is indicated by the purple dashed line in panel (c). The blue contours show the corresponding PV cuts of \cch\ cube with a logarithmic step level. The red numbers indicate the ID of the corresponding N fragments that have a reliable velocity estimate.}
        \label{FIGURE:I13080PV}
\end{figure*}

Kinematic analysis of shell expansion could provide clues on the origin of the ATPCF feature if this feature is caused by regulating distributions of pre-existing fragments. The expansion of shell is not easy to reveal directly from simple mappings, such as moment maps, because of the sky projection bias. Figure~\ref{FIGURE:I13080PV} presents the velocity field of \hfourtyalpha\ and \htcop\ fragments of a splendid example from our sources showing an expansion feature of the shell. The position-velocity (PV) cuts of the \htcop\ and \cch\ cubes along the east-west direction shown in panel (c) present a V-shaped structure with a turning point at offset of 40\arcsec, suggesting a shell expansion velocity of $>2~\rm km s^{-1}$ \citep{Arce2011}. This kind of velocity structure is due to the blue-shifted shell located partially in front of \hii\ region or the red-shifted shell located partially in the back of \hii\ region. Therefore, the estimated expansion velocity is just a lower limit due to projection. The velocity fields of the other eight regions are presented in the supplemental materials available online. Although a careful analysis of the velocity structures can be done for each region, whether it is meaningful from a statistical point of view is doubtful, especially when considering the intricate sky projection and the various techniques adopted in revealing the expansion feature. Therefore, we need a simple statistical parameter to show systematically differences in the kinematics of different types of fragments.

\begin{figure}
\centering
\includegraphics[width=0.47\textwidth]{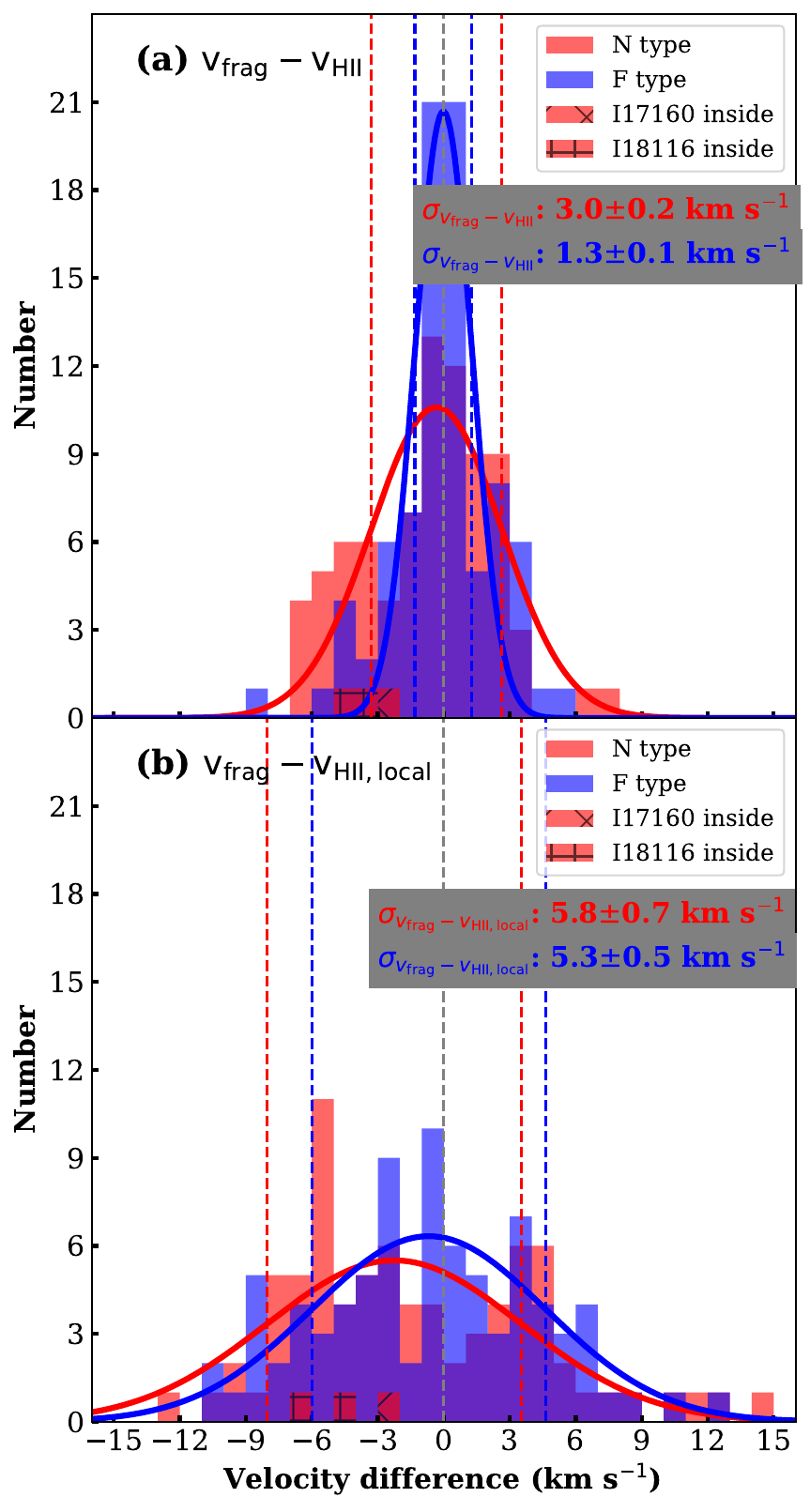}
   \caption{Difference between \hfourtyalpha\ velocity of \hii\ regions and \htcop\ velocity of dense fragments. Panels (a) and (b) show the difference between  \htcop\ velocity of dense fragments and \hfourtyalpha\ system velocity of \hii\ regions (${\rm v_{frag}}-{\rm v_{H~\textsc{ii}}}$) and \hfourtyalpha\ velocity for the ionized gas closest to the corresponding fragments (${\rm v_{frag}}-{\rm v_{H~\textsc{ii},local}}$), respectively. ``I17160 inside'' and ``I18116 inside'' histograms represent the N fragments in the directions of the interior of compact \hii\ regions I17160$-$3707 and I18116$-$1646, respectively. Red and blue curves show the fitted Gaussian of N and F fragment histograms, respectively. Red and blue numbers and vertical lines indicate the Gaussian dispersion of the corresponding ${\rm v_{frag}}-{\rm v_{H~\textsc{ii}}}$ and ${\rm v_{frag}}-{\rm v_{H~\textsc{ii},local}}$ distributions.}
   \label{FIGURE:COMPARISON_DIFF}
\end{figure}

We propose here that the systematic difference in velocity between fragments and \hii\ region (${\rm v_{frag}}-{\rm v_{H~\textsc{ii}}}$) provides important clues about the expansion and its kinematic perturbation on fragment. For dense fragments that are not at the edges of \hii\ regions but are still embedded in the same natal clump, their velocity should have a smaller difference from the system velocity of \hii\ region compared to the fragments at the edges of \hii\ region that are disturbed by the expanding \hii\ region and shell. Figure~\ref{FIGURE:COMPARISON_DIFF} shows the histograms of ${\rm v_{frag}}-{\rm v_{H~\textsc{ii}}}$ for N and F fragments, respectively. The system velocity of \hii\ region $\rm v_{H~\textsc{ii}}$ is derived from a Gaussian fit to the averaged \hfourtyalpha\ spectra of the entire emission region. As shown in Fig.~\ref{FIGURE:COMPARISON_DIFF}, the Gaussian fittings to the histograms of ${\rm v_{frag}}-{\rm v_{H~\textsc{ii}}}$ give a larger histogram dispersion for N fragments and the difference ($\sim1.7$~\kms) is much greater than the measure error of 0.1-0.2~\kms. It suggests that a kinematics perturbation dominated by outward movement driven by the expansion of \hii\ region is prevalent for the N fragments from a statistical point of view.  For I17160$-$3707 and I18116$-$1646, there are few fragments, with a velocity of 3 to 6~\kms\ in comparison to the \hii\ regions, located at an direction close to the center rather than the edges of the \hii\ regions. These special fragments are at the higher end of the observed $|{\rm v_{frag}}-{\rm v_{H~\textsc{ii}}}|$, probably because their outward movement almost follows the line of sight.

\begin{figure*}
\centering
\includegraphics[width=0.85\textwidth]{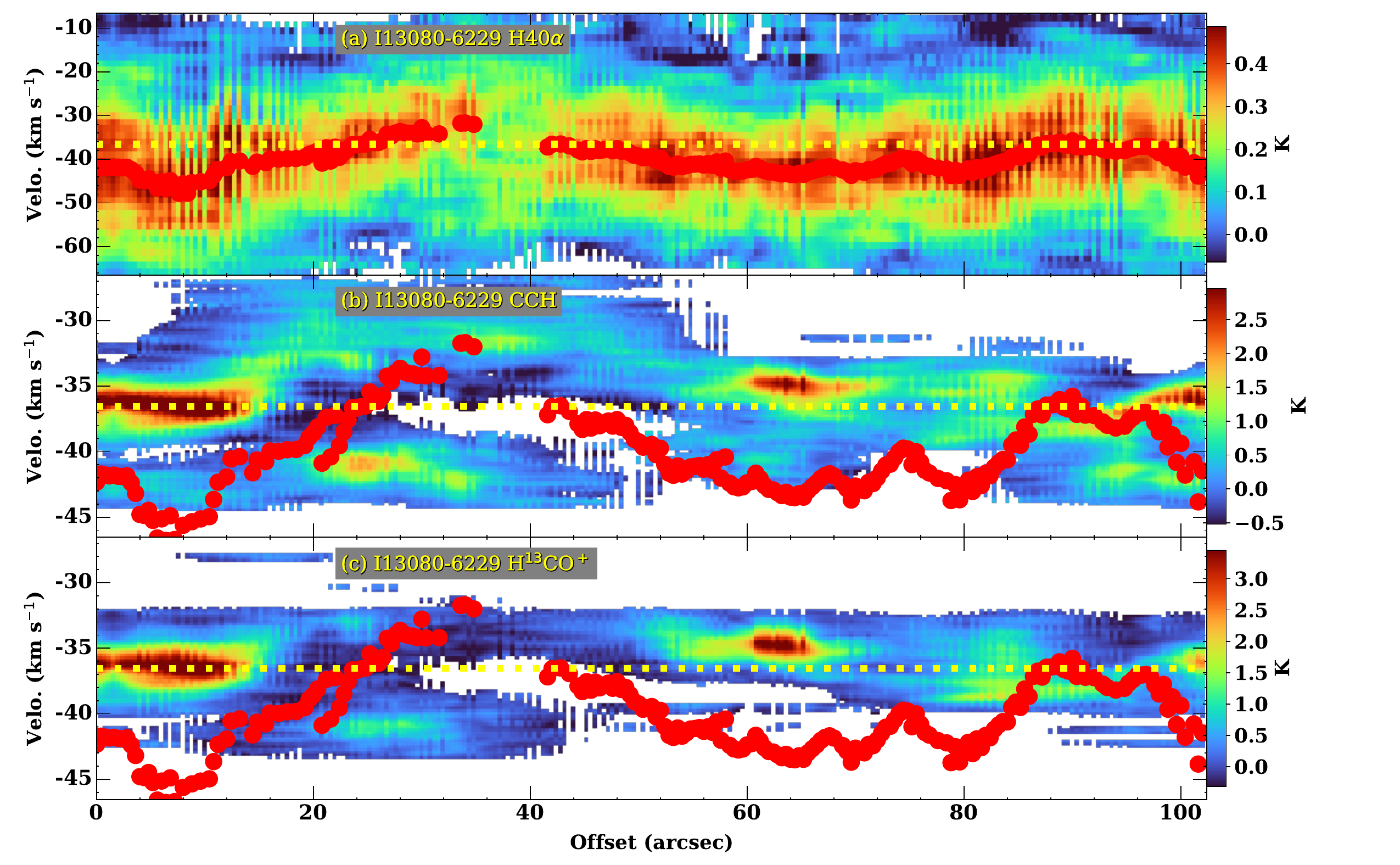}
   \caption{Position-velocity cuts along the edges of \hfourtyalpha\ emission region of I13080$-$6229. Panels (a), (b), and (c) show the cut of \hfourtyalpha, \cchline, and \htcopone\ cubes, respectively. The red dots represent the velocity centers of \hfourtyalpha\ derived from a position-by-position Gaussian fit of the \hfourtyalpha\ position-velocity cut with a threshold of 3$\sigma$. The yellow dotted lines mark the \hfourtyalpha\ system velocity of \hii\ region.}
   \label{FIGURE:I13080-6229_EDGE_PV}
\end{figure*}

Another interesting phenomenon behind the above investigation is the difference in velocity between the N fragments and the \textit{local} ionized gas closest to these fragments (${\rm v_{frag}}-{\rm v_{H~\textsc{ii},local}}$), as shown in panel (b) of Fig.~\ref{FIGURE:COMPARISON_DIFF}. The histograms shed light on the complicated mechanisms behind ${\rm v_{frag}}-{\rm v_{H~\textsc{ii},local}}$. The N fragment ${\rm v_{frag}}-{\rm v_{H~\textsc{ii},local}}$ has a distribution poorly described with Gaussian compared to that of the F fragment. The histogram dispersion difference ($\sim0.5$~\kms) is also at the same level of measurement error ($\sim0.5$~\kms). Figure~\ref{FIGURE:I13080-6229_EDGE_PV} presents the PV cuts along the edges of the \hfourtyalpha\ emission region for I13080$-$6229. The value of ${\rm v_{frag}}-{\rm v_{H~\textsc{ii},local}}$ changes with the edges of the \hii\ region, reflecting the complicated kinematics of the \textit{local} ionized gas with respect to the nearby molecular gas. Photoevaporation and shocks are possible mechanisms to create this difference \citep{Trevino2016}. Furthermore, different models of \hii\ regions, such as bow shock and champagne flow, also present different modes of velocity for the \textit{local} ionized gas \citep{Tenorio1979,vanBuren1990,veena2017}.  A deeper analysis of the ionized gas kinematics is beyond the scope of this paper, and we plan to present it in a subsequent work. We only stress here that the difference in velocity between the fragments and the \textit{local} ionized gas near the fragments ${\rm v_{frag}}-{\rm v_{H~\textsc{ii},local}}$ is not appropriate to be used in our comparison because they are dominated by much more intricate kinematics.

\section{Origin of fragments in molecular shell} \label{SECTION:FRAGMENTATION} 
The origin of fragments in the molecular shells is crucial to understanding the spatial distribution and kinematics of these fragments, as mentioned in Section~\ref{SUBSECTION:FRAGMENTATION-DISTRIBUTION}. Furthermore, these fragments in shells prefer formation of higher-mass stars compared to those outside the shell, and thus ascertaining formation of these shell fragments will help us to understand the important role of the compact \hii\ region and its feedback in HMSF from formation of dense core-scale structures.

\subsection{Fragmentation model for shells swept up by \hii\ regions}   \label{SUBSECTION:CCMODEL}
The expansion of \hii\ regions sweeps up a molecular shell that accumulates more and more mass over time. If the density amplification of the shell dominates over the heating and changes of the magnetic field and turbulence, the shell is expected to experience fragmentation. The produced gas fragments may collapse and then form stars. This process is called as collect \& collapse process \citep[C\&C,][]{Elmegreen1977}. \citet{Whitworth1994} analytically proved that dense fragments condensed from the swept-up molecular shell are preferential sites for HMSF. The work of \citet{Whitworth1994} motivates us to test whether the shell fragments of our compact \hii\ regions have a C\&C origin because of the relative larger mass of our shell fragments. Based on a \hii\ region expansion raw of $R\sim K t^{\alpha}$ where $K$ is a constant and $\alpha = 4/7$ (see equation~\ref{EQU:TDYN}), \citet{Whitworth1994} derived the time when a shell begins to fragment $t_{\rm frag}$, the corresponding radius of shell $r_{\rm frag}$, the mass and separation of the condensed fragments $M_{\rm frag}$ and $D_{\rm frag}$ at $t_{\rm frag}$:
\begin{equation} \label{EQUATION:CC_TIME}
    t_{\rm frag} \sim 1.56\left(\frac{c_{\rm s}}{\rm 0.2~km s^{-1}}\right)^{7/11} \left(\frac{\dot{N}_{\rm ion}}{\rm 10^{49}~s^{-1}}\right)^{-1/11} \left(\frac{n_{\rm ini}}{\rm 10^3~cm^{-3}}\right)^{-5/11} {\rm Myr}
\end{equation}

\begin{equation} \label{EQUATION:CC_RADIUS}
    r_{\rm frag} \sim 5.8\left(\frac{c_{\rm s}}{\rm 0.2~km s^{-1}}\right)^{4/11} \left(\frac{\dot{N}_{\rm ion}}{\rm 10^{49}~s^{-1}}\right)^{1/11} \left(\frac{n_{\rm ini}}{\rm 10^3~cm^{-3}}\right)^{-6/11}{\rm pc}
\end{equation}

\begin{equation} \label{EQUATION:CC_MASS}
    M_{\rm frag} \sim 23\left(\frac{c_{\rm s}}{\rm 0.2~km s^{-1}}\right)^{40/11} \left(\frac{\dot{N}_{\rm ion}}{\rm 10^{49}~s^{-1}}\right)^{-1/11} \left(\frac{n_{\rm ini}}{\rm 10^3~cm^{-3}}\right)^{-5/11}{\rm M_{\odot}}
\end{equation}

\begin{equation} \label{EQUATION:CC_SEPARATION}
    D_{\rm frag} \sim 0.83\left(\frac{c_{\rm s}}{\rm 0.2~km s^{-1}}\right)^{18/11} \left(\frac{\dot{N}_{\rm ion}}{\rm 10^{49}~s^{-1}}\right)^{-1/11} \left(\frac{n_{\rm ini}}{\rm 10^3~cm^{-3}}\right)^{-5/11}{\rm pc},
\end{equation}
where $c_{\rm s}$ is the isothermal sound speed for the shocked gas in shell. \citet{Whitworth1994} assumed sound speed of the ionized gas $c_{\rm i} \sim 10~\rm km~s^{-1}$ and  recombination coefficient $\alpha_{\ast} \sim 2\times10^{-13}~{\rm cm}^{3}~{\rm s}^{-1}$ in the simplification of equations~\ref{EQUATION:CC_TIME} to \ref{EQUATION:CC_SEPARATION}.

These equations have been applied to a number of molecular shells around \textit{evolved large} \hii~regions, to ascertain whether shell fragmentation produces the cores / clumps in the shells. Typically, the studied cases are classical \hii\ regions with a radius from few parsecs such as Sh2$-$212 \citep[$\sim2.0~\rm pc$,][]{Deharveng2008}, N49 \citep[$\sim2.0~\rm pc$,][]{Zavagno2010}, S235 \citep[$\sim2.0~\rm pc$,][]{Kirsanova2014}, RCW 79 \citep[$\sim6.4~\rm pc$,][]{Zavagno2006}, to even nearly ten parsecs \citep{Zhou2020}. \citet{Deharveng2005, Deharveng2010} listed numerous such evolved \hii\ regions with the shell structure. The advantages of using these large and evolved \hii\ regions to investigate the shell fragmentation model are not only that they are easily resolved by single-dish observations, but also that they provide more time for fragmentation to occur in the molecular shell.

The power indices in equations \ref{EQUATION:CC_TIME} to \ref{EQUATION:CC_SEPARATION} show that the fragmentation parameters are dominated by $c_{\rm s}$ and $n_{\rm ini}$. \citet{Whitworth1994} concluded that with an assumed initial cloud density $n_{\rm ini} \sim 10^3~{\rm cm}^{-3}$, the shell is expected to fragment when its column density of hydrogen nuclei reaches 4 to $8\times10^{21}$~\cm. According to equations~\ref{EQUATION:CC_TIME} and \ref{EQUATION:CC_RADIUS}, the fragmentation of shell will occur earlier if the initial density $n_{\rm ini}$ is higher, indicating that it is valuable to test the shell fragmentation model for our \hii\ regions because the shells of our early compact \hii\ regions are embedded in a denser environment with a density of $> 10^4~{\rm cm}^{-3}$.

\subsection{Comparisons with model of \citet{Whitworth1994} and limitations} \label{SUBSECTION:CCLIMITATION}
To constrain the conditions for shell fragmentation using equations~\ref{EQUATION:CC_TIME} to \ref{EQUATION:CC_SEPARATION}, we use the estimated properties of the clumps and \hii\ regions in Section~\ref{SECTION:HII-REGION}. Each fragmentation parameter is calculated using two types of \nini\ that are measured by ATLASGAL in Section~\ref{SECTION:HII-REGION}. The $c_{\rm s}$ of the shocked gas in the shell is estimated with the clump-scale $c_{\rm s}$ derived from the ATLASGAL dust temperature \dustt\ \citep{Urquhart2018}, which is probably a lower limit of $c_{\rm s}$ because the shell is expected to have a higher temperature compared to the entire clump. In the simplifications of equations~\ref{EQUATION:CC_TIME} to \ref{EQUATION:CC_SEPARATION},  \citet{Whitworth1994} adopts a $\alpha_{\ast}$ that is slightly different from the one we used in the calculation of \hfourtyalpha\ emission $\alpha_{\rm B}$, but this difference is negligible for the calculation of shell fragmentation because it only leads to an error of $\sim2\%$. Interstellar dust plays an important role in the equilibrium between ionization and recombination because it absorbs a large fraction of the Lyman continuum at the early evolutionary stage of the \hii\ region. With VLA observations of a sample of compact \hii\ regions, \citet{Garay1993} found that the typical absorption fraction is approximately 55\%.  Dust absorption causes an underestimate of the spectral type of ionizing stars derived from the observed \hfourtyalpha\ but does not significantly change our fragmentation analysis. Taking into account an extreme case where $90\%$ $\dot{N}_{\rm ion}$ are absorbed \citep{Brand2011}, the resulting difference in fragmentation parameters is only $\sim20\%$ according to the weak dependence of \IonRate\ shown in equations~\ref{EQUATION:CC_TIME} to \ref{EQUATION:CC_SEPARATION}.

\begin{figure*}
\centering
\includegraphics[width=0.95\textwidth]{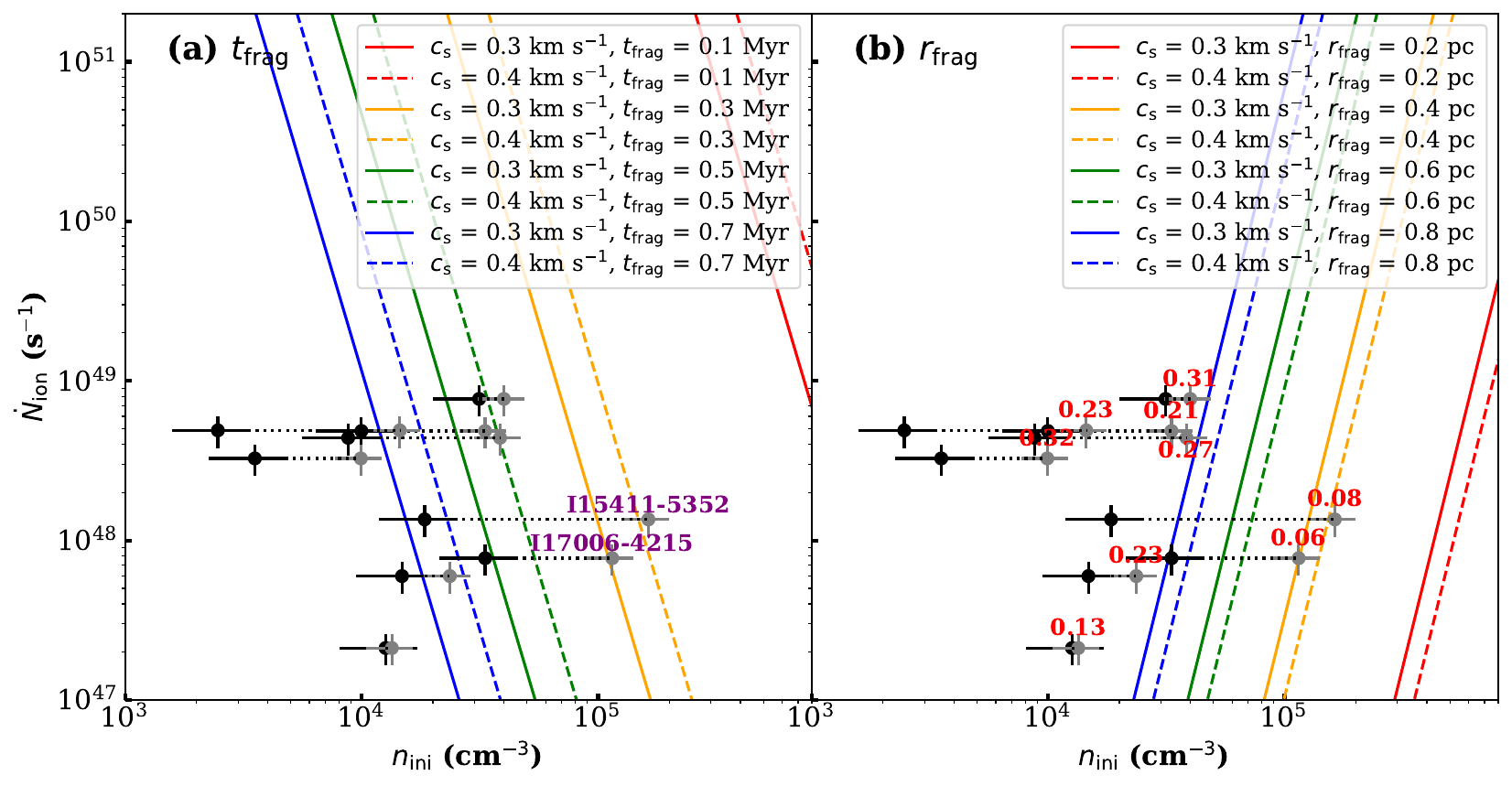}
       \caption{Time and radius for starting shell fragmentation. Panels (a) and (b) show the $n_{\rm ini}$-$\dot{N}_{\rm ion}$ relations with certain $t_{\rm frag}$ or $r_{\rm frag}$, respectively. Lines of different colors represent different conditions to start fragmentation at $t_{\rm frag}$ or $r_{\rm frag}$. For example, if shell fragmentation starts at 0.3~Myr when $c_{\rm s} = 0.3$~\kms, the minimum required \nini\ and \IonRate\ should follow the yellow solid line in panel (a). The black and gray dots represent the parameters derived in \textit{case i} and \textit{case ii}, respectively (see Section~\ref{SECTION:HII-REGION}). The numbers in panel (b) are the observed radius of the corresponding \hii\ region $r_{\rm H40\alpha,eff}$ in units of pc.}
        \label{FIGURE:CC}
\end{figure*}

The shell fragmentation parameters predicted for our regions are listed in Table~\ref{TABLE:CC}. To explore the conditions for shell fragmentation, we show the relations of \nini, \IonRate, $t_{\rm frag}$, and $r_{\rm frag}$ in Fig.~\ref{FIGURE:CC}. The color lines in Fig.~\ref{FIGURE:CC} mark the time and radius required for shell fragmentation under certain conditions of \nini, \IonRate, and $c_{\rm s}$. On the one hand, most regions require $t_{\rm frag} \sim 0.5$ to 0.7~Myr for shell fragmentation as shown in Fig.~\ref{FIGURE:CC}, slightly older than the typical lifetime of UC\hii\ regions $\sim0.3$~Myr given by \citet{Wood1989b} and seems to be in line with the lifetime estimate of our early compact \hii\ regions $\gtrsim0.3$~Myr (Section~\ref{SECTION:HII-REGION}). On the other hand, the initial shell radii required for fragmentation $r_{\rm frag}$ shown in panel (b) of Fig.~\ref{FIGURE:CC}, $\sim0.8$~pc, are much larger than the measured radii of \hii\ regions $\sim0.2~\rm pc$.

We should note that $t_{\rm frag}$ and $r_{\rm frag}$ here are substantially underestimated. First, $c_{\rm s}$ of shocked shell is replaced by the clump-scale $c_{\rm s}$ in shell fragmentation
equations although weak relations with a power index of 7/22 or 2/11 between temperature and $t_{\rm frag}$ or $r_{\rm frag}$ suggest that the resultant error is trivial. Second, the shell fragmentation analysis of \citet{Whitworth1994} has limitations inherited from the used expansion model of \hii\ region. A classical monotonic expansion law $R\sim K t^{\alpha}$ where $K$ is a constant and $\alpha = 4/7$ is used both in the fragmentation model of \citet{Whitworth1994} and in the estimation of dynamical age \tdyn\ (equation~\ref{EQU:TDYN}). The typical lifetime of UC\hii\ region independently estimated from the statistics of massive stars \citep{Wood1989b} is much longer than the dynamic age estimated using $R\sim K t^{4/7}$, suggesting that the averaged expansion before the compact stage is much slower. The expansion before the compact stage is highly variable and sometimes even non-monotonic \citep{Peters2010a, Peters2010b}. Several additional confinement mechanisms are used to explain the slower expansion and lifetime problem, such as additional pressure from ambient gas \citep{Garcia1996,Xie1996} and complicated interplay with infall materials \citep{Peters2010a, Peters2010b}. Physically, a slower expansion corresponds to a smaller time-averaged $K$ or/and $\alpha$, and also a longer time to witness shell fragmentation (detailed $K$-$t_{\rm frag}$ and $\alpha$-$t_{\rm frag}$ relations can be found in Appendix B of \citet{Whitworth1994}). Therefore, $t_{\rm frag}$ is underestimated because it is derived from an expansion law faster than the actual expansion history. Furthermore, the underestimate is substantial when considering the magnitude difference between the dynamical age estimated from the classical expansion law and the real lifetime of UC\hii\ region.

The last factor that can also cause an underestimated $t_{\rm frag}$ is the cometary shape of our \hii\ regions. The cometary morphology suggests that numerous ionized gas and ionizing photons could easily escape from the confinement of the molecular shell, leading to underestimate the dynamical age of \hii\ region and the time for shell fragmentation. According to equation~\ref{EQUATION:CC_TIME}, a loss of available ionizing photons causes a longer time for shell fragmentation to occur. Less ionizing photons participate in powering the formation and evolution of molecular shells for cometary \hii\ regions, thus reducing the possibility of shell fragmentation in general.

Taking into account the uncertainties and limitations discussed above, we propose that $t_{\rm frag}$ and $r_{\rm frag}$ are generally larger than the lifetime and size of our early compact \hii\ regions, which is against the shell fragmentation origin of fragments. Even considering the higher density situation (\textit{case ii}) in which fragmentation will occur earlier, most of the compact \hii\ regions studied here are still too young for shell fragmentation. Figure~\ref{FIGURE:CC} indicates that an extremely dense environment with $n_{\rm ini}\gtrsim10^5~\rm cm^{-3}$ is required if these dense fragments are the result of shell fragmentation. Actually, this kind of high density reaches the regimes of 0.1-pc-scale dense cores in candidate high-mass starless clumps \citep{Li2019, Sanhueza2019, PaperII, Morii2023} and therefore cannot be the global initial density at the clump scale in general situation.

\begin{table*}
\caption{\label{TABLE:CC}fragmentation parameters.} 
\begin{threeparttable}
\setlength{\tabcolsep}{4.0pt}
\renewcommand{\arraystretch}{1.0}
\centering
\begin{tabular}{crrrrrrrrrrrr}
\hline
\hline
name        & \multicolumn{2}{c}{$t_{\rm frag}$}  &   \multicolumn{2}{c}{$r_{\rm frag}$} & \multicolumn{2}{c}{$M_{\rm frag}$}   & \multicolumn{2}{c}{$D_{\rm frag}$}        &  \multicolumn{2}{c}{$M_{\rm Jeans}$} &  \multicolumn{2}{c}{$L_{\rm Jeans}$} \\
     &  \multicolumn{2}{c}{Myr}   &   \multicolumn{2}{c}{pc}           
&  \multicolumn{2}{c}{\msun}           &   \multicolumn{2}{c}{pc}                  &  \multicolumn{2}{c}{\msun}  &     \multicolumn{2}{c}{pc}   \\ 
\hline
          & \textit{i} & \textit{ii} & \textit{i} & \textit{ii} & \textit{i} & \textit{ii} & \textit{i} & \textit{ii} & \textit{i} & \textit{ii} & \textit{i} & \textit{ii} \\
\hline
I13080$-$6229 & 0.83(0.14) & 0.48(0.05) & 1.89(0.37) & 0.98(0.12) & 62.5(15.3) & 36.1(7.6) & 0.76(0.14) & 0.44(0.06) & 22.7(5.3) & 12.4(2.3) & 0.50(0.09) & 0.27(0.03) \\
I15411$-$5352 & 0.68(0.11) & 0.25(0.03) & 1.18(0.23) & 0.36(0.04) & 43.3(10.6) & 16.1(3.4) & 0.59(0.11) & 0.22(0.03) & 14.1(3.3) & 4.8(0.9)  & 0.35(0.06) & 0.12(0.01) \\
I15570$-$5227  & 1.30(0.22) & 0.81(0.09) & 3.11(0.62) & 1.77(0.22) & 75.9(18.6) & 47.5(9.9) & 1.10(0.20) & 0.69(0.09) & 29.4(6.9) & 17.6(3.3) & 0.77(0.14) & 0.46(0.06) \\
I15584$-$5247 &  0.75(0.13) & 0.60(0.07) & 1.18(0.23) & 0.91(0.11) & 33.1(8.1)  & 26.8(5.6) & 0.57(0.11) & 0.46(0.06) & 10.9(2.6) & 8.7(1.6)  & 0.34(0.06) & 0.27(0.03) \\
I16362$-$4639 &  0.88(0.15) & 0.86(0.09) & 1.17(0.23) & 1.13(0.14) & 39.4(9.7)  & 38.3(8.0) & 0.68(0.13) & 0.66(0.09) & 11.9(2.8) & 11.6(2.2) & 0.37(0.07) & 0.36(0.04) \\
I17006$-$4215 &  0.53(0.09) & 0.30(0.03) & 0.80(0.16) & 0.41(0.05) & 29.3(7.2)  & 16.7(3.5) & 0.44(0.08) & 0.25(0.03) & 9.1(2.1)  & 4.9(0.9)  & 0.25(0.05) & 0.13(0.02) \\
I17160$-$3707 &  0.45(0.07) & 0.40(0.04) & 1.02(0.20) & 0.89(0.11) & 25.7(6.3)  & 23.0(4.8) & 0.37(0.07) & 0.33(0.04) & 9.8(2.3)  & 8.7(1.6)  & 0.26(0.05) & 0.23(0.03) \\
I18116$-$1646 &  0.88(0.15) & 0.45(0.05) & 2.00(0.40) & 0.89(0.11) & 65.8(16.2) & 33.6(7.0) & 0.81(0.15) & 0.41(0.05) & 23.9(5.6) & 11.4(2.1) & 0.53(0.10) & 0.25(0.03) \\
I18317$-$0757 &  1.51(0.25) & 0.67(0.07) & 3.97(0.79) & 1.51(0.19) & 95.6(23.5) & 42.7(8.9) & 1.30(0.24) & 0.58(0.08) & 38.4(9.0) & 15.9(3.0) & 0.95(0.18) & 0.39(0.05) \\
\hline
   \end{tabular}
      \begin{tablenotes}
      \item[] \textit{i} and \textit{ii} are derived from ATLASGAL clump properties measured by extraction algorithms \texttt{Gaussclump} and \texttt{SExtractor}, respectively. \textit{i} and \textit{ii} are suitable to trace the global properties and the dense parts of the clump, respectively, as described in Section~\ref{SECTION:HII-REGION}.
      \end{tablenotes}
      \end{threeparttable}
\end{table*}

\subsection{Pre-existing fragments produced from a non-triggered but spontaneous fragmentation process}  \label{SUBSECTION:CCMODELDISCUSSION}

\begin{figure}
\centering
\includegraphics[width=0.47\textwidth]{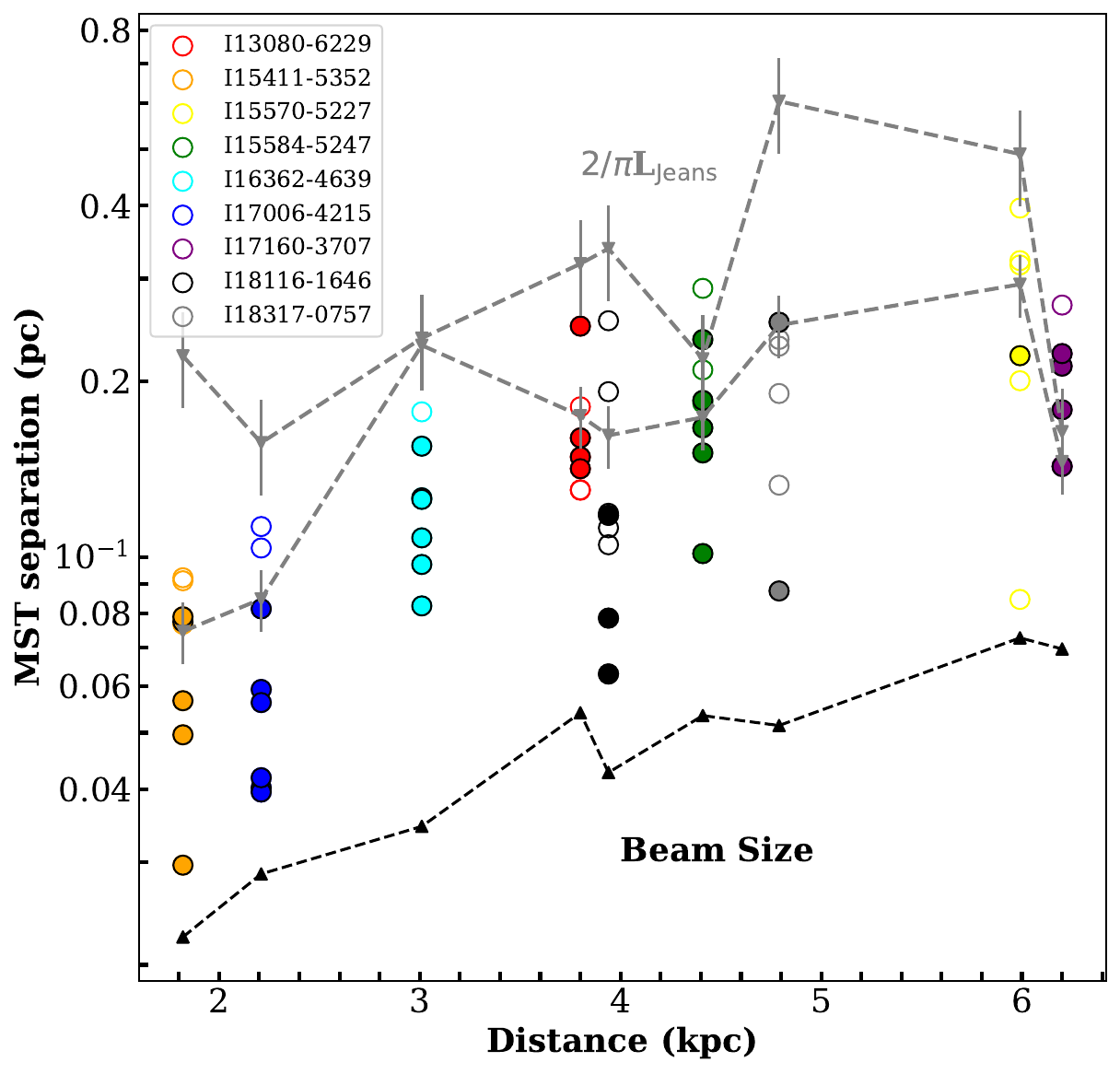}
       \caption{\texttt{MST} separations for the N fragments in shells. The black and gray dashed lines indicate the corresponding beam size and the projection-corrected thermal Jeans length $2/\pi L_{\rm Jeans}$, respectively.  Jeans lengths $L_{\rm Jeans}$ are estimated in two situations (\textit{i} and \textit{ii}) to characterize global properties (upper gray line) and compact parts (lower gray line) of the clumps. $2/\pi$ is the projection correction factor \citep{PaperII}. Solid dots and blank dots represent fragments that are in the same cluster of fragments and that are in the different cluster of fragments. The fragment cluster is identified by \texttt{SCIMES} \citep[Spectral Clustering for Interstellar Molecular Emission Segmentation,][]{Colombo2015} in order to show the potential hierarchical structures between the clump scale and the fragment scale in the mapped regions.}
        \label{FIGURE:SEP}
\end{figure}

What is the origin of the dense fragments observed in the molecular shell of the compact \hii\ regions when they are very unlikely to be produced by shell fragmentation?  The initial mass $M_{\rm frag}$ and separation $D_{\rm frag}$ of the fragments when the shell begins to fragment are much larger than the mass and separation of the observed fragments, as listed in Table~\ref{TABLE:CC}. This result is also against the origin of shell fragmentation and motivates us to compare with the predictions of the Jeans fragmentation from clump to fragment because the mass and separation may match in this situation. With the assumption of an infinite and homogeneous initial gas dominated by thermal motions \citep{Palau2015, PaperII},  we show the predicted separations of the Jeans fragmentation $L_{\rm Jeans}$ in Fig.~\ref{FIGURE:SEP}. $L_{\rm Jeans}$ is estimated using two types of \nini\ mentioned in Section~\ref{SECTION:HII-REGION} to partially shed light on fragmentation in the global clump environment and the dense regions within the clump, respectively. Note that the estimated $L_{\rm Jeans}$ is based on the current characteristics of the clump and probably does not accurately reflect the initial condition when the clump began to fragment. $L_{\rm Jeans}$ is probably overestimated if \nini\ does not change much because the current temperature of the clump is higher than that of the clump that just started to fragment in the infrared quiescent stage. We compare the $L_{\rm Jeans}$  with the \texttt{MST} separations of these shell fragments (lime lines in Fig.~\ref{FIGURE:NORMAL2}). The typical \texttt{MST} separations are $\sim0.05~\rm pc$ and $\sim0.2~\rm pc$ for the closest and most distant regions, and the minimum separation depends in part on resolution limitations.  Most fragment separations are similar to or shorter than $L_{\rm Jeans}$ as shown in Fig.~\ref{FIGURE:SEP}, suggesting that fragments in shells can be explained as a result of clump-to-fragment Jeans fragmentation predominant by thermal motions. A fragmentation dominated by turbulence is very unlikely because the corresponding fragmentation mass of $\gtrsim 100$~\msun\ and length of $\gtrsim 1$~pc derived from the clump-scale velocity dispersion in Table~\ref{TABLE:SAMPLE} are much larger than the observed mass and separations of fragments.

\subsection{Role of stellar winds in shell fragmentation}
Stellar winds ejected from massive stars can directly deposit kinetic energy in the surroundings to aid the formation and expansion of \hii\ region and its shell, but whether they are dominant over the ionization feedback is highly controversial \citep{Pabst2020}. Several simulations, such as \citet{Dale2014}, \citet{Haid2018}, and \citet{Ali2022}, studied the combined effect of photoionization and winds and then concluded that photoionization is the main driver, but \citet{Dale2014} suggested that stellar winds can still play a significant role in shaping the morphology of low-mass clouds. 

\citet{Geen2020} developed a simple algebraic model to depict the expansion of photoionized \hii\ regions that accompany the feedback of photoionization, stellar wind, and radiation in a natal cloud with a density profile of $r^{-2}$. These authors derived a factor, $C_{\rm w}$, to show the relative importance of the stellar wind and photoionization in driving expansion of \hii\ region:
\begin{equation} \label{EQUATION:CW}
\begin{aligned}
    C_{\rm w} =~& 0.009 31 \left(\frac{\dot{p}_{\rm w}}{\rm 10^{28}~g~cm~s^{-2}}\right)^{3/2} \left(\frac{\dot{N}_{\rm ion}}{\rm 10^{49}~s^{-1}}\right)^{-3/4} \\
                & \times \left(\frac{M_{\rm 0}}{\rm 100~M_{\odot}}\right)^{-3/8}  \left(\frac{{\Sigma}_{\rm 0}}{\rm 100~M_{\odot}~pc ^{-2}}\right)^{3/8}  \left(\frac{{c}_{\rm i}}{\rm 10~km~s^{-1}}\right)^{-3},
\end{aligned}
\end{equation}
where $\dot{p}_{\rm w}$ is the momentum deposition rate of stellar wind. $M_{\rm 0}$ and $\Sigma_{\rm 0}$ are the total mass of the natal cloud within the typical radius $r_{\rm 0}$ and the surface density of the natal cloud at $r_{\rm 0}$, respectively. Following a density profile of the natal cloud $n\left(r\right) = n_{\rm 0} \left(r/r_{\rm 0}\right) ^{-2}$, $M_{\rm 0}$ and $\Sigma_{\rm 0}$ are corresponded to the typical density $n_{\rm 0}$ and radius $r_{\rm 0}$ of the natal cloud by \citep[see details in][]{Geen2020}
\begin{equation}
M_{\rm 0} = 1.4 \times 4 \pi m_{\rm H} n_{0} r_{0}^3,~
\Sigma_{\rm 0} = 1.4 \pi m_{\rm H} n_{0} r_{0}.
\end{equation}
$C_{\rm w} < 1$ and $C_{\rm w} > 1$ represent situations in which photoionizations and stellar winds are the main dynamical drivers of the \hii\ region, respectively.

We now derive the typical $C_{\rm w}$ for our compact \hii\ regions, to ascertain the roles of stellar winds in the dynamics of the studied shells and compact \hii\ regions.  We consider two typical cases in which the natal cloud is a core-scale structure with $r_{\rm 0} = 0.05~\rm pc$ and $n_{\rm 0} = 10^6~\rm cm^{-3}$ or the natal cloud is a clump-scale structure with $r_{\rm 0} = 0.5~\rm pc$ and $n_{\rm 0} = 10^4~\rm cm^{-3}$. The first case corresponds to the natal clouds of \hii\ regions at the HC stage, while the second case is representative natal cloud of the compact \hii\ regions in our sample. The momentum deposition rate of  stellar wind is given as $\dot{p}_{\rm w} = {\rm v}_{\rm w} \dot{M}_{\rm w}$ and where ${\rm v}_{\rm w}$ and $\dot{M}_{\rm w}$ are the velocity and the mass loss rate of stellar wind, respectively. We assume typical values ${\rm v}_{\rm w} = 2000~{\rm km~s^{-1}}$ and $\dot{M}_{\rm w} = 10^{-6}~{\rm M_{\odot}~yr^{-1}}$ \citep{Vink2001} because there is no measure of ${\rm v}_{\rm w}$ and $\dot{M}_{\rm w}$ for massive stars in our regions.  With $\dot{N}_{\rm ion} = 10^{48}~\rm s^{-1}$ which is probably a lower limit due to dust absorption, we have $C_{\rm w}$ of 0.4 and 0.08 for these two cases. This simple calculation shows that the stellar wind plays a non-negligible role in the dynamics of HC\hii\ regions, but its importance probably decreases to less than 10\% when \hii\ regions evolve to UC and compact stages.  \citet{Geen2021} further confirmed in their detailed radiative magnetohydrodynamic simulations that the contribution of stellar winds to the total radial momentum of the expanding cloud is at 10\% level of the photoionization feedback in the first Myr of massive star lifetime, and therefore stellar winds on the dynamics of the early compact \hii\ regions are generally unimportant. 

Based on the unimportant role of stellar winds in the dynamics of the compact \hii\ region proven above, we naturally propose that stellar winds are also generally trivial for the evolution and fragmentation of molecular shells driven by these compact \hii\ regions. In extreme cases such as one order higher of the mass loss rate for stellar wind, the contribution of stellar wind could be dominant over that of photoionization. \citet{Whitworth1994} also derived the fragmentation time and radius for the molecular shell mainly powered by stellar winds:
\begin{equation} \label{EQUATION:CC_WIND__TIME}
    t_{\rm frag,w} \sim 0.9~{\rm Myr}~\left(\frac{c_{\rm s}}{\rm 0.2~km~s^{-1}}\right)^{5/8} \left(\frac{L_{\rm w}}{\rm 10^{37}~erg~s^{-1}}\right)^{-1/8} \left(\frac{n_{\rm ini}}{\rm 10^3~cm^{-3}}\right)^{-1/2},
\end{equation}

\begin{equation} \label{EQUATION:CC_WIND_RADIUS}
    r_{\rm frag,w} \sim 9.6~{\rm pc}~\left(\frac{c_{\rm s}}{\rm 0.2~km~s^{-1}}\right)^{3/8} \left(\frac{L_{\rm w}}{\rm 10^{37}~erg~s^{-1}}\right)^{1/8} \left(\frac{n_{\rm ini}}{\rm 10^3~cm^{-3}}\right)^{-1/2},
\end{equation}
where $L_{\rm w} = 1/2 \dot{M}_{\rm w} {\rm v}_{\rm w}^2$ is mechanical luminosity of stellar wind. Adopting an extreme mass loss rate of $10^{-5}~{\rm M_{\odot}~yr^{-1}}$, the estimated $t_{\rm frag,w}$ and $r_{\rm frag,w}$ are around 0.4~Myr and 3.7~pc when $c_{\rm s}$ is set as the typical value of 0.32 \kms\ for our compact \hii\ regions. The $t_{\rm frag,w}$ and $r_{\rm frag,w}$ are also largely underestimated due to the limitations of stellar wind bubble expansion model used in \citet{Whitworth1994}, similar to the situations of \hii\ region expansion model. It shows that the fragments in the shell of compact \hii\ regions are also hard to be explained as shell fragmentation powered by extreme stellar winds.\\

Hereto, we have justified that dense fragments located in the molecular shell of early compact \hii\ regions do not come from shell fragmentation. Instead, they are likely to arise from a spontaneous Jeans fragmentation process prior to the development of \hii\ regions. This conclusion does not change when considering uncertainties such as the morphology of \hii\ region, dust absorption, and the effect of stellar winds. The expanding non-fragmented shell encountered the pre-existing fragments produced by initial clump fragmentation at the previous time, and then the shell swept up the fragments. The fragment kinematics was then disturbed by expansion and some of the irradiated fragments were photoevaporated. This process not only forms the features of expansion and velocity difference seen in Section~\ref{SUBSECTION:EXPANSION} but also shapes the spatial distribution of the fragments to produce an ATPCF that typically peaks at the edges of the \hii\ region, as seen in Section~\ref{SUBSECTION:FRAGMENTATION-DISTRIBUTION}.

\section{discussion}
\label{SECTION:SCENARIO}

\subsection{Mass growth of fragments and core formation efficiency}
In the above section, we proposed that fragments are likely to form by the Jeans fragmentation of their natal clumps. The fact that the mass of F fragments ($\sim 10$~\msun) is much lower than that of N fragments ($\sim 25$~\msun) suggests a different mass growth process since the initial Jeans fragmentation. This difference is partially related to the feedback from \hii\ regions and the special environment of the swept-up molecular shell. The shell provides a mass reservoir, which is denser and more bounded than other locations in the natal clump, to be more easily accreted by the pre-existing fragments. Under this explanation, there should be a more significant mass growth with the evolution of the \hii\ region for the N fragment due to its higher accretion rate. 

\begin{figure}
\centering
\includegraphics[width=0.47\textwidth]{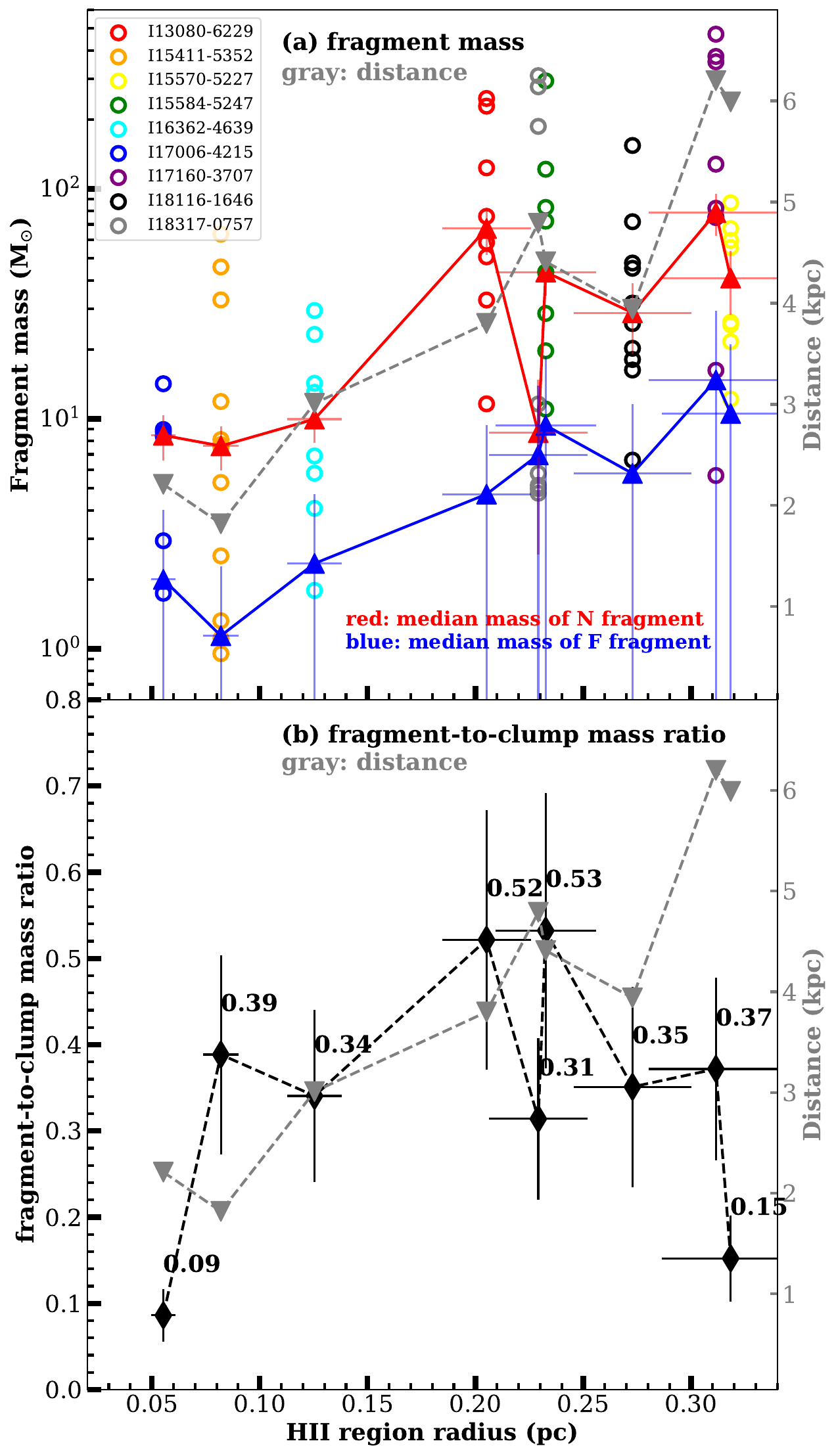}
   \caption{Fragment mass, fragment-to-clump mass ratio, and their relations with radius of corresponding \hii\ regions. The distance to the Earth for each region is indicated by gray dashed lines (corresponding to right y-axis) in panels (a) and (b). Panel (a) The red and blue lines indicate the variation of N and F fragments median mass with radius of \hii\ region, respectively. Dots only show the N fragments. Panel (b) Mass ratio of total N fragments to natal clump (black lines and triangles). The numbers indicate the corresponding values of mass ratio. }
        \label{FIGURE:MASS_GROWTH}
\end{figure}

Panel (a) of Fig.~\ref{FIGURE:MASS_GROWTH} shows the relationship between the mass of fragments and the size of corresponding \hii\ region. The positive correlations between the individual mass of fragment and the size of \hii\ region presented in Fig.~\ref{FIGURE:MASS_GROWTH} probably indicate a growth of the fragment mass during expansion of the \hii\ regions \citep{Zhou2023}, but the difference in this growth trend is not significant between the N and F fragments. The sample size of this work is probably too small to distinguish the difference. Furthermore, distance bias in our sample can also produce a positive correlation between fragment mass and size of \hii\ region as shown in Fig.~\ref{FIGURE:MASS_GROWTH}. Therefore, we cannot confirm or reject the scenario in which N fragments have a higher mass accretion rate from our limited observations \citep{Yi2021}. From a physical point of view, a higher accretion rate is possible for N fragments. With the assumption that the fragment in a shell gains a mass of 10~\msun\ in the 0.1~Myr-duration of \hii\ region expansion, the associated accretion rate is of the order of $10^{-4}$~\msun~year$^{-1}$ and comparable to the high end of the accretion rate at the core scale in some observations \citep{Contreras2018, Wells2024}. 

Another explanation for the mass difference between the N and F fragments in addition to the increased accretion scenario is the coalescence of swept-up shell's material and pre-existing fragments after their encounter. It can contribute additional mass to some of the N fragments, but it is hard to distinguish this from the increased accretion scenario using current data sets. A recent survey toward ATOMS sample using ACA carried out by \citet{Xu2024b} finds that the dense gas fraction of ACA-detected structures to the ATLASGAL clump increases with evolution, providing indirect evidence of shell collection during expansion of \hii\ region.  A higher resolution and sensitivity survey that is able to resolve the infall (or accretion) streams embedded in the fragments without the distance bias is the key to reveal the accretion and massive nature of the fragments in the shells. 

The mass ratio of the total shell fragments to the natal clump is less biased by distance because distance has a similar effect on the extraction and property estimation for both fragment and clump. Using the fragment mass derived from \htcop\ and the natal clump mass derived from ATLASGAL \citep{Urquhart2018}, we show the mass ratio of the total N fragment to the natal clump in panel (b) of Fig.~\ref{FIGURE:MASS_GROWTH}. The median ratio is around 0.3, with maximum and minimum values of around 0.5 and 0.1, respectively. It probably suggests very efficient star formation in these regions, but note that the different mass probes for fragment and clump (\htcop\ and dust continuum) can bias the absolute value. We are more interested in the evolution of this mass ratio, which is less affected by mass probe bias. The correlation between the mass ratio and the size of \hii\ region is not simply positive or negative, as shown in Fig.~\ref{FIGURE:MASS_GROWTH}. It indicates that the natal clump of a compact \hii\ region probably also gains mass from a larger environment when the fragments in a shell experience mass growth, causing the clump mass to increase and the mass ratio to decrease or become stable, similar to the processes described in the global hierarchical collapse model \citep{Vazquez-Semadeni2017}. A simple cross-match with the candidate hub-filament systems identified from the Herschel images by \citet{Kumar2020} further confirms that at least six of our compact \hii\ regions are embedded in the dense hub, indicating the existence of larger-scale mass accretion.

\subsection{Dense gas fragments in an evolutionary scenario from IRDC stage to compact \hii\ region stage}

\begin{figure*}
\centering
\includegraphics[width=0.95\textwidth]{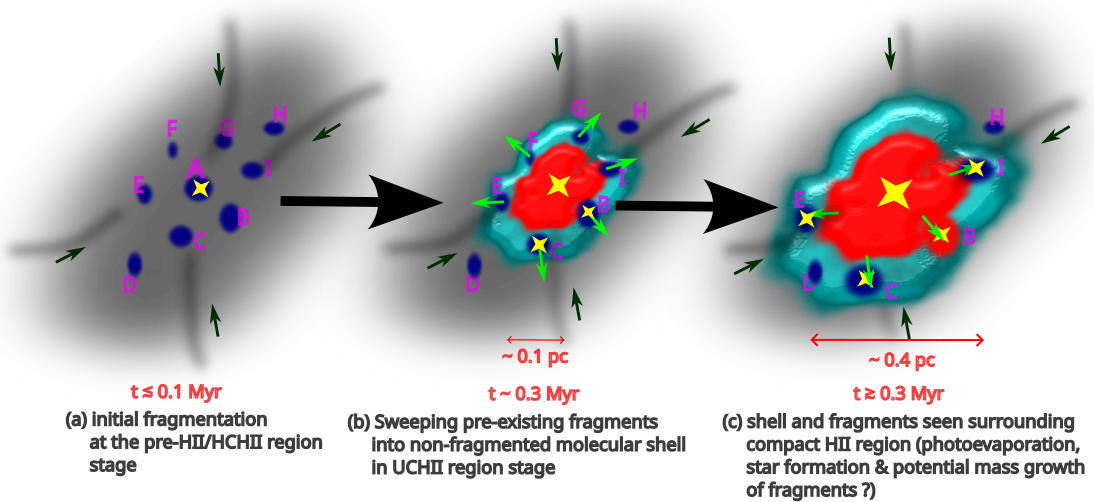}
       \caption{Proposed scenario for the evolution of fragments and associated molecular shell in early compact \hii\ regions. The small black arrows indicate the potential inflows because we find most of the natal clumps of our compact \hii\ regions are observed in the hubs. The green arrows indicate the outward displacement of the pre-existing fragments due to feedback from the \hii\ regions. Some low-mass pre-existing fragments (such as F and G) are photoevaporated during expansion from the UC to the compact stage.}
        \label{FIGURE:MORPHOLOGY}
\end{figure*}

Based on observations of the early compact \hii\ regions in this work and previous studies of UC\hii\ regions and their precursors, we propose a simplified evolutionary scenario to describe how dense gas fragments interact with the ionization feedback up to the early compact \hii\ region stage. A diagram of this scenario, including three stages, is shown in Fig.~\ref{FIGURE:MORPHOLOGY}. The time scale of each stage can be marked by typical ages of the HC, UC, and compact \hii\ regions, which are $\lesssim0.1$, $\sim0.3$ and $\gtrsim0.3$~Myr, respectively \citep{Motte2018}.

\begin{itemize}
    \item (a) Initial fragmentation at the pre-\hii\ / HC\hii\ region stage. At the earliest stage of high-mass star formation, the pc-scale infrared-quiescent massive clump undergoes hierarchical thermal Jeans fragmentation with a typical separation of $\sim0.1~\rm pc$ for the first-level fragmentation and $\sim0.03~\rm pc$ for the hierarchical second-level fragmentation \citep{Svoboda2019, PaperII}.  Jeans fragments continue to gain masses from the surroundings and grow as relatively massive fragments \citep{LiuHL2023,Xu2024}. The fragments produced at this early stage do not show mass segregation \citep{Sanhueza2019,Morii2023}, revealing that the relatively massive fragments that can evolve to HC\hii\ regions \citep{Motte2018,Yang2019} have no preferred location compared to low-mass fragments. We analyze the ATPCF for the most massive dust cores and their surrounding cores in the 70-\micron\ dark massive clumps of \citet{Morii2023} and find that the radial distribution of the surrounding cores with respect to the most massive core is quite random, further indicating the random distribution of fragments in the initial stage if we assume that the most massive core in the 70-\micron\ dark massive clump will evolve to \hii\ region later.
    
    The development of a HC\hii\ region cleans a tiny molecular cavity, which is detected in some observations \citep{Hunter2008,Klaassen2018}. The ionization front (IF) at the HC stage is quite local. For example, using masers \citet{Moscadelli2018, Moscadelli2021} detected a possible fast expansion signature for the ionized gas in the HC\hii\ region G24.78$+$0.08 that is still embedded in the parental fragment. Interestingly, in HC\hii\ region W49 N:A2 \citet{Miyawaki2023} detected a small ionized gas ring (radius $\sim700$~AU) expanding with a velocity of $\sim13$~\kms\ and these authors proposed that it could be the remnant of the accretion disk. We suggest that the ionization feedback at the HC\hii\ region stage only interacts with its parental fragment.
    
   \item (b) Sweeping pre-existing fragments into non-fragmented molecular shell in the UC\hii\ region stage. The molecular shell in this stage becomes dense and large enough to be evident in millimeter interferometric observations even few studies reveal that shell can be detected in earlier stages. In ALMA observations \citet{Fernandez2021} reported a dense dust shell with a radius of $\sim5000~\rm AU$ and a mass of $\sim100$~\msun\ in HC\hii\ region G5.89$-$0.39. Considering that the typical scale of UC\hii\ regions and the Jeans length of infrared-quiescent massive clump are both $\sim0.1~\rm pc$, it is expected that as the expansion the IF arrives and interacts with nearby pre-existing fragments at the UC\hii\ region stage \citep{Trevino2016} if these fragments are located in situ over time up to the UC\hii\ region stage. The expansion at HC and UC stages is highly variable and even not necessary monotonic due to complicated interplay with inhomogeneous and structured ambient molecular gas, as discussed in Section~\ref{SUBSECTION:CCLIMITATION}. The anisotropic inflow and accretion may play an important role on shaping early expansion \citep{Peters2010b, Komesh2024}. Six of our regions reside in the hubs that may connect larger-scale mass inflow to the compact \hii\ regions \citepalias{ATOMSXI}. Recent survey on filamentary structures embedded in 100 massive clumps by \citet{Wells2024} finds a strengthen trend of mass inflow from quiescent stage to \hii\ region, underscoring its important role in the expansion of early \hii\ region.

   The pre-existing fragments are swept into the non-fragmented molecular shell at UC\hii\ region stage and then their kinematics start to be disturbed by expansion. The ram pressure of the shell is possibly too weak to drag the swept-up pre-existing fragments to move together with the shell. The acceleration $a_{\rm ram}$ powered by shell ram pressure $P_{\rm ram}$ is 
   \begin{equation}
       a_{\rm ram} = \frac{P_{\rm ram} \pi r_{\rm core}^2}{M_{\rm core}}  \sim  \frac{\rho_{\rm shell}{\rm v_{\rm shell}^2} \pi r_{\rm core}^2}{\frac{4}{3} \pi r_{\rm core}^3 \rho_{\rm core}} \sim \frac{3\rho_{\rm shell} {\rm v_{\rm shell}^2}}{4\rho_{\rm core}r_{\rm core}},
    \end{equation}
    where $\rho_{\rm core}$, $\rho_{\rm shell}$ and $\rm v_{\rm shell}$ are fragment mass density, shell mass density and velocity, respectively. The interaction time of the ram pressure is about $w_{\rm shell}/{\rm v_{shell}}$ and $w_{\rm shell}$ here is the width of the shell. Thus, the velocity of fragment after interaction with shell is around ${\rm v_{\rm ram}} \sim a_{\rm ram} t_{\rm ram} \sim 3 \rho_{\rm shell} w_{\rm shell} {\rm v_{\rm shell}}/4\rho_{\rm core} r_{\rm core}$. The shell-to-fragment density ratio determines ${\rm v_{\rm ram}}$. Although the shell component is not decoupled and measured from \htcop\ emission, the shell density should be much smaller than the fragment density if the masses swept up by expansion are in a shell with a width similar to fragment size. Therefore, the ram pressure of the shell is likely unable to hold the fragments to move together. The fragments will pass through the shell and then they are exposed to ionization radiation and photoevaporated, forming protruding structures of the shell.

    The fragments that are massive enough to survive from photoevaporation may confine the IF of the UC\hii\ region. From an observational point of view, the diverse and complicated morphology of UC\hii\ regions and their molecular shells \citep{Wood1989} sheds light on the pre-existing nature of fragments that now interact with the IF because a homogeneous environment favors the formation of a spherical \hii\ region, but an environment with a number of randomly distributed pre-existing fragments will confine the IF to form a ``random''-shape UC\hii\ region. The ionized gas pressure at UC\hii\ stage is one to two order of magnitude higher than current measured pressure when typical \nelectron\ of UC\hii\ region is $\sim10^4$~cm$^{-3}$. Furthermore, the fragments protruding from shell may also be accelerated to a few kilometers per second by ``rocket effect'' of the photoevaporation flows \citep{Kahn1954,Oort1955,Walch2013,Saha2022}. All these dynamical factors, combined with the photoevaporation of relative low-mass fragments, make shells in compact \hii\ region stage without many protrusions such as pillars, at the same time the shells and the corresponding fragments are likely velocity-coherent structures in the compact stages as shown by Figs~\ref{FIGURE:I13080PV} and \ref{FIGURE:I13080-6229_EDGE_PV}.

    \item (c) Shell and fragments seen surrounding compact \hii\ region. The UC\hii\ region continues to expand to the compact stage, reaching the regime of our observed regions. The shell expansion feature can be found directly in some of our regions. The velocity difference between the fragment molecular gas and the entire ionized gas statistically confirms the kinematics disturbance. During expansion of the \hii\ region, the spatial distribution of fragments within the natal clump transitions from more or less ``random'' to peak at the edges of the \hii\ regions, partly due to photoevaporation and outward displacement of some pre-existing fragments. Therefore, the observed spatial coherence between fragments and shell is actually a relic of the initial fragmentation of natal clump.

     In this scenario, star formation activities inside the shell continue in dense fragments,  but they are affected by the special environment created by ionization feedback and the swept-up shell. Fragments in the shell provide a condition of higher surface density, mass, and turbulence that favors the formation of higher-mass stars compared to the fragments outside the shell.  The ionized gas with much higher pressure probably strongly compresses the pre-existing fragments to let them have a higher surface density. Turbulence driven by the inhomogeneous ionized gas leads to the turbulent nature of fragment. The massive nature of shell fragments may result from a hybrid effect of a stronger mass accretion in the shell and coalescence with the shell materials. 
     
    \citet{Wang2016} presented a bona fide example shell that interacts with the IF of the 0.3 pc-scale compact \hii\ region IRAS 22134$+$5834. These authors found that this compact \hii\ region presents an expanding \cch\ shell associated with \chtoh\ emission that traces the shock gas produced by the interaction front.  It indicates that mass collection between the IF and the shock front is still ongoing at the compact stage \citep{Elmegreen1977}. Smaller condensations with different evolutionary stages, from prestellar to protostellar, are found for the dense cores of this exemplar compact \hii\ region shell when observed with higher resolution, indicating that the entire process of star formation is under the influence of ionization feedback.\\

     \end{itemize}

  
 The evolutionary scenario proposed here can be the basis for studying star formation in a complicated environment, especially for the starburst regions of the Milky Way, which always accompany the interaction between ionized gas and dense molecular gas. An exemplar case is the evolution of the core mass function (CMF) in W43-MM2\&MM3 mini-starburst regions.  As part of the ALMA large program ALMA-IMF \citep{Motte2022,Ginsburg2022}, \citet{Pouteau2023} separated the W43-MM2\&MM3 into a total of six subregions and then classified them further into three evolutionary stages (pre-burst, main-burst, and post-burst) based on the number of dense cores and outflows, plus the presence of UC\hii\ region. These authors found that the CMF top-heavy characteristic appears from the main-burst stage up to the post-burst stage in which an UC\hii\ region has developed well. The evolution of the top-heavy CMF proposed by \citet{Pouteau2023} agrees qualitatively well with our proposed scenario because we found that the molecular shells swept up by the early \hii\ regions host fragments that are more massive than other locations in the same natal clump. In future observations, it will be very interesting to analyze the CMF for dust cores in swept-up shells and its evolution with \hii\ regions, to ascertain the origins of top-heavy CMF in more detail \citep{Armante2024}.

\section{Conclusions and prospects}
\label{SECTIONS:CONCLUSIONS} 
We have systematically analyzed dense gas in nine early compact \hii\ regions using ATOMS data. The goal is to explore the interaction between \hii\ regions and dense molecular gas to understand its potential effect on the star formation process. The main findings are:
\begin{itemize}
    \item The selected nine \hii\ regions have just passed the ultra compact stages and are now at the early time of the following compact stage, according to the properties measured by our ATOMS observations. The molecular shells swept up by these compact \hii\ regions are inhomogeneous gas structures composed of several to a dozen dense gas fragments ($<0.1$~pc), as seen in \htcop\ emission.
    \item Most (if not all) of these dense gas fragments in the shell favor the formation of higher-mass stars compared to the other fragments outside the shell but within the same pc-scale natal clump. The former has a higher surface density ($\Sigma\sim1~\rm g~cm^2$, $\sim0.4~\rm g~cm^2$ larger), mass ($M\sim25$~\msun, $\sim15$~\msun\ larger) and turbulence ($\sigma\sim0.7~\rm km~s^{-1}$, $\sim0.2~\rm km~s^{-1}$ greater). 
    \item The spatial distributions and kinematics of the fragments are deeply regulated by photoevaporation and expansion of the \hii\ region.
    \item The dense gas fragments in the molecular shell are probably formed by the Jeans fragmentation of the natal clump before the shell sweeps them up, rather than by the shell fragmentation or so-called collect \& collapse process \citep{Elmegreen1977}.

\end{itemize}

Taken the above results together, we propose the following scenario: the non-fragmented, swept-up molecular shell is expected to encounter pre-existing fragments around the ultra-compact stage of the \hii\ region and then some of the low-mass fragments are photoevaporated. Other relative massive fragments can survive and be kinetically disturbed by expansion of \hii\ region, to form the expansion velocity structure and to reorganize the spatial distribution of the fragments. The higher surface density and turbulent characteristics of the fragments in the shell may be due to compression and injection of turbulence in the \hii\ region, but the cause behind the massive nature of these fragments in the shell remains unclear according to current observations. Potential explanations include a higher accretion rate for the fragments in the shell or / and the coalescence of the swept-up shell materials with fragments.

The results presented in this work are pivotal in understanding the impact of ionization feedback on high-mass star formation within complicated early compact \hii\ regions. Subsequent observations in ALMA Band 6 \citep{LiuXC2023} of the ATOMS \hii\ regions investigated here will enable a thorough analysis of dust cores and star formation activities in the molecular shell swept-up by these early compact \hii\ regions. They will improve our understanding of the interaction between \hii\ region and star formation in great detail, especially for the question of whether the external \hii\ region influences the core-scale mass accretion process.

\section*{Acknowledgements}
This work has been supported by the National Key R\&D Program of China (2022YFA1603100), the National Science Foundation of China (12041305, 12033005),  the China Manned Space Project (CMS-CSST-2021-A09, CMS-CSST-2021-B06), the Tianchi Talent Program of Xinjiang Uygur Autonomous Region, and the China-Chile Joint Research Fund (CCJRF No. 2211). CCJRF is provided by Chinese Academy of Sciences South America Center for Astronomy (CASSACA) and established by National Astronomical Observatories, Chinese Academy of Sciences (NAOC) and Chilean Astronomy Society (SOCHIAS) to support China-Chile collaborations in astronomy.

SZ gratefully acknowledges support by the CAS-ANID project CAS220003. AZ thanks the support of the Institut Universitaire de France. GG gratefully acknowledges support by the ANID BASAL project FB210003. PS was partially supported by a Grant-in-Aid for Scientific Research (KAKENHI Number JP22H01271 and JP23H01221) of JSPS. LB gratefully acknowledges support by the ANID BASAL project FB210003. This work was performed in part at the Jet Propulsion Laboratory, California Institute of Technology, under contract with the National Aeronautics and Space Administration.

This paper makes use of the following ALMA data: ADS/JAO.ALMA\#2019.1.00685.S. ALMA is a partnership of ESO (representing its member states), NSF (USA), and NINS (Japan), together with NRC (Canada), MOST and ASIAA (Taiwan), and KASI (Republic of Korea), in cooperation with the Republic of Chile. The Joint ALMA Observatory is operated by ESO, AUI/NRAO, and NAOJ.

\section*{Data Availability}
The data underlying this article will be shared on request to the corresponding author.



\bibliographystyle{mnras}
\bibliography{example} 

\begin{thebibliography}{}
\makeatletter
\relax
\def\mn@urlcharsother{\let\do\@makeother \do\$\do\&\do\#\do\^\do\_\do\%\do\~}
\def\mn@doi{\begingroup\mn@urlcharsother \@ifnextchar [ {\mn@doi@} {\mn@doi@[]}}
\def\mn@doi@[#1]#2{\def\@tempa{#1}\ifx\@tempa\@empty \href {http://dx.doi.org/#2} {doi:#2}\else \href {http://dx.doi.org/#2} {#1}\fi \endgroup}
\def\mn@eprint#1#2{\mn@eprint@#1:#2::\@nil}
\def\mn@eprint@arXiv#1{\href {http://arxiv.org/abs/#1} {{\tt arXiv:#1}}}
\def\mn@eprint@dblp#1{\href {http://dblp.uni-trier.de/rec/bibtex/#1.xml} {dblp:#1}}
\def\mn@eprint@#1:#2:#3:#4\@nil{\def\@tempa {#1}\def\@tempb {#2}\def\@tempc {#3}\ifx \@tempc \@empty \let \@tempc \@tempb \let \@tempb \@tempa \fi \ifx \@tempb \@empty \def\@tempb {arXiv}\fi \@ifundefined {mn@eprint@\@tempb}{\@tempb:\@tempc}{\expandafter \expandafter \csname mn@eprint@\@tempb\endcsname \expandafter{\@tempc}}}

\bibitem[\protect\citeauthoryear{{Ali}, {Bending}  \& {Dobbs}}{{Ali} et~al.}{2022}]{Ali2022}
{Ali} A.~A.,  {Bending} T. J.~R.,   {Dobbs} C.~L.,  2022, \mn@doi [\mnras] {10.1093/mnras/stac025}, \href {https://ui.adsabs.harvard.edu/abs/2022MNRAS.510.5592A} {510, 5592}

\bibitem[\protect\citeauthoryear{{Arce}, {Borkin}, {Goodman}, {Pineda}  \& {Beaumont}}{{Arce} et~al.}{2011}]{Arce2011}
{Arce} H.~G.,  {Borkin} M.~A.,  {Goodman} A.~A.,  {Pineda} J.~E.,   {Beaumont} C.~N.,  2011, \mn@doi [\apj] {10.1088/0004-637X/742/2/105}, \href {https://ui.adsabs.harvard.edu/abs/2011ApJ...742..105A} {742, 105}

\bibitem[\protect\citeauthoryear{{Armante} et~al.,}{{Armante} et~al.}{2024}]{Armante2024}
{Armante} M.,  et~al., 2024, \mn@doi [\aap] {10.1051/0004-6361/202347595}, \href {https://ui.adsabs.harvard.edu/abs/2024A&A...686A.122A} {686, A122}

\bibitem[\protect\citeauthoryear{{Baker} \& {Menzel}}{{Baker} \& {Menzel}}{1938}]{Barker1938}
{Baker} J.~G.,  {Menzel} D.~H.,  1938, \mn@doi [\apj] {10.1086/143959}, \href {https://ui.adsabs.harvard.edu/abs/1938ApJ....88...52B} {88, 52}

\bibitem[\protect\citeauthoryear{{Bending}, {Dobbs}  \& {Bate}}{{Bending} et~al.}{2020}]{Bending2020}
{Bending} T. J.~R.,  {Dobbs} C.~L.,   {Bate} M.~R.,  2020, \mn@doi [\mnras] {10.1093/mnras/staa1293}, \href {https://ui.adsabs.harvard.edu/abs/2020MNRAS.495.1672B} {495, 1672}

\bibitem[\protect\citeauthoryear{{Bertoldi}}{{Bertoldi}}{1989}]{Bertoldi1989}
{Bertoldi} F.,  1989, \mn@doi [\apj] {10.1086/168055}, \href {https://ui.adsabs.harvard.edu/abs/1989ApJ...346..735B} {346, 735}

\bibitem[\protect\citeauthoryear{{Bertoldi} \& {McKee}}{{Bertoldi} \& {McKee}}{1992}]{Bertoldi1992}
{Bertoldi} F.,  {McKee} C.~F.,  1992, \mn@doi [\apj] {10.1086/171638}, \href {https://ui.adsabs.harvard.edu/abs/1992ApJ...395..140B} {395, 140}

\bibitem[\protect\citeauthoryear{{Beuther} et~al.,}{{Beuther} et~al.}{2022}]{Beuther2022}
{Beuther} H.,  et~al., 2022, \mn@doi [\aap] {10.1051/0004-6361/202142689}, \href {https://ui.adsabs.harvard.edu/abs/2022A&A...659A..77B} {659, A77}

\bibitem[\protect\citeauthoryear{{Brand}, {Massi}, {Zavagno}, {Deharveng}  \& {Lefloch}}{{Brand} et~al.}{2011}]{Brand2011}
{Brand} J.,  {Massi} F.,  {Zavagno} A.,  {Deharveng} L.,   {Lefloch} B.,  2011, \mn@doi [\aap] {10.1051/0004-6361/201015389}, \href {https://ui.adsabs.harvard.edu/abs/2011A&A...527A..62B} {527, A62}

\bibitem[\protect\citeauthoryear{{Chevance}, {Krumholz}, {McLeod}, {Ostriker}, {Rosolowsky}  \& {Sternberg}}{{Chevance} et~al.}{2023}]{Chevance2023}
{Chevance} M.,  {Krumholz} M.~R.,  {McLeod} A.~F.,  {Ostriker} E.~C.,  {Rosolowsky} E.~W.,   {Sternberg} A.,  2023, in {Inutsuka} S.,  {Aikawa} Y.,  {Muto} T.,  {Tomida} K.,   {Tamura} M.,  eds,  Astronomical Society of the Pacific Conference Series Vol. 534, Protostars and Planets VII. p.~1 (\mn@eprint {arXiv} {2203.09570}), \mn@doi{10.48550/arXiv.2203.09570}

\bibitem[\protect\citeauthoryear{{Churchwell}}{{Churchwell}}{2002}]{Churchwell2002}
{Churchwell} E.,  2002, \mn@doi [\araa] {10.1146/annurev.astro.40.060401.093845}, \href {https://ui.adsabs.harvard.edu/abs/2002ARA&A..40...27C} {40, 27}

\bibitem[\protect\citeauthoryear{{Colombo}, {Rosolowsky}, {Ginsburg}, {Duarte-Cabral}  \& {Hughes}}{{Colombo} et~al.}{2015}]{Colombo2015}
{Colombo} D.,  {Rosolowsky} E.,  {Ginsburg} A.,  {Duarte-Cabral} A.,   {Hughes} A.,  2015, \mn@doi [\mnras] {10.1093/mnras/stv2063}, \href {https://ui.adsabs.harvard.edu/abs/2015MNRAS.454.2067C} {454, 2067}

\bibitem[\protect\citeauthoryear{{Contreras} et~al.,}{{Contreras} et~al.}{2018}]{Contreras2018}
{Contreras} Y.,  et~al., 2018, \mn@doi [\apj] {10.3847/1538-4357/aac2ec}, \href {https://ui.adsabs.harvard.edu/abs/2018ApJ...861...14C} {861, 14}

\bibitem[\protect\citeauthoryear{{Csengeri} et~al.,}{{Csengeri} et~al.}{2014}]{Csengeri2014}
{Csengeri} T.,  et~al., 2014, \mn@doi [\aap] {10.1051/0004-6361/201322434}, \href {https://ui.adsabs.harvard.edu/abs/2014A&A...565A..75C} {565, A75}

\bibitem[\protect\citeauthoryear{{Cuadrado}, {Goicoechea}, {Pilleri}, {Cernicharo}, {Fuente}  \& {Joblin}}{{Cuadrado} et~al.}{2015}]{Cuadrado2015}
{Cuadrado} S.,  {Goicoechea} J.~R.,  {Pilleri} P.,  {Cernicharo} J.,  {Fuente} A.,   {Joblin} C.,  2015, \mn@doi [\aap] {10.1051/0004-6361/201424568}, \href {https://ui.adsabs.harvard.edu/abs/2015A&A...575A..82C} {575, A82}

\bibitem[\protect\citeauthoryear{{Dale}, {Ngoumou}, {Ercolano}  \& {Bonnell}}{{Dale} et~al.}{2014}]{Dale2014}
{Dale} J.~E.,  {Ngoumou} J.,  {Ercolano} B.,   {Bonnell} I.~A.,  2014, \mn@doi [\mnras] {10.1093/mnras/stu816}, \href {https://ui.adsabs.harvard.edu/abs/2014MNRAS.442..694D} {442, 694}

\bibitem[\protect\citeauthoryear{{Dame}}{{Dame}}{2011}]{Dame2011}
{Dame} T.~M.,  2011, \mn@doi [arXiv e-prints] {10.48550/arXiv.1101.1499}, \href {https://ui.adsabs.harvard.edu/abs/2011arXiv1101.1499D} {p. arXiv:1101.1499}

\bibitem[\protect\citeauthoryear{{Deharveng}, {Zavagno}  \& {Caplan}}{{Deharveng} et~al.}{2005}]{Deharveng2005}
{Deharveng} L.,  {Zavagno} A.,   {Caplan} J.,  2005, \mn@doi [\aap] {10.1051/0004-6361:20041946}, \href {https://ui.adsabs.harvard.edu/abs/2005A&A...433..565D} {433, 565}

\bibitem[\protect\citeauthoryear{{Deharveng}, {Lefloch}, {Kurtz}, {Nadeau}, {Pomar{\`e}s}, {Caplan}  \& {Zavagno}}{{Deharveng} et~al.}{2008}]{Deharveng2008}
{Deharveng} L.,  {Lefloch} B.,  {Kurtz} S.,  {Nadeau} D.,  {Pomar{\`e}s} M.,  {Caplan} J.,   {Zavagno} A.,  2008, \mn@doi [\aap] {10.1051/0004-6361:20079233}, \href {https://ui.adsabs.harvard.edu/abs/2008A&A...482..585D} {482, 585}

\bibitem[\protect\citeauthoryear{{Deharveng} et~al.,}{{Deharveng} et~al.}{2010}]{Deharveng2010}
{Deharveng} L.,  et~al., 2010, \mn@doi [\aap] {10.1051/0004-6361/201014422}, \href {https://ui.adsabs.harvard.edu/abs/2010A&A...523A...6D} {523, A6}

\bibitem[\protect\citeauthoryear{{Dobbs}, {Bending}, {Pettitt}  \& {Bate}}{{Dobbs} et~al.}{2022}]{Dobbs2022}
{Dobbs} C.~L.,  {Bending} T.~J.~R.,  {Pettitt} A.~R.,   {Bate} M.~R.,  2022, \mn@doi [\mnras] {10.1093/mnras/stab3036}, \href {https://ui.adsabs.harvard.edu/abs/2022MNRAS.509..954D} {509, 954}

\bibitem[\protect\citeauthoryear{{Dyson} \& {Williams}}{{Dyson} \& {Williams}}{1997}]{Dyson1997}
{Dyson} J.~E.,  {Williams} D.~A.,  1997, {The physics of the interstellar medium}, \mn@doi{10.1201/9780585368115.
}

\bibitem[\protect\citeauthoryear{{Elmegreen}}{{Elmegreen}}{1998}]{Elmegreen1998}
{Elmegreen} B.~G.,  1998, in {Woodward} C.~E.,  {Shull} J.~M.,   {Thronson} Harley~A. J.,  eds,  Astronomical Society of the Pacific Conference Series Vol. 148, Origins. p.~150 (\mn@eprint {arXiv} {astro-ph/9712352}), \mn@doi{10.48550/arXiv.astro-ph/9712352}

\bibitem[\protect\citeauthoryear{{Elmegreen} \& {Lada}}{{Elmegreen} \& {Lada}}{1977}]{Elmegreen1977}
{Elmegreen} B.~G.,  {Lada} C.~J.,  1977, \mn@doi [\apj] {10.1086/155302}, \href {https://ui.adsabs.harvard.edu/abs/1977ApJ...214..725E} {214, 725}

\bibitem[\protect\citeauthoryear{{Eswaraiah} et~al.,}{{Eswaraiah} et~al.}{2020}]{Eswaraiah2020}
{Eswaraiah} C.,  et~al., 2020, \mn@doi [\apj] {10.3847/1538-4357/ab83f2}, \href {https://ui.adsabs.harvard.edu/abs/2020ApJ...897...90E} {897, 90}

\bibitem[\protect\citeauthoryear{{Fa{\'u}ndez}, {Bronfman}, {Garay}, {Chini}, {Nyman}  \& {May}}{{Fa{\'u}ndez} et~al.}{2004}]{Faundez2004}
{Fa{\'u}ndez} S.,  {Bronfman} L.,  {Garay} G.,  {Chini} R.,  {Nyman} L.~{\r{A}}.,   {May} J.,  2004, \mn@doi [\aap] {10.1051/0004-6361:20035755}, \href {https://ui.adsabs.harvard.edu/abs/2004A&A...426...97F} {426, 97}

\bibitem[\protect\citeauthoryear{{Fern{\'a}ndez-L{\'o}pez} et~al.,}{{Fern{\'a}ndez-L{\'o}pez} et~al.}{2021}]{Fernandez2021}
{Fern{\'a}ndez-L{\'o}pez} M.,  et~al., 2021, \mn@doi [\apj] {10.3847/1538-4357/abf2b6}, \href {https://ui.adsabs.harvard.edu/abs/2021ApJ...913...29F} {913, 29}

\bibitem[\protect\citeauthoryear{{Fukushima} \& {Yajima}}{{Fukushima} \& {Yajima}}{2022}]{Fukushima2022}
{Fukushima} H.,  {Yajima} H.,  2022, \mn@doi [\mnras] {10.1093/mnras/stac244}, \href {https://ui.adsabs.harvard.edu/abs/2022MNRAS.511.3346F} {511, 3346}

\bibitem[\protect\citeauthoryear{{Gao} \& {Solomon}}{{Gao} \& {Solomon}}{2004}]{Gao2004}
{Gao} Y.,  {Solomon} P.~M.,  2004, \mn@doi [\apj] {10.1086/382999}, \href {https://ui.adsabs.harvard.edu/abs/2004ApJ...606..271G} {606, 271}

\bibitem[\protect\citeauthoryear{{Garay}, {Rodriguez}, {Moran}  \& {Churchwell}}{{Garay} et~al.}{1993}]{Garay1993}
{Garay} G.,  {Rodriguez} L.~F.,  {Moran} J.~M.,   {Churchwell} E.,  1993, \mn@doi [\apj] {10.1086/173396}, \href {https://ui.adsabs.harvard.edu/abs/1993ApJ...418..368G} {418, 368}

\bibitem[\protect\citeauthoryear{{Garcia-Segura} \& {Franco}}{{Garcia-Segura} \& {Franco}}{1996}]{Garcia1996}
{Garcia-Segura} G.,  {Franco} J.,  1996, \mn@doi [\apj] {10.1086/177769}, \href {https://ui.adsabs.harvard.edu/abs/1996ApJ...469..171G} {469, 171}

\bibitem[\protect\citeauthoryear{{Geen}, {Pellegrini}, {Bieri}  \& {Klessen}}{{Geen} et~al.}{2020}]{Geen2020}
{Geen} S.,  {Pellegrini} E.,  {Bieri} R.,   {Klessen} R.,  2020, \mn@doi [\mnras] {10.1093/mnras/stz3491}, \href {https://ui.adsabs.harvard.edu/abs/2020MNRAS.492..915G} {492, 915}

\bibitem[\protect\citeauthoryear{{Geen}, {Bieri}, {Rosdahl}  \& {de Koter}}{{Geen} et~al.}{2021}]{Geen2021}
{Geen} S.,  {Bieri} R.,  {Rosdahl} J.,   {de Koter} A.,  2021, \mn@doi [\mnras] {10.1093/mnras/staa3705}, \href {https://ui.adsabs.harvard.edu/abs/2021MNRAS.501.1352G} {501, 1352}

\bibitem[\protect\citeauthoryear{{Ginsburg} et~al.,}{{Ginsburg} et~al.}{2022}]{Ginsburg2022}
{Ginsburg} A.,  et~al., 2022, \mn@doi [\aap] {10.1051/0004-6361/202141681}, \href {https://ui.adsabs.harvard.edu/abs/2022A&A...662A...9G} {662, A9}

\bibitem[\protect\citeauthoryear{{Goicoechea} et~al.,}{{Goicoechea} et~al.}{2016}]{Goicoechea2016}
{Goicoechea} J.~R.,  et~al., 2016, \mn@doi [\nat] {10.1038/nature18957}, \href {https://ui.adsabs.harvard.edu/abs/2016Natur.537..207G} {537, 207}

\bibitem[\protect\citeauthoryear{{Gonz{\'a}lez-Samaniego} \& {Vazquez-Semadeni}}{{Gonz{\'a}lez-Samaniego} \& {Vazquez-Semadeni}}{2020}]{Gonzalez2020}
{Gonz{\'a}lez-Samaniego} A.,  {Vazquez-Semadeni} E.,  2020, \mn@doi [\mnras] {10.1093/mnras/staa2921}, \href {https://ui.adsabs.harvard.edu/abs/2020MNRAS.499..668G} {499, 668}

\bibitem[\protect\citeauthoryear{{Gordon} \& {Sorochenko}}{{Gordon} \& {Sorochenko}}{2002}]{Gordon2002}
{Gordon} M.~A.,  {Sorochenko} R.~L.,  2002, {Radio Recombination Lines. Their Physics and Astronomical Applications}.
~ Vol. 282, \mn@doi{10.1007/978-0-387-09604-9, }

\bibitem[\protect\citeauthoryear{{Haid}, {Walch}, {Seifried}, {W{\"u}nsch}, {Dinnbier}  \& {Naab}}{{Haid} et~al.}{2018}]{Haid2018}
{Haid} S.,  {Walch} S.,  {Seifried} D.,  {W{\"u}nsch} R.,  {Dinnbier} F.,   {Naab} T.,  2018, \mn@doi [\mnras] {10.1093/mnras/sty1315}, \href {https://ui.adsabs.harvard.edu/abs/2018MNRAS.478.4799H} {478, 4799}

\bibitem[\protect\citeauthoryear{{Haworth}, {Harries}  \& {Acreman}}{{Haworth} et~al.}{2012}]{Haworth2012}
{Haworth} T.~J.,  {Harries} T.~J.,   {Acreman} D.~M.,  2012, \mn@doi [\mnras] {10.1111/j.1365-2966.2012.21838.x}, \href {https://ui.adsabs.harvard.edu/abs/2012MNRAS.426..203H} {426, 203}

\bibitem[\protect\citeauthoryear{{Hennebelle} et~al.,}{{Hennebelle} et~al.}{2022}]{Hennebelle2022}
{Hennebelle} P.,  et~al., 2022, \mn@doi [\aap] {10.1051/0004-6361/202243803}, \href {https://ui.adsabs.harvard.edu/abs/2022A&A...668A.147H} {668, A147}

\bibitem[\protect\citeauthoryear{{Herrington}, {Dobbs}  \& {Bending}}{{Herrington} et~al.}{2023}]{Herrington2023}
{Herrington} N.~P.,  {Dobbs} C.~L.,   {Bending} T. J.~R.,  2023, \mn@doi [\mnras] {10.1093/mnras/stad923}, \href {https://ui.adsabs.harvard.edu/abs/2023MNRAS.521.5712H} {521, 5712}

\bibitem[\protect\citeauthoryear{{Hoare}, {Kurtz}, {Lizano}, {Keto}  \& {Hofner}}{{Hoare} et~al.}{2007}]{Hoare2007}
{Hoare} M.~G.,  {Kurtz} S.~E.,  {Lizano} S.,  {Keto} E.,   {Hofner} P.,  2007, in {Reipurth} B.,  {Jewitt} D.,   {Keil} K.,  eds, Protostars and Planets V. p.~181 (\mn@eprint {arXiv} {astro-ph/0603560}), \mn@doi{10.48550/arXiv.astro-ph/0603560}

\bibitem[\protect\citeauthoryear{{Hopkins}, {Kere{\v{s}}}, {O{\~n}orbe}, {Faucher-Gigu{\`e}re}, {Quataert}, {Murray}  \& {Bullock}}{{Hopkins} et~al.}{2014}]{Hopkins2014}
{Hopkins} P.~F.,  {Kere{\v{s}}} D.,  {O{\~n}orbe} J.,  {Faucher-Gigu{\`e}re} C.-A.,  {Quataert} E.,  {Murray} N.,   {Bullock} J.~S.,  2014, \mn@doi [\mnras] {10.1093/mnras/stu1738}, \href {https://ui.adsabs.harvard.edu/abs/2014MNRAS.445..581H} {445, 581}

\bibitem[\protect\citeauthoryear{{Hunter}, {Brogan}, {Indebetouw}  \& {Cyganowski}}{{Hunter} et~al.}{2008}]{Hunter2008}
{Hunter} T.~R.,  {Brogan} C.~L.,  {Indebetouw} R.,   {Cyganowski} C.~J.,  2008, \mn@doi [\apj] {10.1086/588016}, \href {https://ui.adsabs.harvard.edu/abs/2008ApJ...680.1271H} {680, 1271}

\bibitem[\protect\citeauthoryear{{Jackson} et~al.,}{{Jackson} et~al.}{2013}]{Jackson2013}
{Jackson} J.~M.,  et~al., 2013, \mn@doi [\pasa] {10.1017/pasa.2013.37}, \href {https://ui.adsabs.harvard.edu/abs/2013PASA...30...57J} {30, e057}

\bibitem[\protect\citeauthoryear{{Kahn}}{{Kahn}}{1954}]{Kahn1954}
{Kahn} F.~D.,  1954, \bain, \href {https://ui.adsabs.harvard.edu/abs/1954BAN....12..187K} {12, 187}

\bibitem[\protect\citeauthoryear{{Kalcheva}, {Hoare}, {Urquhart}, {Kurtz}, {Lumsden}, {Purcell}  \& {Zijlstra}}{{Kalcheva} et~al.}{2018}]{Kalcheva2018}
{Kalcheva} I.~E.,  {Hoare} M.~G.,  {Urquhart} J.~S.,  {Kurtz} S.,  {Lumsden} S.~L.,  {Purcell} C.~R.,   {Zijlstra} A.~A.,  2018, \mn@doi [\aap] {10.1051/0004-6361/201832734}, \href {https://ui.adsabs.harvard.edu/abs/2018A&A...615A.103K} {615, A103}

\bibitem[\protect\citeauthoryear{{Kauffmann} \& {Pillai}}{{Kauffmann} \& {Pillai}}{2010}]{Kauffmann2010}
{Kauffmann} J.,  {Pillai} T.,  2010, \mn@doi [\apjl] {10.1088/2041-8205/723/1/L7}, \href {https://ui.adsabs.harvard.edu/abs/2010ApJ...723L...7K} {723, L7}

\bibitem[\protect\citeauthoryear{{Kendrew} et~al.,}{{Kendrew} et~al.}{2016}]{Kendrew2016}
{Kendrew} S.,  et~al., 2016, \mn@doi [\apj] {10.3847/0004-637X/825/2/142}, \href {https://ui.adsabs.harvard.edu/abs/2016ApJ...825..142K} {825, 142}

\bibitem[\protect\citeauthoryear{{Kim}, {Wyrowski}, {Urquhart}, {P{\'e}rez-Beaupuits}, {Pillai}, {Tiwari}  \& {Menten}}{{Kim} et~al.}{2020}]{Kim2020}
{Kim} W.~J.,  {Wyrowski} F.,  {Urquhart} J.~S.,  {P{\'e}rez-Beaupuits} J.~P.,  {Pillai} T.,  {Tiwari} M.,   {Menten} K.~M.,  2020, \mn@doi [\aap] {10.1051/0004-6361/202039059}, \href {https://ui.adsabs.harvard.edu/abs/2020A&A...644A.160K} {644, A160}

\bibitem[\protect\citeauthoryear{{Kirsanova}, {Wiebe}, {Sobolev}, {Henkel}  \& {Tsivilev}}{{Kirsanova} et~al.}{2014}]{Kirsanova2014}
{Kirsanova} M.~S.,  {Wiebe} D.~S.,  {Sobolev} A.~M.,  {Henkel} C.,   {Tsivilev} A.~P.,  2014, \mn@doi [\mnras] {10.1093/mnras/stt1991}, \href {https://ui.adsabs.harvard.edu/abs/2014MNRAS.437.1593K} {437, 1593}

\bibitem[\protect\citeauthoryear{{Kirsanova}, {Punanova}, {Semenov}  \& {Vasyunin}}{{Kirsanova} et~al.}{2021}]{Kirsanova2021}
{Kirsanova} M.~S.,  {Punanova} A.~F.,  {Semenov} D.~A.,   {Vasyunin} A.~I.,  2021, \mn@doi [\mnras] {10.1093/mnras/stab2361}, \href {https://ui.adsabs.harvard.edu/abs/2021MNRAS.507.3810K} {507, 3810}

\bibitem[\protect\citeauthoryear{{Klaassen} et~al.,}{{Klaassen} et~al.}{2018}]{Klaassen2018}
{Klaassen} P.~D.,  et~al., 2018, \mn@doi [\aap] {10.1051/0004-6361/201731727}, \href {https://ui.adsabs.harvard.edu/abs/2018A&A...611A..99K} {611, A99}

\bibitem[\protect\citeauthoryear{{Komesh} et~al.,}{{Komesh} et~al.}{2024}]{Komesh2024}
{Komesh} T.,  et~al., 2024, \mn@doi [\apj] {10.3847/1538-4357/ad3e7b}, \href {https://ui.adsabs.harvard.edu/abs/2024ApJ...967...15K} {967, 15}

\bibitem[\protect\citeauthoryear{{Krumholz}}{{Krumholz}}{2017}]{Krumholz2017}
{Krumholz} M.~R.,  2017, {Star Formation}, \mn@doi{10.1142/10091.
}

\bibitem[\protect\citeauthoryear{{Krumholz} \& {McKee}}{{Krumholz} \& {McKee}}{2008}]{Krumholz2008}
{Krumholz} M.~R.,  {McKee} C.~F.,  2008, \mn@doi [\nat] {10.1038/nature06620}, \href {https://ui.adsabs.harvard.edu/abs/2008Natur.451.1082K} {451, 1082}

\bibitem[\protect\citeauthoryear{{Kumar}, {Palmeirim}, {Arzoumanian}  \& {Inutsuka}}{{Kumar} et~al.}{2020}]{Kumar2020}
{Kumar} M.~S.~N.,  {Palmeirim} P.,  {Arzoumanian} D.,   {Inutsuka} S.~I.,  2020, \mn@doi [\aap] {10.1051/0004-6361/202038232}, \href {https://ui.adsabs.harvard.edu/abs/2020A&A...642A..87K} {642, A87}

\bibitem[\protect\citeauthoryear{{Kurtz}}{{Kurtz}}{2005}]{Kurtz2005}
{Kurtz} S.,  2005, in {Cesaroni} R.,  {Felli} M.,  {Churchwell} E.,   {Walmsley} M.,  eds, ~ Vol. 227, Massive Star Birth: A Crossroads of Astrophysics. pp 111--119, \mn@doi{10.1017/S1743921305004424}

\bibitem[\protect\citeauthoryear{{Landy} \& {Szalay}}{{Landy} \& {Szalay}}{1993}]{Landy1993}
{Landy} S.~D.,  {Szalay} A.~S.,  1993, \mn@doi [\apj] {10.1086/172900}, \href {https://ui.adsabs.harvard.edu/abs/1993ApJ...412...64L} {412, 64}

\bibitem[\protect\citeauthoryear{{Larson}}{{Larson}}{1981}]{Larson1981}
{Larson} R.~B.,  1981, \mn@doi [\mnras] {10.1093/mnras/194.4.809}, \href {https://ui.adsabs.harvard.edu/abs/1981MNRAS.194..809L} {194, 809}

\bibitem[\protect\citeauthoryear{{Li}, {Zhang}, {Pillai}, {Stephens}, {Wang}  \& {Li}}{{Li} et~al.}{2019}]{Li2019}
{Li} S.,  {Zhang} Q.,  {Pillai} T.,  {Stephens} I.~W.,  {Wang} J.,   {Li} F.,  2019, \mn@doi [\apj] {10.3847/1538-4357/ab464e}, \href {https://ui.adsabs.harvard.edu/abs/2019ApJ...886..130L} {886, 130}

\bibitem[\protect\citeauthoryear{{Liu} et~al.,}{{Liu} et~al.}{2017}]{Liu2017}
{Liu} T.,  et~al., 2017, \mn@doi [\apj] {10.3847/1538-4357/aa8d73}, \href {https://ui.adsabs.harvard.edu/abs/2017ApJ...849...25L} {849, 25}

\bibitem[\protect\citeauthoryear{{Liu} et~al.,}{{Liu} et~al.}{2020a}]{ATOMSI}
{Liu} T.,  et~al., 2020a, \mn@doi [\mnras] {10.1093/mnras/staa1577}, \href {https://ui.adsabs.harvard.edu/abs/2020MNRAS.496.2790L} {496, 2790}

\bibitem[\protect\citeauthoryear{{Liu} et~al.,}{{Liu} et~al.}{2020b}]{ATOMSII}
{Liu} T.,  et~al., 2020b, \mn@doi [\mnras] {10.1093/mnras/staa1501}, \href {https://ui.adsabs.harvard.edu/abs/2020MNRAS.496.2821L} {496, 2821}

\bibitem[\protect\citeauthoryear{{Liu} et~al.,}{{Liu} et~al.}{2021}]{ATOMSIII}
{Liu} H.-L.,  et~al., 2021, \mn@doi [\mnras] {10.1093/mnras/stab1352}, \href {https://ui.adsabs.harvard.edu/abs/2021MNRAS.505.2801L} {505, 2801}

\bibitem[\protect\citeauthoryear{{Liu} et~al.,}{{Liu} et~al.}{2022}]{ATOMSIX}
{Liu} H.-L.,  et~al., 2022, \mn@doi [\mnras] {10.1093/mnras/stac378}, \href {https://ui.adsabs.harvard.edu/abs/2022MNRAS.511.4480L} {511, 4480}

\bibitem[\protect\citeauthoryear{{Liu} et~al.,}{{Liu} et~al.}{2023}]{LiuHL2023}
{Liu} H.-L.,  et~al., 2023, \mn@doi [\mnras] {10.1093/mnras/stad047}, \href {https://ui.adsabs.harvard.edu/abs/2023MNRAS.tmp..140L} {}

\bibitem[\protect\citeauthoryear{{Liu} et~al.,}{{Liu} et~al.}{2024}]{LiuXC2023}
{Liu} X.,  et~al., 2024, \mn@doi [Research in Astronomy and Astrophysics] {10.1088/1674-4527/ad0d5c}, \href {https://ui.adsabs.harvard.edu/abs/2024RAA....24b5009L} {24, 025009}

\bibitem[\protect\citeauthoryear{{Mazumdar}, {Wyrowski}, {Colombo}, {Urquhart}, {Thompson}  \& {Menten}}{{Mazumdar} et~al.}{2021}]{Mazumdar2021}
{Mazumdar} P.,  {Wyrowski} F.,  {Colombo} D.,  {Urquhart} J.~S.,  {Thompson} M.~A.,   {Menten} K.~M.,  2021, \mn@doi [\aap] {10.1051/0004-6361/202040205}, \href {https://ui.adsabs.harvard.edu/abs/2021A&A...650A.164M} {650, A164}

\bibitem[\protect\citeauthoryear{{McMullin}, {Waters}, {Schiebel}, {Young}  \& {Golap}}{{McMullin} et~al.}{2007}]{Mcmullin2007}
{McMullin} J.~P.,  {Waters} B.,  {Schiebel} D.,  {Young} W.,   {Golap} K.,  2007, {CASA Architecture and Applications}.
p.~127

\bibitem[\protect\citeauthoryear{{Miettinen}}{{Miettinen}}{2014}]{Miettinen2014}
{Miettinen} O.,  2014, \mn@doi [\aap] {10.1051/0004-6361/201322596}, \href {https://ui.adsabs.harvard.edu/abs/2014A&A...562A...3M} {562, A3}

\bibitem[\protect\citeauthoryear{{Milam}, {Savage}, {Brewster}, {Ziurys}  \& {Wyckoff}}{{Milam} et~al.}{2005}]{Milam2005}
{Milam} S.~N.,  {Savage} C.,  {Brewster} M.~A.,  {Ziurys} L.~M.,   {Wyckoff} S.,  2005, \mn@doi [\apj] {10.1086/497123}, \href {https://ui.adsabs.harvard.edu/abs/2005ApJ...634.1126M} {634, 1126}

\bibitem[\protect\citeauthoryear{{Miyawaki}, {Hayashi}  \& {Hasegawa}}{{Miyawaki} et~al.}{2023}]{Miyawaki2023}
{Miyawaki} R.,  {Hayashi} M.,   {Hasegawa} T.,  2023, \mn@doi [\pasj] {10.1093/pasj/psac105}, \href {https://ui.adsabs.harvard.edu/abs/2023PASJ...75..225M} {75, 225}

\bibitem[\protect\citeauthoryear{{Molinari}, {Merello}, {Elia}, {Cesaroni}, {Testi}  \& {Robitaille}}{{Molinari} et~al.}{2016}]{Molinari2016}
{Molinari} S.,  {Merello} M.,  {Elia} D.,  {Cesaroni} R.,  {Testi} L.,   {Robitaille} T.,  2016, \mn@doi [\apjl] {10.3847/2041-8205/826/1/L8}, \href {https://ui.adsabs.harvard.edu/abs/2016ApJ...826L...8M} {826, L8}

\bibitem[\protect\citeauthoryear{{Mookerjea}}{{Mookerjea}}{2022}]{Mookerjea2022}
{Mookerjea} B.,  2022, \mn@doi [\apj] {10.3847/1538-4357/ac4258}, \href {https://ui.adsabs.harvard.edu/abs/2022ApJ...926....4M} {926, 4}

\bibitem[\protect\citeauthoryear{{Morii} et~al.,}{{Morii} et~al.}{2023}]{Morii2023}
{Morii} K.,  et~al., 2023, \mn@doi [\apj] {10.3847/1538-4357/acccea}, \href {https://ui.adsabs.harvard.edu/abs/2023ApJ...950..148M} {950, 148}

\bibitem[\protect\citeauthoryear{{Moscadelli} et~al.,}{{Moscadelli} et~al.}{2018}]{Moscadelli2018}
{Moscadelli} L.,  et~al., 2018, \mn@doi [\aap] {10.1051/0004-6361/201832680}, \href {https://ui.adsabs.harvard.edu/abs/2018A&A...616A..66M} {616, A66}

\bibitem[\protect\citeauthoryear{{Moscadelli}, {Cesaroni}, {Beltr{\'a}n}  \& {Rivilla}}{{Moscadelli} et~al.}{2021}]{Moscadelli2021}
{Moscadelli} L.,  {Cesaroni} R.,  {Beltr{\'a}n} M.~T.,   {Rivilla} V.~M.,  2021, \mn@doi [\aap] {10.1051/0004-6361/202140829}, \href {https://ui.adsabs.harvard.edu/abs/2021A&A...650A.142M} {650, A142}

\bibitem[\protect\citeauthoryear{{Motte}, {Bontemps}  \& {Louvet}}{{Motte} et~al.}{2018}]{Motte2018}
{Motte} F.,  {Bontemps} S.,   {Louvet} F.,  2018, \mn@doi [\araa] {10.1146/annurev-astro-091916-055235}, \href {https://ui.adsabs.harvard.edu/abs/2018ARA&A..56...41M} {56, 41}

\bibitem[\protect\citeauthoryear{{Motte} et~al.,}{{Motte} et~al.}{2022}]{Motte2022}
{Motte} F.,  et~al., 2022, \mn@doi [\aap] {10.1051/0004-6361/202141677}, \href {https://ui.adsabs.harvard.edu/abs/2022A&A...662A...8M} {662, A8}

\bibitem[\protect\citeauthoryear{{Mottram} et~al.,}{{Mottram} et~al.}{2011}]{Mottram2011}
{Mottram} J.~C.,  et~al., 2011, \mn@doi [\apjl] {10.1088/2041-8205/730/2/L33}, \href {https://ui.adsabs.harvard.edu/abs/2011ApJ...730L..33M} {730, L33}

\bibitem[\protect\citeauthoryear{{Naidoo}}{{Naidoo}}{2019}]{Naidoo2019}
{Naidoo} K.,  2019, \mn@doi [The Journal of Open Source Software] {10.21105/joss.01721}, \href {https://ui.adsabs.harvard.edu/abs/2019JOSS....4.1721N} {4, 1721}

\bibitem[\protect\citeauthoryear{{Nakano} et~al.,}{{Nakano} et~al.}{2017}]{Nakano2017}
{Nakano} M.,  et~al., 2017, \mn@doi [\pasj] {10.1093/pasj/psw120}, \href {https://ui.adsabs.harvard.edu/abs/2017PASJ...69...16N} {69, 16}

\bibitem[\protect\citeauthoryear{{Nandakumar}, {Veena}, {Vig}, {Tej}, {Ghosh}  \& {Ojha}}{{Nandakumar} et~al.}{2016}]{Nandakumar2016}
{Nandakumar} G.,  {Veena} V.~S.,  {Vig} S.,  {Tej} A.,  {Ghosh} S.~K.,   {Ojha} D.~K.,  2016, \mn@doi [\aj] {10.3847/0004-6256/152/5/146}, \href {https://ui.adsabs.harvard.edu/abs/2016AJ....152..146N} {152, 146}

\bibitem[\protect\citeauthoryear{{Oort} \& {Spitzer}}{{Oort} \& {Spitzer}}{1955}]{Oort1955}
{Oort} J.~H.,  {Spitzer} Lyman J.,  1955, \mn@doi [\apj] {10.1086/145958}, \href {https://ui.adsabs.harvard.edu/abs/1955ApJ...121....6O} {121, 6}

\bibitem[\protect\citeauthoryear{{Pabst} et~al.,}{{Pabst} et~al.}{2020}]{Pabst2020}
{Pabst} C.~H.~M.,  et~al., 2020, \mn@doi [\aap] {10.1051/0004-6361/202037560}, \href {https://ui.adsabs.harvard.edu/abs/2020A&A...639A...2P} {639, A2}

\bibitem[\protect\citeauthoryear{{Palau} et~al.,}{{Palau} et~al.}{2015}]{Palau2015}
{Palau} A.,  et~al., 2015, \mn@doi [\mnras] {10.1093/mnras/stv1834}, \href {https://ui.adsabs.harvard.edu/abs/2015MNRAS.453.3785P} {453, 3785}

\bibitem[\protect\citeauthoryear{{Palmeirim} et~al.,}{{Palmeirim} et~al.}{2017}]{Palmeirim2017}
{Palmeirim} P.,  et~al., 2017, \mn@doi [\aap] {10.1051/0004-6361/201629963}, \href {https://ui.adsabs.harvard.edu/abs/2017A&A...605A..35P} {605, A35}

\bibitem[\protect\citeauthoryear{{Pandey} et~al.,}{{Pandey} et~al.}{2022}]{Pandey2022}
{Pandey} R.,  et~al., 2022, \mn@doi [\apj] {10.3847/1538-4357/ac41c3}, \href {https://ui.adsabs.harvard.edu/abs/2022ApJ...926...25P} {926, 25}

\bibitem[\protect\citeauthoryear{{Patel}, {Urquhart}, {Yang}, {Moore}, {Thompson}, {Menten}  \& {Csengeri}}{{Patel} et~al.}{2024}]{Patel2024}
{Patel} A.~L.,  {Urquhart} J.~S.,  {Yang} A.~Y.,  {Moore} T.,  {Thompson} M.~A.,  {Menten} K.~M.,   {Csengeri} T.,  2024, \mn@doi [\mnras] {10.1093/mnras/stae1910}, \href {https://ui.adsabs.harvard.edu/abs/2024MNRAS.tmp.1866P} {}

\bibitem[\protect\citeauthoryear{{Peters}, {Banerjee}, {Klessen}, {Mac Low}, {Galv{\'a}n-Madrid}  \& {Keto}}{{Peters} et~al.}{2010a}]{Peters2010a}
{Peters} T.,  {Banerjee} R.,  {Klessen} R.~S.,  {Mac Low} M.-M.,  {Galv{\'a}n-Madrid} R.,   {Keto} E.~R.,  2010a, \mn@doi [\apj] {10.1088/0004-637X/711/2/1017}, \href {https://ui.adsabs.harvard.edu/abs/2010ApJ...711.1017P} {711, 1017}

\bibitem[\protect\citeauthoryear{{Peters}, {Mac Low}, {Banerjee}, {Klessen}  \& {Dullemond}}{{Peters} et~al.}{2010b}]{Peters2010b}
{Peters} T.,  {Mac Low} M.-M.,  {Banerjee} R.,  {Klessen} R.~S.,   {Dullemond} C.~P.,  2010b, \mn@doi [\apj] {10.1088/0004-637X/719/1/831}, \href {https://ui.adsabs.harvard.edu/abs/2010ApJ...719..831P} {719, 831}

\bibitem[\protect\citeauthoryear{{Peters} et~al.,}{{Peters} et~al.}{2017}]{Peters2017}
{Peters} T.,  et~al., 2017, \mn@doi [\mnras] {10.1093/mnras/stw3216}, \href {https://ui.adsabs.harvard.edu/abs/2017MNRAS.466.3293P} {466, 3293}

\bibitem[\protect\citeauthoryear{{Pineda} et~al.,}{{Pineda} et~al.}{2022}]{Pineda2022}
{Pineda} J.~E.,  et~al., 2022, \mn@doi [arXiv e-prints] {10.48550/arXiv.2205.03935}, \href {https://ui.adsabs.harvard.edu/abs/2022arXiv220503935P} {p. arXiv:2205.03935}

\bibitem[\protect\citeauthoryear{{Pouteau} et~al.,}{{Pouteau} et~al.}{2023}]{Pouteau2023}
{Pouteau} Y.,  et~al., 2023, \mn@doi [\aap] {10.1051/0004-6361/202244776}, \href {https://ui.adsabs.harvard.edu/abs/2023A&A...674A..76P} {674, A76}

\bibitem[\protect\citeauthoryear{{Qin} et~al.,}{{Qin} et~al.}{2022}]{ATOMSVIII}
{Qin} S.-L.,  et~al., 2022, \mn@doi [\mnras] {10.1093/mnras/stac219}, \href {https://ui.adsabs.harvard.edu/abs/2022MNRAS.511.3463Q} {511, 3463}

\bibitem[\protect\citeauthoryear{{Rathjen} et~al.,}{{Rathjen} et~al.}{2021}]{Rathjen2021}
{Rathjen} T.-E.,  et~al., 2021, \mn@doi [\mnras] {10.1093/mnras/stab900}, \href {https://ui.adsabs.harvard.edu/abs/2021MNRAS.504.1039R} {504, 1039}

\bibitem[\protect\citeauthoryear{{Rebolledo} et~al.,}{{Rebolledo} et~al.}{2020}]{Rebolledo2020}
{Rebolledo} D.,  et~al., 2020, \mn@doi [\apj] {10.3847/1538-4357/ab6d76}, \href {https://ui.adsabs.harvard.edu/abs/2020ApJ...891..113R} {891, 113}

\bibitem[\protect\citeauthoryear{{Redaelli}, {Bovino}, {Giannetti}, {Sabatini}, {Caselli}, {Wyrowski}, {Schleicher}  \& {Colombo}}{{Redaelli} et~al.}{2021}]{Redaelli2021}
{Redaelli} E.,  {Bovino} S.,  {Giannetti} A.,  {Sabatini} G.,  {Caselli} P.,  {Wyrowski} F.,  {Schleicher} D.~R.~G.,   {Colombo} D.,  2021, \mn@doi [\aap] {10.1051/0004-6361/202140694}, \href {https://ui.adsabs.harvard.edu/abs/2021A&A...650A.202R} {650, A202}

\bibitem[\protect\citeauthoryear{{Redaelli}, {Bovino}, {Sanhueza}, {Morii}, {Sabatini}, {Caselli}, {Giannetti}  \& {Li}}{{Redaelli} et~al.}{2022}]{Redaelli2022}
{Redaelli} E.,  {Bovino} S.,  {Sanhueza} P.,  {Morii} K.,  {Sabatini} G.,  {Caselli} P.,  {Giannetti} A.,   {Li} S.,  2022, \mn@doi [\apj] {10.3847/1538-4357/ac85b4}, \href {https://ui.adsabs.harvard.edu/abs/2022ApJ...936..169R} {936, 169}

\bibitem[\protect\citeauthoryear{{Riener}, {Kainulainen}, {Henshaw}, {Orkisz}, {Murray}  \& {Beuther}}{{Riener} et~al.}{2019}]{Riener2019}
{Riener} M.,  {Kainulainen} J.,  {Henshaw} J.~D.,  {Orkisz} J.~H.,  {Murray} C.~E.,   {Beuther} H.,  2019, \mn@doi [\aap] {10.1051/0004-6361/201935519}, \href {https://ui.adsabs.harvard.edu/abs/2019A&A...628A..78R} {628, A78}

\bibitem[\protect\citeauthoryear{{Rosolowsky}, {Pineda}, {Kauffmann}  \& {Goodman}}{{Rosolowsky} et~al.}{2008}]{Rosolowsky2008}
{Rosolowsky} E.~W.,  {Pineda} J.~E.,  {Kauffmann} J.,   {Goodman} A.~A.,  2008, \mn@doi [\apj] {10.1086/587685}, \href {https://ui.adsabs.harvard.edu/abs/2008ApJ...679.1338R} {679, 1338}

\bibitem[\protect\citeauthoryear{{Saha}, {Maheswar}, {Ojha}, {Baug}  \& {Neha}}{{Saha} et~al.}{2022}]{Saha2022}
{Saha} P.,  {Maheswar} G.,  {Ojha} D.~K.,  {Baug} T.,   {Neha} S.,  2022, \mn@doi [\mnras] {10.1093/mnrasl/slac074}, \href {https://ui.adsabs.harvard.edu/abs/2022MNRAS.515L..67S} {515, L67}

\bibitem[\protect\citeauthoryear{{Salda{\~n}o}, {Rubio}, {Cappa}  \& {G{\'o}mez}}{{Salda{\~n}o} et~al.}{2019}]{Saldano2019}
{Salda{\~n}o} H.~P.,  {Rubio} M.,  {Cappa} C.~E.,   {G{\'o}mez} M.,  2019, \mn@doi [\mnras] {10.1093/mnras/stz1409}, \href {https://ui.adsabs.harvard.edu/abs/2019MNRAS.487.2881S} {487, 2881}

\bibitem[\protect\citeauthoryear{{Sanhueza}, {Jackson}, {Foster}, {Garay}, {Silva}  \& {Finn}}{{Sanhueza} et~al.}{2012}]{Sanhueza2012}
{Sanhueza} P.,  {Jackson} J.~M.,  {Foster} J.~B.,  {Garay} G.,  {Silva} A.,   {Finn} S.~C.,  2012, \mn@doi [\apj] {10.1088/0004-637X/756/1/60}, \href {https://ui.adsabs.harvard.edu/abs/2012ApJ...756...60S} {756, 60}

\bibitem[\protect\citeauthoryear{{Sanhueza} et~al.,}{{Sanhueza} et~al.}{2019}]{Sanhueza2019}
{Sanhueza} P.,  et~al., 2019, \mn@doi [\apj] {10.3847/1538-4357/ab45e9}, \href {https://ui.adsabs.harvard.edu/abs/2019ApJ...886..102S} {886, 102}

\bibitem[\protect\citeauthoryear{{Schneider} et~al.,}{{Schneider} et~al.}{2020}]{Schneider2020}
{Schneider} N.,  et~al., 2020, \mn@doi [\pasp] {10.1088/1538-3873/aba840}, \href {https://ui.adsabs.harvard.edu/abs/2020PASP..132j4301S} {132, 104301}

\bibitem[\protect\citeauthoryear{{Schuller} et~al.,}{{Schuller} et~al.}{2009}]{Schuller2009}
{Schuller} F.,  et~al., 2009, \mn@doi [\aap] {10.1051/0004-6361/200811568}, \href {https://ui.adsabs.harvard.edu/abs/2009A&A...504..415S} {504, 415}

\bibitem[\protect\citeauthoryear{{Schuller} et~al.,}{{Schuller} et~al.}{2017}]{Schuller2017}
{Schuller} F.,  et~al., 2017, \mn@doi [\aap] {10.1051/0004-6361/201628933}, \href {https://ui.adsabs.harvard.edu/abs/2017A&A...601A.124S} {601, A124}

\bibitem[\protect\citeauthoryear{{Sewi{\l}o}, {Churchwell}, {Kurtz}, {Goss}  \& {Hofner}}{{Sewi{\l}o} et~al.}{2011}]{Sewilo2011}
{Sewi{\l}o} M.,  {Churchwell} E.,  {Kurtz} S.,  {Goss} W.~M.,   {Hofner} P.,  2011, \mn@doi [\apjs] {10.1088/0067-0049/194/2/44}, \href {https://ui.adsabs.harvard.edu/abs/2011ApJS..194...44S} {194, 44}

\bibitem[\protect\citeauthoryear{{Shen} et~al.,}{{Shen} et~al.}{2024}]{Shen2024}
{Shen} H.,  et~al., 2024, \mn@doi [A&A] {10.1051/0004-6361/202347972}, 689, A140

\bibitem[\protect\citeauthoryear{{Sherman}}{{Sherman}}{2012}]{Sherman2012}
{Sherman} R.~A.,  2012, \mn@doi [\apj] {10.1088/0004-637X/760/1/58}, \href {https://ui.adsabs.harvard.edu/abs/2012ApJ...760...58S} {760, 58}

\bibitem[\protect\citeauthoryear{{St{\'e}phan}, {Schilke}, {Le Bourlot}, {Schmiedeke}, {Choudhury}, {Godard}  \& {S{\'a}nchez-Monge}}{{St{\'e}phan} et~al.}{2018}]{Stephan2018}
{St{\'e}phan} G.,  {Schilke} P.,  {Le Bourlot} J.,  {Schmiedeke} A.,  {Choudhury} R.,  {Godard} B.,   {S{\'a}nchez-Monge} {\'A}.,  2018, \mn@doi [\aap] {10.1051/0004-6361/201730639}, \href {https://ui.adsabs.harvard.edu/abs/2018A&A...617A..60S} {617, A60}

\bibitem[\protect\citeauthoryear{{Sternberg}, {Hoffmann}  \& {Pauldrach}}{{Sternberg} et~al.}{2003}]{Sternberg2003}
{Sternberg} A.,  {Hoffmann} T.~L.,   {Pauldrach} A.~W.~A.,  2003, \mn@doi [\apj] {10.1086/379506}, \href {https://ui.adsabs.harvard.edu/abs/2003ApJ...599.1333S} {599, 1333}

\bibitem[\protect\citeauthoryear{{Suin}, {Zavagno}, {Colman}, {Hennebelle}, {Verliat}  \& {Russeil}}{{Suin} et~al.}{2024}]{Suin2024}
{Suin} P.,  {Zavagno} A.,  {Colman} T.,  {Hennebelle} P.,  {Verliat} A.,   {Russeil} D.,  2024, \mn@doi [\aap] {10.1051/0004-6361/202347527}, \href {https://ui.adsabs.harvard.edu/abs/2024A&A...682A..76S} {682, A76}

\bibitem[\protect\citeauthoryear{{Svoboda} et~al.,}{{Svoboda} et~al.}{2019}]{Svoboda2019}
{Svoboda} B.~E.,  et~al., 2019, \mn@doi [\apj] {10.3847/1538-4357/ab40ca}, \href {https://ui.adsabs.harvard.edu/abs/2019ApJ...886...36S} {886, 36}

\bibitem[\protect\citeauthoryear{{Tahani} et~al.,}{{Tahani} et~al.}{2023}]{Tahani2023}
{Tahani} M.,  et~al., 2023, \mn@doi [\apj] {10.3847/1538-4357/acac81}, \href {https://ui.adsabs.harvard.edu/abs/2023ApJ...944..139T} {944, 139}

\bibitem[\protect\citeauthoryear{{Takekoshi} et~al.,}{{Takekoshi} et~al.}{2019}]{Takekoshi2019}
{Takekoshi} T.,  et~al., 2019, \mn@doi [\apj] {10.3847/1538-4357/ab3a55}, \href {https://ui.adsabs.harvard.edu/abs/2019ApJ...883..156T} {883, 156}

\bibitem[\protect\citeauthoryear{{Takemura} et~al.,}{{Takemura} et~al.}{2023}]{Takemura2023}
{Takemura} H.,  et~al., 2023, \mn@doi [\apjs] {10.3847/1538-4365/aca4d4}, \href {https://ui.adsabs.harvard.edu/abs/2023ApJS..264...35T} {264, 35}

\bibitem[\protect\citeauthoryear{{Tenorio-Tagle}}{{Tenorio-Tagle}}{1979}]{Tenorio1979}
{Tenorio-Tagle} G.,  1979, \aap, \href {https://ui.adsabs.harvard.edu/abs/1979A&A....71...59T} {71, 59}

\bibitem[\protect\citeauthoryear{{Thompson}, {Urquhart}, {Moore}  \& {Morgan}}{{Thompson} et~al.}{2012}]{Thompson2012}
{Thompson} M.~A.,  {Urquhart} J.~S.,  {Moore} T.~J.~T.,   {Morgan} L.~K.,  2012, \mn@doi [\mnras] {10.1111/j.1365-2966.2011.20315.x}, \href {https://ui.adsabs.harvard.edu/abs/2012MNRAS.421..408T} {421, 408}

\bibitem[\protect\citeauthoryear{{Tremblin} et~al.,}{{Tremblin} et~al.}{2014a}]{Tremblin2014}
{Tremblin} P.,  et~al., 2014a, \mn@doi [\aap] {10.1051/0004-6361/201322700}, \href {https://ui.adsabs.harvard.edu/abs/2014A&A...564A.106T} {564, A106}

\bibitem[\protect\citeauthoryear{{Tremblin} et~al.,}{{Tremblin} et~al.}{2014b}]{Tremblin2014b}
{Tremblin} P.,  et~al., 2014b, \mn@doi [\aap] {10.1051/0004-6361/201423959}, \href {https://ui.adsabs.harvard.edu/abs/2014A&A...568A...4T} {568, A4}

\bibitem[\protect\citeauthoryear{{Trevi{\~n}o-Morales} et~al.,}{{Trevi{\~n}o-Morales} et~al.}{2016}]{Trevino2016}
{Trevi{\~n}o-Morales} S.~P.,  et~al., 2016, \mn@doi [\aap] {10.1051/0004-6361/201628899}, \href {https://ui.adsabs.harvard.edu/abs/2016A&A...593L..12T} {593, L12}

\bibitem[\protect\citeauthoryear{{Urquhart} et~al.,}{{Urquhart} et~al.}{2014}]{Urquhart2014}
{Urquhart} J.~S.,  et~al., 2014, \mn@doi [\aap] {10.1051/0004-6361/201424126}, \href {https://ui.adsabs.harvard.edu/abs/2014A&A...568A..41U} {568, A41}

\bibitem[\protect\citeauthoryear{{Urquhart} et~al.,}{{Urquhart} et~al.}{2018}]{Urquhart2018}
{Urquhart} J.~S.,  et~al., 2018, \mn@doi [\mnras] {10.1093/mnras/stx2258}, \href {https://ui.adsabs.harvard.edu/abs/2018MNRAS.473.1059U} {473, 1059}

\bibitem[\protect\citeauthoryear{{Urquhart} et~al.,}{{Urquhart} et~al.}{2019}]{Urquhart2019}
{Urquhart} J.~S.,  et~al., 2019, \mn@doi [\mnras] {10.1093/mnras/stz154}, \href {https://ui.adsabs.harvard.edu/abs/2019MNRAS.484.4444U} {484, 4444}

\bibitem[\protect\citeauthoryear{{Urquhart} et~al.,}{{Urquhart} et~al.}{2022}]{Urquhart2022}
{Urquhart} J.~S.,  et~al., 2022, \mn@doi [\mnras] {10.1093/mnras/stab3511}, \href {https://ui.adsabs.harvard.edu/abs/2022MNRAS.510.3389U} {510, 3389}

\bibitem[\protect\citeauthoryear{{V{\'a}zquez-Semadeni}, {Gonz{\'a}lez-Samaniego}  \& {Col{\'\i}n}}{{V{\'a}zquez-Semadeni} et~al.}{2017}]{Vazquez-Semadeni2017}
{V{\'a}zquez-Semadeni} E.,  {Gonz{\'a}lez-Samaniego} A.,   {Col{\'\i}n} P.,  2017, \mn@doi [\mnras] {10.1093/mnras/stw3229}, \href {https://ui.adsabs.harvard.edu/abs/2017MNRAS.467.1313V} {467, 1313}

\bibitem[\protect\citeauthoryear{{Veena}, {Vig}, {Tej}, {Kantharia}  \& {Ghosh}}{{Veena} et~al.}{2017}]{veena2017}
{Veena} V.~S.,  {Vig} S.,  {Tej} A.,  {Kantharia} N.~G.,   {Ghosh} S.~K.,  2017, \mn@doi [\mnras] {10.1093/mnras/stw2997}, \href {https://ui.adsabs.harvard.edu/abs/2017MNRAS.465.4219V} {465, 4219}

\bibitem[\protect\citeauthoryear{{Verliat}, {Hennebelle}, {Gonz{\'a}lez}, {Lee}  \& {Geen}}{{Verliat} et~al.}{2022}]{Verliat2022}
{Verliat} A.,  {Hennebelle} P.,  {Gonz{\'a}lez} M.,  {Lee} Y.-N.,   {Geen} S.,  2022, \mn@doi [\aap] {10.1051/0004-6361/202141765}, \href {https://ui.adsabs.harvard.edu/abs/2022A&A...663A...6V} {663, A6}

\bibitem[\protect\citeauthoryear{{Vink}, {de Koter}  \& {Lamers}}{{Vink} et~al.}{2001}]{Vink2001}
{Vink} J.~S.,  {de Koter} A.,   {Lamers} H.~J.~G.~L.~M.,  2001, \mn@doi [\aap] {10.1051/0004-6361:20010127}, \href {https://ui.adsabs.harvard.edu/abs/2001A&A...369..574V} {369, 574}

\bibitem[\protect\citeauthoryear{{Walch}}{{Walch}}{2023}]{Walch2023}
{Walch} S.,  2023, in Physics and Chemistry of Star Formation: The Dynamical ISM Across Time and Spatial Scales. p.~97

\bibitem[\protect\citeauthoryear{{Walch}, {Whitworth}, {Bisbas}, {W{\"u}nsch}  \& {Hubber}}{{Walch} et~al.}{2013}]{Walch2013}
{Walch} S.,  {Whitworth} A.~P.,  {Bisbas} T.~G.,  {W{\"u}nsch} R.,   {Hubber} D.~A.,  2013, \mn@doi [\mnras] {10.1093/mnras/stt1115}, \href {https://ui.adsabs.harvard.edu/abs/2013MNRAS.435..917W} {435, 917}

\bibitem[\protect\citeauthoryear{{Wall}, {Mac Low}, {McMillan}, {Klessen}, {Portegies Zwart}  \& {Pellegrino}}{{Wall} et~al.}{2020}]{Wall2020}
{Wall} J.~E.,  {Mac Low} M.-M.,  {McMillan} S. L.~W.,  {Klessen} R.~S.,  {Portegies Zwart} S.,   {Pellegrino} A.,  2020, \mn@doi [\apj] {10.3847/1538-4357/abc011}, \href {https://ui.adsabs.harvard.edu/abs/2020ApJ...904..192W} {904, 192}

\bibitem[\protect\citeauthoryear{{Wang} et~al.,}{{Wang} et~al.}{2016}]{Wang2016}
{Wang} Y.,  et~al., 2016, \mn@doi [\aap] {10.1051/0004-6361/201526637}, \href {https://ui.adsabs.harvard.edu/abs/2016A&A...587A..69W} {587, A69}

\bibitem[\protect\citeauthoryear{{Wells} et~al.,}{{Wells} et~al.}{2024}]{Wells2024}
{Wells} M.~R.~A.,  et~al., 2024, \mn@doi [\aap] {10.1051/0004-6361/202449794}, \href {https://ui.adsabs.harvard.edu/abs/2024A&A...690A.185W} {690, A185}

\bibitem[\protect\citeauthoryear{{Whitworth}, {Bhattal}, {Chapman}, {Disney}  \& {Turner}}{{Whitworth} et~al.}{1994}]{Whitworth1994}
{Whitworth} A.~P.,  {Bhattal} A.~S.,  {Chapman} S.~J.,  {Disney} M.~J.,   {Turner} J.~A.,  1994, \mn@doi [\mnras] {10.1093/mnras/268.1.291}, \href {https://ui.adsabs.harvard.edu/abs/1994MNRAS.268..291W} {268, 291}

\bibitem[\protect\citeauthoryear{{Wong} et~al.,}{{Wong} et~al.}{2022}]{Wong2022}
{Wong} T.,  et~al., 2022, \mn@doi [\apj] {10.3847/1538-4357/ac723a}, \href {https://ui.adsabs.harvard.edu/abs/2022ApJ...932...47W} {932, 47}

\bibitem[\protect\citeauthoryear{{Wood} \& {Churchwell}}{{Wood} \& {Churchwell}}{1989a}]{Wood1989}
{Wood} D. O.~S.,  {Churchwell} E.,  1989a, \mn@doi [\apjs] {10.1086/191329}, \href {https://ui.adsabs.harvard.edu/abs/1989ApJS...69..831W} {69, 831}

\bibitem[\protect\citeauthoryear{{Wood} \& {Churchwell}}{{Wood} \& {Churchwell}}{1989b}]{Wood1989b}
{Wood} D. O.~S.,  {Churchwell} E.,  1989b, \mn@doi [\apj] {10.1086/167390}, \href {https://ui.adsabs.harvard.edu/abs/1989ApJ...340..265W} {340, 265}

\bibitem[\protect\citeauthoryear{{Xie}, {Mundy}, {Vogel}  \& {Hofner}}{{Xie} et~al.}{1996}]{Xie1996}
{Xie} T.,  {Mundy} L.~G.,  {Vogel} S.~N.,   {Hofner} P.,  1996, \mn@doi [\apjl] {10.1086/310401}, \href {https://ui.adsabs.harvard.edu/abs/1996ApJ...473L.131X} {473, L131}

\bibitem[\protect\citeauthoryear{{Xu} et~al.,}{{Xu} et~al.}{2024a}]{Xu2024b}
{Xu} F.,  et~al., 2024a, \mn@doi [Research in Astronomy and Astrophysics] {10.1088/1674-4527/ad3dc3}, \href {https://ui.adsabs.harvard.edu/abs/2024RAA....24f5011X} {24, 065011}

\bibitem[\protect\citeauthoryear{{Xu} et~al.,}{{Xu} et~al.}{2024b}]{Xu2024}
{Xu} F.,  et~al., 2024b, \mn@doi [\apjs] {10.3847/1538-4365/acfee5}, \href {https://ui.adsabs.harvard.edu/abs/2024ApJS..270....9X} {270, 9}

\bibitem[\protect\citeauthoryear{{Yang}, {Thompson}, {Tian}, {Bihr}, {Beuther}  \& {Hindson}}{{Yang} et~al.}{2019}]{Yang2019}
{Yang} A.~Y.,  {Thompson} M.~A.,  {Tian} W.~W.,  {Bihr} S.,  {Beuther} H.,   {Hindson} L.,  2019, \mn@doi [\mnras] {10.1093/mnras/sty2811}, \href {https://ui.adsabs.harvard.edu/abs/2019MNRAS.482.2681Y} {482, 2681}

\bibitem[\protect\citeauthoryear{{Yang} et~al.,}{{Yang} et~al.}{2021}]{Yang2021}
{Yang} A.~Y.,  et~al., 2021, \mn@doi [\aap] {10.1051/0004-6361/202038608}, \href {https://ui.adsabs.harvard.edu/abs/2021A&A...645A.110Y} {645, A110}

\bibitem[\protect\citeauthoryear{{Yi}, {Lee}, {Kim}, {Liu}, {Lim}, {Tatematsu}  \& {Jcmt Large Program ''Scope'' Collaboration}}{{Yi} et~al.}{2021}]{Yi2021}
{Yi} H.-W.,  {Lee} J.-E.,  {Kim} K.-T.,  {Liu} T.,  {Lim} B.,  {Tatematsu} K.,   {Jcmt Large Program ''Scope'' Collaboration} 2021, \mn@doi [\apjs] {10.3847/1538-4365/abec4a}, \href {https://ui.adsabs.harvard.edu/abs/2021ApJS..254...14Y} {254, 14}

\bibitem[\protect\citeauthoryear{{Zakardjian} et~al.,}{{Zakardjian} et~al.}{2023}]{Zakardjian2023}
{Zakardjian} A.,  et~al., 2023, \mn@doi [\aap] {10.1051/0004-6361/202244520}, \href {https://ui.adsabs.harvard.edu/abs/2023A&A...678A.171Z} {678, A171}

\bibitem[\protect\citeauthoryear{{Zavagno}, {Deharveng}, {Comer{\'o}n}, {Brand}, {Massi}, {Caplan}  \& {Russeil}}{{Zavagno} et~al.}{2006}]{Zavagno2006}
{Zavagno} A.,  {Deharveng} L.,  {Comer{\'o}n} F.,  {Brand} J.,  {Massi} F.,  {Caplan} J.,   {Russeil} D.,  2006, \mn@doi [\aap] {10.1051/0004-6361:20053952}, \href {https://ui.adsabs.harvard.edu/abs/2006A&A...446..171Z} {446, 171}

\bibitem[\protect\citeauthoryear{{Zavagno} et~al.,}{{Zavagno} et~al.}{2010}]{Zavagno2010}
{Zavagno} A.,  et~al., 2010, \mn@doi [\aap] {10.1051/0004-6361/201014587}, \href {https://ui.adsabs.harvard.edu/abs/2010A&A...518L.101Z} {518, L101}

\bibitem[\protect\citeauthoryear{{Zhang} et~al.,}{{Zhang} et~al.}{2020}]{PaperI}
{Zhang} S.,  et~al., 2020, \mn@doi [\aap] {10.1051/0004-6361/201936792}, \href {https://ui.adsabs.harvard.edu/abs/2020A&A...637A..40Z} {637, A40}

\bibitem[\protect\citeauthoryear{{Zhang} et~al.,}{{Zhang} et~al.}{2021}]{PaperII}
{Zhang} S.,  et~al., 2021, \mn@doi [\aap] {10.1051/0004-6361/202038421}, \href {https://ui.adsabs.harvard.edu/abs/2021A&A...646A..25Z} {646, A25}

\bibitem[\protect\citeauthoryear{{Zhang} et~al.,}{{Zhang} et~al.}{2022}]{ATOMSIV}
{Zhang} C.,  et~al., 2022, \mn@doi [\mnras] {10.1093/mnras/stab2733}, \href {https://ui.adsabs.harvard.edu/abs/2022MNRAS.510.4998Z} {510, 4998}

\bibitem[\protect\citeauthoryear{{Zhang} et~al.,}{{Zhang} et~al.}{2023a}]{ATOMSXIII}
{Zhang} S.,  et~al., 2023a, \mn@doi [\mnras] {10.1093/mnras/stad011}, \href {https://ui.adsabs.harvard.edu/abs/2023MNRAS.520..322Z} {520, 322}

\bibitem[\protect\citeauthoryear{{Zhang} et~al.,}{{Zhang} et~al.}{2023b}]{ATOMSXIV}
{Zhang} C.,  et~al., 2023b, \mn@doi [\mnras] {10.1093/mnras/stad190}, \href {https://ui.adsabs.harvard.edu/abs/2023MNRAS.520.3245Z} {520, 3245}

\bibitem[\protect\citeauthoryear{{Zhou} et~al.,}{{Zhou} et~al.}{2020}]{Zhou2020}
{Zhou} J.,  et~al., 2020, \mn@doi [\apj] {10.3847/1538-4357/ab94c0}, \href {https://ui.adsabs.harvard.edu/abs/2020ApJ...897...74Z} {897, 74}

\bibitem[\protect\citeauthoryear{{Zhou} et~al.,}{{Zhou} et~al.}{2021}]{ATOMSVI}
{Zhou} J.-W.,  et~al., 2021, \mn@doi [\mnras] {10.1093/mnras/stab2801}, \href {https://ui.adsabs.harvard.edu/abs/2021MNRAS.508.4639Z} {508, 4639}

\bibitem[\protect\citeauthoryear{{Zhou} et~al.,}{{Zhou} et~al.}{2022}]{ATOMSXI}
{Zhou} J.-W.,  et~al., 2022, \mn@doi [\mnras] {10.1093/mnras/stac1735}, \href {https://ui.adsabs.harvard.edu/abs/2022MNRAS.514.6038Z} {514, 6038}

\bibitem[\protect\citeauthoryear{{Zhou} et~al.,}{{Zhou} et~al.}{2024}]{Zhou2023}
{Zhou} J.~W.,  et~al., 2024, \mn@doi [\aap] {10.1051/0004-6361/202348108}, \href {https://ui.adsabs.harvard.edu/abs/2024A&A...682A.173Z} {682, A173}

\bibitem[\protect\citeauthoryear{{de la Fuente}, {Porras}, {Trinidad}, {Kurtz}, {Kemp}, {Tafoya}, {Franco}  \& {Rodr{\'\i}guez-Rico}}{{de la Fuente} et~al.}{2020}]{delaFuente2020}
{de la Fuente} E.,  {Porras} A.,  {Trinidad} M.~A.,  {Kurtz} S.~E.,  {Kemp} S.~N.,  {Tafoya} D.,  {Franco} J.,   {Rodr{\'\i}guez-Rico} C.,  2020, \mn@doi [\mnras] {10.1093/mnras/stz3482}, \href {https://ui.adsabs.harvard.edu/abs/2020MNRAS.492..895D} {492, 895}

\bibitem[\protect\citeauthoryear{{van Buren}, {Mac Low}, {Wood}  \& {Churchwell}}{{van Buren} et~al.}{1990}]{vanBuren1990}
{van Buren} D.,  {Mac Low} M.-M.,  {Wood} D. O.~S.,   {Churchwell} E.,  1990, \mn@doi [\apj] {10.1086/168645}, \href {https://ui.adsabs.harvard.edu/abs/1990ApJ...353..570V} {353, 570}

\bibitem[\protect\citeauthoryear{{van der Tak}, {Black}, {Sch{\"o}ier}, {Jansen}  \& {van Dishoeck}}{{van der Tak} et~al.}{2007}]{vanderTak2007}
{van der Tak} F.~F.~S.,  {Black} J.~H.,  {Sch{\"o}ier} F.~L.,  {Jansen} D.~J.,   {van Dishoeck} E.~F.,  2007, \mn@doi [\aap] {10.1051/0004-6361:20066820}, \href {https://ui.adsabs.harvard.edu/abs/2007A&A...468..627V} {468, 627}

\makeatother
\end{thebibliography}









\bsp	
\label{lastpage}
\clearpage
\noindent
$^{1}$Kavli Institute for Astronomy and Astrophysics, Peking University, 5 Yiheyuan Road, Haidian District, Beĳing 100871, China\\
$^{2}$Departamento de Astronom\'{i}a, Universidad de Chile, Las Condes, 7591245 Santiago, Chile\\
$^{3}$Shanghai Astronomical Observatory, Chinese Academy of Sciences, 80 Nandan Road, Shanghai 200030, China\\
$^{4}$Key Laboratory for Research in Galaxies and Cosmology, Chinese Academy of Sciences, 80 Nandan Road, Shanghai 200030, China\\
$^{5}$Aix Marseille Univ, CNRS, CNES, LAM, Marseille, France\\
$^{6}$Institut Universitaire de France, Paris, 1 rue Descartes, 75231 Paris Cedex 05, France\\
$^{7}$Chinese Academy of Sciences South America Center for Astronomy, National Astronomical Observatories, CAS, Beĳing 100101, China\\
$^{8}$School of Physics and Astronomy, Yunnan University, Kunming, 650091, China\\
$^{9}$Department of Astronomy, School of Physics, Peking University, Beijing 100871, China\\
$^{10}$National Astronomical Observatory of Japan, National Institutes of Natural Sciences, 2-21-1 Osawa, Mitaka, Tokyo 181-8588, Japan\\
$^{11}$Astronomical Science Program, The Graduate University for Advanced Studies, SOKENDAI, 2-21-1 Osawa, Mitaka, Tokyo 181-8588, Japan\\
$^{12}$Indian Institute of Astrophysics, II Block, Koramangala, Bengaluru 560034, India\\
$^{13}$Max-Planck-Institut f\"{u}r Radioastronomie, Auf dem H\"{u}gel 69, 53121 Bonn, Germany\\
$^{14}$Max Planck Institute for Astronomy, Konigstuhl 17, D-69117 Heidelberg, Germany\\
$^{15}$Jet Propulsion Laboratory, California Institute of Technology, 4800 Oak Grove Drive, Pasadena CA 91109, USA\\
$^{16}$School of Physics and Astronomy, Sun Yat-sen University, 2 Daxue Road, Tangjia, Zhuhai, Guangdong Province, People's Republic of China\\
$^{17}$Department of Mathematical Sciences, University of South Africa, Cnr Christian de Wet Rd and Pioneer Avenue, Florida Park, 1709, Roodepoort, South Africa\\
$^{18}$Department of Physics and Astronomy, Faculty of Physical Sciences, University of Nigeria, Carver Building, 1 University Road, Nsukka 410001, Nigeria\\
$^{19}$Korea Astronomy and Space Science Institute, 776 Daedeokdaero, Yuseong-gu, Daejeon 34055, Republic of Korea\\
$^{20}$University of Science and Technology, Korea (UST), 217 Gajeong-ro, Yuseong-gu, Daejeon 34113, Republic of Korea\\
$^{21}$Physical Research Laboratory, Navrangpura, Ahmedabad - 380 009, India\\

\end{document}